\documentclass[12pt]{article}
\usepackage{url}
\usepackage{amssymb,amsmath, amsfonts}
\usepackage{cases}
\usepackage{graphicx}
\usepackage{natbib}
\usepackage[includeheadfoot,left=1.in,right=1.in,top=1.in,bottom=1.in,bindingoffset=0.in,nohead]{geometry}
\usepackage[onehalfspacing]{setspace}
\usepackage{booktabs}
\usepackage{appendix}
\usepackage[pdfborder={0 0 0}]{hyperref}
\usepackage{float}
\usepackage{xcolor}
\usepackage{subcaption}
\usepackage[justification=centering]{caption}
\usepackage{multirow} 
\usepackage{enumerate}

\begin{document}
	  
\date{}
\title{On Robust Inference in Time Series Regression}
\author{Richard T. Baillie\\Michigan State University \\King's College, University of London
\and Francis X. Diebold\\University of Pennsylvania \\and NBER \vspace{3mm}
\and George Kapetanios\\King's College, University of London
\and Kun Ho Kim\\Concordia University \vspace{3mm}
\and Aaron Mora\\University of South Carolina}

\smallskip

\smallskip

\maketitle

\begin{center}
	
 This draft, \today
 
 \footnotesize  Earlier drafts at \url{https://arxiv.org/abs/2203.04080}

\end{center}

\thispagestyle{empty}

\smallskip

\smallskip

{\footnotesize \begin{spacing}{1}
\noindent \textbf{Abstract:}  Least squares regression with heteroskedasticity consistent standard errors (``OLS-HC regression") has proved very useful in cross section environments. However, several major difficulties, which are generally overlooked, must be confronted when transferring the HC technology to time series environments via heteroskedasticity and autocorrelation consistent standard errors (``OLS-HAC regression"). First, in plausible time-series environments, OLS parameter estimates can be inconsistent, so that OLS-HAC inference fails even asymptotically. Second, most economic time series have  autocorrelation, which renders OLS parameter estimates  inefficient. Third,  autocorrelation similarly renders conditional predictions based on OLS parameter estimates   inefficient. Finally, the structure of popular HAC covariance matrix estimators is ill-suited for capturing the autoregressive autocorrelation typically present in economic time series, which produces large size distortions and reduced power in HAC-based hypothesis testing, in all but the largest samples.  We show that all four problems are largely avoided by the use of a simple and easily-implemented dynamic regression procedure, which we call DURBIN.  We demonstrate the advantages of DURBIN with detailed simulations covering a range of practical issues.

\scriptsize

\smallskip

\smallskip
 
\noindent \textbf{Acknowledgments:} For detailed comments we are greatly indebted to the editor, co-editor, and two referees. In addition we gratefully acknowledge useful discussions and/or comments from Rob Engle, Domenico Giannone, Jim Hamilton, Daniel Lewis, Nour Meddahi, Ulrich M{\"u}ller, Serena Ng, Lasse Pedersen, Pierre Perron, Peter Phillips, Mikkel Plagborg-Moller, Peter Schmidt, George Tauchen,  Tim Volgelsang, Mark Watson, Ken West, and Jeff Wooldridge.  We are also grateful to seminar participants at Michigan State University and the University of Pennsylvania, and  conference participants at the 2022 NBER Summer Institute, the 2023 joint meetings of the Royal Economic Society and Scottish Economic Society, and the 2023 Copenhagen Conference on Advances in Financial Econometrics. All remaining errors or misunderstandings are ours alone.

\smallskip

\smallskip


\smallskip

\smallskip	

\noindent \textbf{Contact:}  {baillie@msu.edu}, {fdiebold@sas.upenn.edu}, {george.kapetanios@kcl.ac.uk}, {kunho.kim@concordia.ca}, \\moramela@mailbox.sc.edu

\smallskip

\smallskip

\end{spacing}
}

\clearpage


{\normalsize \setcounter{page}{1} \thispagestyle{empty} }

\section{Introduction}

For nearly a century, regression with heteroskedastic and/or autocorrelated disturbances has featured prominently in empirical economics research. For many decades, attention centered on modeling the heteroskedasticity or autocorrelation in the context of feasible generalized least squares (FGLS) estimation.

The dominant estimation approach in recent decades, however, is ordinary least squares (OLS) with standard errors adjusted to achieve valid asymptotic inference without taking a stand on the form of heteroskedasticity or autocorrelation. The idea traces to the classic contribution of \cite{White1980}, who considered OLS regression with heteroskedasticity consistent (HC) standard errors (``OLS-HC regression") in cross-sectional environments, where sample sizes are typically very large, little or no information is available regarding the form of any possible heteroskedasticity, and serial correlation is irrelevant. In such environments HC standard errors are appropriate and justly emphasized (e.g. \citet{angrist2008mostly}).

In an elegant extension, \cite{newey1987simple} generalize White's estimator from cross sections to time series, with possible heteroskedasticity \textit{and} serial correlation, by replacing White's covariance matrix estimator with an appropriate time-series analog based on an estimator of a spectral density at frequency zero.\footnote{The Newey-West estimator collapses to the \cite{White1980} estimator if serial correlation is absent, but appropriately incorporates  serial correlation in the calculation of robust standard errors when serial correlation is present.} Such OLS regression with  heteroskedasticity and autocorrelation consistent (HAC) standard errors (``OLS-HAC regression") has become extremely popular in time series environments.

 In this paper we argue, however, that, in contrast to cross section OLS-HC regression, time series OLS-HAC regression as typically implemented is likely to be problematic, for a variety of reasons:

\begin{enumerate}
\item \label{one} In plausible time-series environments, OLS parameter estimates can be inconsistent, so that OLS-HAC inference fails even asymptotically.


\noindent And moreover, even when OLS parameter estimates are consistent:

\item \label{two} OLS parameter estimates can be highly inefficient in the presence of  serial correlation, compared to estimators that account for the serial correlation.
 
\item \label{three} OLS-HAC regression discards valuable predictive information in serially-correlated disturbances and hence produces sub-optimal (inefficient) forecasts, whereas accurate out-of-sample prediction is often a central concern in time series econometrics.

\item \label{four} Newey-West-style HAC covariance matrix estimators are ill-suited for capturing the autoregressive autocorrelation typically present in economic time series, which can produce large size distortions, and large power reductions even when the size is not distorted.
\end{enumerate}

\noindent Claim \ref{one} is not widely appreciated, with the exception of \cite{glsp}, whose results and approach complement ours.\footnote{By now parts of our paper and theirs are entangled. A preliminary version of our paper was presented at the 2016 NBER-NSF Time Series Conference at Columbia University. Our first-draft working paper was released in March 2022, with no knowledge of their work-in-progress. Their first-draft working paper was released in September 2022, with knowledge of ours. Our second-draft working paper was released in June 2022, with knowledge of theirs. This third draft of our paper was released on \today.} Claim \ref{two} is well known, but its importance in finite samples is ignored when using OLS-HAC regression. Claim \ref{three} is obvious, but again ignored when using OLS-HAC regression. Claim \ref{four} is appreciated and has motivated several important refinements of the Newey-West HAC covariance matrix estimator (e.g.,\cite{andrews1991heteroskedasticity}, \cite{kiefer2002heteroskedasticity}, \cite{lazarus2018har}), as well as use of spectral density estimators that differ from the Newey-West lag-window estimator (e.g., \cite{muller2014hac}). However, those refinements have been only partially successful.

Against the background of the above claims \ref{one}-\ref{four}, which we will substantiate in detail, we proceed to make a constructive contribution. We propose an alternative to OLS-HAC regression based on so-called ``Durbin regressions" (\cite{durbin1970testing}). Working in a very general environment that includes most dynamic specifications of interest as special cases, we show that the new procedure simultaneously addresses claims \ref{one}-\ref{four} above. Indeed, the Durbin regression procedure performs well in all situations, dominating the traditional OLS-HAC and FGLS procedures.

Our paper proceeds as follows. In section \ref{frameworks} we introduce the basic data-generating process and estimators, including not only traditional OLS-HAC regression and our Durbin regression, but also traditional FGLS and a recently-proposed modified FGLS procedure. In section \ref{generalized} we present a generalized modeling framework.  In section \ref{mc} we present extensive simulation evidence. We conclude in section \ref{concl}, and we present supplementary results in three Appendices.

\section{Data Generating Process and Estimators}
\label{frameworks}

Traditional OLS-HAC regression  focuses exclusively on OLS parameter estimation, assuming consistency and surrendering on efficiency. But, as we emphasize in this section, even OLS consistency cannot be assumed without significant loss of generality. Moreover, aspects of the consistency and efficiency of OLS and various competitors, under various conditions, are nuanced and not widely appreciated. Hence in this section we begin by reviewing aspects of OLS consistency and efficiency in comparison to competitors -- in particular, a new procedure that we propose based on \cite{durbin1970testing} regressions, a new modified FGLS procedure, and traditional FGLS -- in a sequence of progressively-richer dynamic environments.

\subsection{Data-Generating Process} 

We start with the standard data-generating process (DGP) in the OLS-HAC regression literature,%
\begin{equation}
	 \label{dgp}
	y_{t}=x_{t}^{\prime}\beta+u_{t},%
\end{equation}
where $t=1,2,...,T$, $\beta$ \ is a $k$-vector of parameters, $x_{t}$ is a $k$-vector of covariance-stationary covariates and $u_{t}$ is a scalar covariance-stationary disturbance  with  $E(u_{t}u_{t}')=\sigma^{2}\Omega$.\footnote{Because $u_{t}$ is covariance-stationary, it can be serially correlated and/or conditionally heteroskedastic. In this paper we emphasize serial correlation exclusively, because serial correlation is the unique feature of time-series data relative to cross-section data. Cross sections do of course sometimes have a spatial dimension and therefore a natural ordering in space if not in time, and spatial correlation has recently begun to receive attention from a HAC estimation perspective, as in \cite{muller2022spatial}. Spatial HAC estimation is, however, beyond the scope of this paper.}  DGP (\ref{dgp}) is usually augmented with conditions such that OLS is consistent. Then the econometrician generally aims to provide standard error corrections that enable asymptotically valid  inference. Note that such OLS-HAC regression  involves just a static regression of $y_t$ on $x_{t}$, basically imported directly from cross-sectional micro-econometrics, with dynamics allowed only through $u_{t}$. We will later argue that such a framework is uncompelling in time-series environments, but it is the industry standard in OLS-HAC regression, so we maintain it for now.

Crucial insights will flow from adopting a starting point that allows for significant generality regarding possible relationships between $x_{t}$ and $u_{t}$. In particular, consider the Wold representation of the Gaussian vector process $z_{t}=(x_{t}^{\prime},u_{t})^{\prime}$,
\begin{equation} \label{wold}
z_{t}=\sum_{i=0}^{\infty}\Xi_{i}\varepsilon_{t-i}.%
\end{equation}
The  coefficient matrices are $\Xi_{0}=I$ and 
\[
\Xi_{i}=%
\begin{pmatrix}
\xi_{x,i} & \xi_{xu,i}\\
\xi_{ux,i}^{\prime} & \xi_{u,i}%
\end{pmatrix}
,
\]
and ${\varepsilon}_{t}=({\varepsilon}_{x,t}^{\prime},{\varepsilon}_{u,t})^{\prime}$ is a vector white noise innovation process with $E({\varepsilon}_{t})=0$  and  $\ E({\varepsilon}_{t}{\varepsilon}_{s}^{\prime})=0$ for $s \ne t$, and contemporaneous covariance matrix $E(\varepsilon_{t}\varepsilon_{t}^{\prime})=\Sigma$, where

\[
\Sigma=%
\begin{pmatrix}
\Sigma_{x} & \Sigma_{xu}\\
\Sigma_{xu}^{\prime} & \sigma_{u}%
\end{pmatrix}
.
\]
Under mild regularity conditions, the infinite vector moving-average representation \eqref{wold} is equivalent to the infinite vector-autoregressive (VAR) representation\footnote{Such regularity conditions include assumptions on the rate of decline of $\left\Vert \Xi_{i}\right\Vert $ towards zero as $i\rightarrow\infty$, for suitable norm $\left\Vert .\right\Vert $, to control the persistence of the process and to avoid phenomena such as long memory that complicate the analysis. For details see, e.g., \cite{dav2002} and references cited therein.}
\begin{equation}
z_{t}=\sum_{i=1}^{\infty}\Psi_{i}z_{t-i}+\varepsilon_{t}, \label{zz}%
\end{equation}
where%
\[
\Psi_{i}=%
\begin{pmatrix}
{\Psi}_{x,i} & {\Psi}_{xu,i}\\
{\Psi}_{ux,i} & \psi_{u,i}%
\end{pmatrix}
.
\]
This setting encompasses a variety of DGPs, and we will consider the consistency and efficiency properties of different estimators under various restrictions imposed on (\ref{zz}).

We now  proceed to consider various estimation strategies that may be appropriate in the environment given by (\ref{dgp}) and (\ref{zz}). 

\subsection{OLS Parameter Estimation and HAC Covariance Matrix Estimation}
\label{est1}

The  OLS estimator of the regression parameter  is of course
\[
\hat{\beta}=(X^{\prime}X)^{-1}X^{\prime}Y.
\]
If  $\Omega=I$,  the limiting distribution of the OLS estimator is
\[
T^{1/2}\left(  \hat{\beta}_{OLS}-\beta\right)  \rightarrow N\left( 0,\sigma^{2}Q^{-1}\right),
\]
where  $Q=p\lim_{T\rightarrow\infty}\left(  T^{-1}X^{\prime}X\right)$.

Based on the $VAR$ representation (\ref{zz}), we define ``block diagonality" ($BD$) as holding when  ${\Psi}_{ux,i}={\Psi}_{xu,i}=0$, for all $i$, and ${\Sigma}_{xu}=0$. The BD condition implies strong exogeneity, namely that $E(u_s|x_t)=0$ for all $s$ and $t$.\footnote{Strong exogeneity is sometimes called strict exogeneity.} In the $BD$ environment OLS is consistent but asymptotically inefficient, with limiting distribution
\[
T^{1/2}(\hat{\beta}_{OLS}-\beta)\rightarrow N(0,V),
\]
where  $V=Q^{-1}\Omega Q^{-1}$. The key object in $V$ \ is $\Omega$, which is the spectrum of $\ x_{t}u_{t}$ at frequency zero. HAC inference estimates $V$ using 
\[
\widehat{V}=Q^{-1}\widehat{\Omega}Q^{-1},
\]
where $\widehat{\Omega}$ is a consistent estimator of $\Omega$, so that  $\widehat{V}$ is consistent for $V$. Different choices for $\widehat{\Omega}$ therefore define different HAC covariance matrix estimators and are the main issue in implementing OLS-HAC regression, as we discuss subsequently in section \ref{op_OLS}.

\subsection{FGLS Estimation}

If condition BD holds, and if the matrix $\Omega$ is known, then GLS is a consistent and  asymptotically efficient estimator of $\beta$. However,  $\Omega$ is almost always unknown, in which case  attention turns to FGLS as defined by \cite{amemiya1973generalized}, which is again both consistent and asymptotically efficient provided that condition BD holds.\footnote{Recent contributions to the FGLS literature include  \cite{RW2017} for heteroskedastic environments, and  \cite{kapetanios2016semiparametric} for dynamic environments.}

The OLS-HAC regression literature was historically motivated by environments where OLS is consistent for $\beta$, but where condition $BD$ simultaneously fails in such a way that FGLS is inconsistent. Such situations are possible, and we will discuss a classic such situation \citep{hansen1980forward} at some length  in section \ref{HH} below, but they are by no means the only or the most important possibility. Indeed there is much more to investigate when $BD$ fails, as emphasized in the insightful work of \cite{glsp}.

We now consider an alternative estimation procedure that avoids the above discussed OLS-HAC and FGLS complications and \textit{always} delivers consistent (and sometimes fully efficient) estimates of $\beta$, together with reliable asymptotic inference.

\subsection{Durbin Estimation and its Relatives}
\label{1b} 

A natural third approach to estimation and inference, which we will argue is generally preferable to both OLS-HAC and FGLS, is based on the ``\cite{durbin1970testing} regression",  given by%
\begin{equation}
y_{t}={x}_{t}^{\prime}{\beta}+\sum_{j=1}^{\infty}\phi_{j}y_{t-j}+\sum
_{j=1}^{\infty}{x}_{t-j}^{\prime}{\gamma}_{j}+{\varepsilon}_{y,t}, \label{durbin1}
\end{equation}
where ${\varepsilon}_{y,t}$ is serially uncorrelated, and uncorrelated with $y_{t-j}$ and ${x}_{t-j}$ for all $j$. The Durbin regression ``cleans out" disturbance dynamics by its direct inclusion of $y_{t-j}$ and ${x}_{t-j}$, so that standard OLS estimation and inference are trustworthy.  We refer to the Durbin regression, and the associated estimator of $\beta$, as DURBIN. Crucially, note well that the DGP remains \eqref{dgp} and \eqref{zz}; DURBIN is simply a certain procedure (regression) that can be implemented on data from that DGP, just as OLS and FGLS are certain procedures that can be implemented on data from that DGP.

Operationally,  it is of course  necessary to use a finite order   approximation  to the infinite order DURBIN regression  \eqref{durbin1}, 
\begin{equation}
	y_{t}={x}_{t}^{\prime}{\beta}+\sum_{j=1}^{p}\phi_{j}y_{t-j}+\sum
	_{j=1}^{p}{x}_{t-j}^{\prime}{\gamma}_{j}+{\varepsilon}_{y,t}, \label{durbin2}
\end{equation}
with  finite lag order $p$ selected  using a data based  procedure, typically an information criterion, and increasing at a suitable rate. The theoretical validity of such a procedure for producing  valid asymptotic estimation and inference is well known (see, e.g., \cite{lewis1985prediction} or \cite{hannan2012statistical}), and we shall have more to say about it when we later implement DURBIN in the  simulations of section \ref{mc}.

We can also write the finite order DURBIN approximation as 
\begin{equation}\label{dr}
y_{t}=\sum_{j=1}^{p}\phi_{j}y_{t-j}+\sum_{i=1}^{k}\beta_{i}%
x_{i,t}+\sum_{j=1}^{p}\sum_{i=1}^{k}\gamma_{i,j}x_{i,t-j}+{\varepsilon}_{y,t},
\end{equation}
which emphasizes the extent of the parameterization and lag structure. We will later explore in greater detail the relationship between the DGP given by  (\ref{dgp}) and (\ref{zz}) and the DURBIN regression \eqref{dr}, which is effectively one equation of a VAR and appears to be the originator of the autoregressive distributed lag (ADL) model  model, which is widely used in empirical econometric work.

An estimator closely related to DURBIN, recently proposed by \cite{glsp}, is a variation on FGLS. We refer to it as FGLS-D (short for ``FGLS-Durbin"). While FGLS uses a first-stage OLS regression, FGLS-D uses a first-stage DURBIN regression (\ref{dr}). Under $BD$, it follows that FGLS-D is also efficient. However, when $BD$ does not hold, FGLS-D may not be efficient or even consistent, while DURBIN remains consistent.

Given that condition $BD$ may not hold, it is important to consider the implications of its violation for the various methods of estimation and inference. To see the effects of the various sub-conditions embedded in condition $BD$, we will relax it in sequential stages. First, we impose only that ${\Psi}_{ux,i}=0$ for all $i$ and that ${\Sigma}_{xu}=0$, so that $x$ is weakly exogenous (that is, $E(u_s|x_t)=0$, $\forall s, t {\le}s$) but not strongly exogenous.\footnote{Weak exogeneity is sometimes called predeterminedness. See \cite{anna}.}  ${x}_{t}$ now depends on lags of $u_{t}$, but not vice versa. We refer to this restriction as $GEXOG$ (``GLS exogeneity"). Clearly, OLS is now inconsistent, as is FGLS, which uses OLS residuals, while FGLS-D remains consistent and efficient. Importantly, DURBIN remains consistent, even if not fully efficient, throughout.

Second, we impose only ${\Sigma}_{xu}=0$, so that $x$ is neither strongly nor weakly exogenous. We denote this condition by $EBD$ (``error variance block diagonal"). $u_{t}$ now depends on lags of ${x}_{t}$, and the finite-ordered FGLS autoregression for $u_{t}$ is no longer valid. Therefore, neither FGLS nor FGLS-D is consistent. DURBIN, however, remains consistent under $EBD$, and moreover it is also efficient. 
  
To see the consistency and efficiency of DURBIN under  EBD, note that, using (\ref{zz}) and $u_{t}=y_{t}-{x}_{t}^{\prime}{\beta}$, we can write (\ref{dgp}) as
\begin{align}
	y_{t}  &  ={x}_{t}^{\prime}{\beta}+u_t \label{gettoardl}\\
 &  ={x}_{t}^{\prime}{\beta}+\sum_{j=1}^{\infty}\psi_{u,j}u_{t-j}+\sum_{j=1}^{\infty} {\Psi}_{xu,j} {x}_{t-j}+{\varepsilon}_{u,t} \nonumber\\
&  ={x}_{t}^{\prime}{\beta}+\sum_{j=1}^{\infty}\psi_{u,j}\left(  y_{t-j}-{x}_{t-j}^{\prime}{\beta}\right)  +\sum_{j=1}^{\infty} {\Psi}_{xu,j} {x}_{t-j} +{\varepsilon}_{u,t}\nonumber\\
&  ={x}_{t}^{\prime}{\beta}+\sum_{j=1}^{\infty}\psi_{u,j}y_{t-j}+\sum_{j=1}^{\infty}{x}_{t-j}^{\prime}\left(  {\Psi}_{xu,j}'-\psi_{u,j}{\beta}\right)  +{\varepsilon}_{u,t}.\nonumber
\end{align}
Noting that the relationship ${\gamma}_{j}={\Psi}_{xu,j}-\psi_{u,j}{\beta}$ gives a one-to-one mapping between ${\gamma}_{j}$ and ${\Psi}_{xu,j}$, given values for $\psi_{u,j}$ and ${\beta}$, we immediately obtain efficiency for DURBIN.\footnote{Of course, if ${\Psi}_{xu,j}=0$, then DURBIN, which estimates ${\gamma}_{j}$, is over-parameterized, providing a simple argument showing that DURBIN is inefficient under $GEXOG$.}

Finally, we impose no restrictions at all, in which case all methods become inconsistent and the use of instrumentation appears to be the only way forward.

In summary, OLS requires stronger conditions for consistency than the FGLS variants. The FGLS variants, in turn, require stronger conditions for consistency than DURBIN. Hence, overall, DURBIN has attractive consistency features in comparison with OLS and the FGLS variants. On the other hand, when the FGLS variants are consistent, they are also fully efficient. We shall see how such trade-offs resolve themselves in the simulations of section \ref{mc} below.

\section{A Generalized Data-Generating Process} \label{generalized}
\label{general} 
 
We now move from the basic DGP \eqref{dgp} to a generalized version that subsumes all cases of interest. 

\subsection{Data-Generating Process}

Henceforth we work with the data-generating process given by
\begin{equation}
y_{t}=\sum_{j=1}^{p}\phi_{j}y_{t-j}+\sum_{i=1}^{k}\beta_{i}x_{i,t}+\sum
_{j=1}^{p}\sum_{i=1}^{k}\gamma_{i,j}x_{i,t-j}+u_{t}. \label{dgp0}%
\end{equation}
We emphasize that  ${u}_{t}$ may also be a dynamic process, to allow, for example, for missing covariates. In particular, we continue to allow  $(x_{t}^{\prime},u_{t})^{\prime}$ to follow the the vector moving average \eqref{wold}, or equivalently, the vector autoregression (\ref{zz}).   This generalized DGP covers most linear dynamic relationships of conceivable interest. We use $NDY$ (``no dynamics in $y$") to refer to the restriction imposed on the generalized DGP (\ref{dgp0}) to get the basic DGP (\ref{dgp}), namely $\phi_{j}={\gamma}_{i,j}=0$ $\, \forall i,j$.


We also emphasize that \eqref{dgp0} is now the data generating process, and various regressions could be fit to its data realizations in various attempts at estimation and inference for $\beta$. One such regression, for example, is FGLS. Clearly the use of FGLS in environments characterized by the generalized DGP \eqref{dgp0} accounts only for  ${x}_{t}^{\prime}{\beta}$ and therefore ignores all terms involving lags, resulting in misspecification of the conditional mean part of the fitted regression. That is, the only way lagged information is used in FGLS is through estimation of the error covariance matrix, which neglects the problem of misspecification of the conditional mean. 
 
DURBIN is another such regression that can be fit to the generalized DGP \eqref{dgp0}. Indeed DURBIN can perfectly accommodate the generalized DGP, because, in precise parallel to  (\ref{gettoardl}), we have
\begin{align*}
y_{t}  &  =x_{t}^{\prime}\beta+\sum_{j=1}^{\infty}\lambda_{j}y_{t-j}%
+\sum_{j=1}^{\infty}x_{t-j}^{\prime}\theta_{j}+{u}_{t}\\
&  =x_{t}^{\prime}\beta+\sum_{j=1}^{\infty}\lambda_{j}y_{t-j}+\sum
_{j=1}^{\infty}x_{t-j}^{\prime}\theta_{j}+\sum_{j=1}^{\infty}\psi_{u,j}%
u_{t-j}+\sum_{j=1}^{\infty}x_{t-j}^{\prime}\psi_{xu,j}+{\varepsilon}_{u,t}\\
&  =x_{t}^{\prime}\beta+\sum_{j=1}^{\infty}\lambda_{j}y_{t-j}+\sum
_{j=1}^{\infty}x_{t-j}^{\prime}\theta_{j}\\
&  +\sum_{j=1}^{\infty}\psi_{u,j}\left(  y_{t-j}-x_{t-j}^{\prime}\beta
-\sum_{s=1}^{\infty}\lambda_{s}y_{t-j-s}-\sum_{s=1}^{\infty}x_{t-j-s}^{\prime
}\theta_{s}\right)  +\sum_{j=1}^{\infty}x_{t-j}^{\prime}\Psi_{xu,j}%
+{\varepsilon}_{u,t}\\
&  =x_{t}^{\prime}\beta+\sum_{j=0}^{\infty}\left(  \lambda_{j}+\psi_{u,j}%
-\sum_{s=1}^{j-1}\psi_{u,s}\lambda_{j-s}\right)  y_{t-j} +\sum_{j=1}^{\infty
}x_{t-j}^{\prime}\left(  \Psi_{xu,j}-\psi_{u,j}\beta-\sum_{s=1}^{j-1}%
\psi_{u,s}\theta_{j-s}\right)  +\varepsilon_{u,t},
\end{align*}
 which is a DURBIN regression with
\begin{align*}
	\phi_{j}  &  =\lambda_{j}+\psi_{u,j}-\sum_{s=1}^{j-1}\psi_{u,s}\lambda_{j-s}\\
	{\gamma}_{j}  &  ={\psi}_{xu,j}-\psi_{u,j}{\beta-}\sum_{s=1}^{j-1}\psi
	_{u,s}{\theta}_{j-s}.
\end{align*}

The above relationships between the parameters of the generalized DGP (\ref{dgp0}) and the DURBIN regression (\ref{dr}) also  show that the generalized DGP is so richly parameterized that not all parameters are identified through estimation of (\ref{dr}) alone. $\beta$ is always identified and consistently estimable via DURBIN, however, even in cases where OLS, FGLS, and FGLS-D are inconsistent.

\begin{table}[t]
\caption{Estimator Consistency and Efficiency Under Various Conditions}%
\label{ssset}%
\vspace*{-5mm}
\par
\begin{center}%
\begin{tabular}
[c]{ccccc}\hline\hline
\underline{Restriction} & \multicolumn{4}{c}{\underline{Estimator}}\\
& OLS & DURBIN & FGLS & FGLS-D\\\hline
$NDY+BD$ & \checkmark$\times$ & \checkmark$\times$ & \checkmark\checkmark &
\checkmark\checkmark\\
$NDY+GEXOG$ & $\times$ $\times$ & \checkmark$\times$ & $\times$ $\times$ &
\checkmark\checkmark\\
$NDY+EBD$ & $\times$ $\times$ & \checkmark\checkmark & $\times$ $\times$ &
$\times$ $\times$\\
$EBD$ & $\times$ $\times$ & \checkmark\checkmark & $\times$ $\times$ &
$\times$ $\times$\\
$None$ & $\times$ $\times$ & $\times$ $\times$ & $\times$ $\times$ & $\times$
$\times$\\\hline\hline
\end{tabular} 
\end{center}
\par
\begin{spacing}{1.0}  \noindent  \footnotesize  Notes: We show the consistency and efficiency properties of various estimators under various restrictions on the generalized DGP \eqref{dgp0} with $({x_t}, u_t)'$ governed by \eqref{zz}. In each cell of the table, the first checkmark, or lack thereof, relates to consistency  and the second to efficiency.
\end{spacing}	
\end{table}

\subsection{Estimator Comparisons}

In Table \ref{ssset} we summarize the consistency and efficiency properties of all estimators in the leading environments that we have considered, all of which are  specializations of the generalized DGP given by \eqref{dgp0} and \eqref{zz}. Table 1 makes clear the important trade-off between the occasional efficiency of FGLS/FGLS-D and the robust consistency of DURBIN. That is, although FGLS is sometimes efficient when DURBIN is not (under $NDY+BD$ and $NDY+GEXOG$), DURBIN is \textit{always} at least consistent, and FGLS is not. 

Indeed the $EBD$ row of Table \ref{ssset} is starkly revealing, as for example it includes simple and natural DGPs like
\[
y_{t}=x_{t}\beta+\phi y_{t-1}+x_{t-1}\gamma+u_{t}.
\]
The conventional FGLS procedure would be to regress $y_{t}$ on $x_{t}$, and then to regress the residuals on lagged residuals, thereby obtaining the Cochrane-Orcutt filter to apply to the $y_{t}$ and $x_{t}$ series. One strongly suspects, and our subsequent simulations in section \ref{mc} show clearly, that FGLS will perform poorly in this environment unless $ \gamma \approx \beta \phi$, in which case the DGP reduces (approximately) to just a static regression of $y_{t}$ on $x_{t}$ with $AR(1)$ disturbances.\footnote{\label{cf}The restriction  $ \gamma \approx \beta \phi$  is known as the common factor restriction.}

\subsection{Hausman Tests}

Table \ref{ssset} also highlights the potential usefulness of tests for validity of the various restrictions. If for example, one ``knew" that $NDY +GEXOG $ held, then FGLS or FGLS-D would be fully appealing estimators (consistent and efficient) whereas DURBIN would be less appealing (consistent but not efficient). Alternatively, if one knew that instead $NDY + EBD$ held, then FGLS or FGLS-D would be highly \textit{un}appealing (inconsistent) whereas DURBIN would be fully appealing (consistent and efficient).

Hausman tests are available, as follows. Clearly, restrictions on the parameters of (\ref{zz}) determine the comparative desirability of alternative methods of estimation and inference for $\beta$. The key restriction is $BD$.
Under the null hypothesis that $BD$ holds with $u$ serially correlated, OLS is consistent but not efficient, while FGLS is both consistent and efficient. Under the alternative  hypothesis that $BD$ fails, OLS and FGLS are generally both inconsistent but have different limits, which depend on  the parameters of (\ref{zz}). As a result, Hausman tests can be used.

In particular, one may wish to query whether $\beta_1=\beta_2$, where
\[
\label{sim}E(y_{t}|x_{t})=x_{t}^{\prime}\beta_1
\]
and
\[
E(y_{t}|x_{t},x_{t-1},y_{t-1}, x_{t-2},y_{t-2},...)=x_{t}^{\prime}{\beta_2} +
\sum_{j=1}^{\infty}\phi_{j}y_{t-j}+\sum_{j=1}^{\infty}x_{t-j}^{\prime}%
\gamma_{j}.
\]
Under  the null hypothesis, FGLS should be used. Otherwise one should consider using FGLS-D or DURBIN if one is interested in ${\beta_2}$ as would typically be the case, or consider using OLS if for some reason $\beta_1$ is of interest.

Overall, however, we find it preferable simply to use DURBIN under all circumstances, unless there is some compelling reason to do otherwise. There are three reasons:
\begin{enumerate}
	\item An acceptable HAC estimator of the variance of the OLS estimator may not be available when implementing a Hausman test.  Indeed the poor performance of OLS-HAC is the theme of this paper.

\item \label{point2} As regards consistent/efficient estimation, it will be clear from the simulation results in section \ref{mc} below that the MSE cost of using DURBIN when a more efficient estimator is available (i.e., when $BD$ or at least $GEXOG$ holds) is generally small, whereas the MSE cost of \textit{not} using DURBIN can be very large when neither $BD$ nor $GEXOG$ holds. 

\item \label{point3}  As regards consistent inference, it will also be clear from the simulation results in section \ref{mc} below that DURBIN-based inference performs well in all circumstances that we investigate, both in terms of test size and power, in contrast to all other methods that we consider, where inference often fails.

\end{enumerate}

\noindent We will shortly turn to the extensive simulation results alluded to in points \ref{point2} and \ref{point3} above, but first we briefly consider DURBIN vs other estimation approaches in the important context of predictive inference.

\subsection{Predictive Inference}
\label{HH}

As is clear from Table \ref{ssset}, OLS is rarely consistent in time-series situations of interest.  One case where  OLS \textit{is} consistent and  simultaneously FGLS is inconsistent involves multi-step forecast evaluation, where one tests whether a forecast $x_{t}$ is unbiased for $y_{t+k}$. That is, one tests whether
\[
E(y_{t+k}|x_{t})=x_{t},
\]
for $k \ge 1$.
 
One of the earliest analyses of this problem was by \cite{hansen1980forward}, where $y_{t+k}$ represented the $k$-period-ahead spot exchange rate and $x_{t}$ represented the current $k$-period forward rate. The null hypothesis of $\beta=1$ implies moving-average disturbances, producing a violation of strong exogeneity while nevertheless satisfying weak exogeneity. \cite{hansen1980forward} recognized that FGLS can be inconsistent in such a situation, whereas OLS remains consistent, albeit inefficient. They recognized, moreover, that the OLS standard error was inconsistent and therefore required a ``correction" -- and OLS-HAC was born.

Note however, that DURBIN is also perfectly applicable in the Hansen-Hodrick environment, delivering not only consistent standard errors, but also efficient as opposed to merely consistent parameter estimates.\footnote{For a full empirical analysis, see \cite{baillie2023new}.} In particular, under the null of unbiasedness, the error term,
\[
u_{t+k}=y_{t+k}-x_{t},
\]
satisfies $Cov(u_{t+j}u_{t})=0$ for $j>k$, which implies that $u_{t+k}$ can be represented by an $MA(k-1)$ process. Hence we can write
\begin{equation}
y_{t+k}=x_{t}\beta+\theta(L)\varepsilon_{t}, \label{hh1}%
\end{equation}
where $\varepsilon_{t}$ is a white noise process and $\theta(L)$ is a polynomial in the lag operator of order $k- 1$. 

Conceptually, equation (\ref{hh1}) is merely a restricted DURBIN model, because on using the filter
$\theta(L)^{-1}$ we obtain
\begin{equation}
\left\{  \theta(L)^{-1}y_{t+k}\right\}  =\beta\left\{  \theta(L)^{-1}%
x_{t}\right\}  +\varepsilon_{t+k}. \label{xxxxx}%
\end{equation}
The filtered explanatory variable is uncorrelated with current and future innovations, $\varepsilon_{t+k}$, so that estimation of equation \eqref{xxxxx} by OLS will produce consistent and asymptotically efficient estimates of the regression parameters. In practice it is convenient to use the approximation $\theta(L)^{-1}\approx\pi(L)$, where $\pi(L)=\left(  1-\pi_{1}L-...-\pi_{p}L^{p}\right)$ is a $p$th-order lag-operator polynomial with all roots outside the unit circle. DURBIN will then be
\begin{equation}
 \label{r1}
\pi(L)y_{t+k}=\beta\pi(L)x_{t}+\varepsilon_{t+k},
\end{equation}
which is a restricted version of the generalized DGP  (\ref{dgp0}) and can also be estimated by restricted OLS.

\section{Simulation Evidence on Estimation and Testing}
\label{mc}

In this section we examine, via simulation, the sampling properties of the various estimators, the properties of forecasts that use those estimated parameters, and crucially, the size and power of associated hypothesis tests.

\subsection{Simulation Design}
\label{design}

The main simulation results will comprise four data generation processes that impose different assumptions on the generalized DGP given by  (\ref{dgp0}) and (\ref{zz}):

\begin{enumerate}
	\item Autoregressive Disturbances, AR(1) ($NDY + BD$)%
	\begin{align}
		\label{dgp_ar1}y_{t}  &  =\beta x_{t}+u_{t}\nonumber\\%
		\begin{pmatrix}
			x_{t}\\
			u_{t}%
		\end{pmatrix}
		&  =
		\begin{pmatrix}
			0.7 & 0\\
			0 & \rho
		\end{pmatrix}
		\begin{pmatrix}
			x_{t-1}\\
			u_{t-1}%
		\end{pmatrix}
		+
		\begin{pmatrix}
			{\varepsilon}_{x,t}\\
			{\varepsilon}_{u,t}%
		\end{pmatrix}
	\end{align}

	\item Triangular vector autoregression (VAR) on (\ref{zz}) ($NDY + GEXOG$)
	\begin{align}
		\label{dgp_tri_var}y_{t}  &  =\beta x_{t}+u_{t}\nonumber\\%
		\begin{pmatrix}
			x_{t}\\
			u_{t}%
		\end{pmatrix}
		&  =
		\begin{pmatrix}
			\psi_{11} & \psi_{12}\\
			0 & \psi_{22}%
		\end{pmatrix}
		\begin{pmatrix}
			x_{t-1}\\
			u_{t-1}%
		\end{pmatrix}
		+
		\begin{pmatrix}
			{\varepsilon}_{x,t}\\
			{\varepsilon}_{u,t}%
		\end{pmatrix}
	\end{align}

	\item Unrestricted VAR on (\ref{zz}) ($NDY + EBD$)
	\begin{align}
		\label{dgp_unr_var}y_{t}  &  =\beta x_{t}+u_{t}\nonumber\\%
		\begin{pmatrix}
			x_{t}\\
			u_{t}%
		\end{pmatrix}
		&  =
		\begin{pmatrix}
			\psi_{11} & \psi_{12}\\
			\psi_{21} & \psi_{22}%
		\end{pmatrix}
		\begin{pmatrix}
			x_{t-1}\\
			u_{t-1}%
		\end{pmatrix}
		+
		\begin{pmatrix}
			{\varepsilon}_{x,t}\\
			{\varepsilon}_{u,t}%
		\end{pmatrix}
	\end{align}

	\item Dynamic Regression ($EBD$)
	\begin{align}
		\label{dgp_dyn_reg}y_{t}  &  =\beta x_{t}+\rho y_{t-1}-0.5 x_{t-1}%
		+u_{t}\nonumber\\%
		\begin{pmatrix}
			x_{t}\\
			u_{t}%
		\end{pmatrix}
		&  =
		\begin{pmatrix}
			0.7 & 0\\
			0 & 0
		\end{pmatrix}
		\begin{pmatrix}
			x_{t-1}\\
			u_{t-1}%
		\end{pmatrix}
		+
		\begin{pmatrix}
			{\varepsilon}_{x,t}\\
			{\varepsilon}_{u,t}%
		\end{pmatrix}.
	\end{align}
	
\end{enumerate}

In all cases, $(\varepsilon_{x,t} , \varepsilon_{u,t})^{\prime}\sim iid N(0,I)$ with $t=1,...,T$. We explore $T \in\{ 50, 200, 600, 2500\}$, which also spans the relevant range for macroeconomics, where structural change and other considerations tend to keep sample spans to roughly ``the most recent fifty years"; that is, sample sizes of 50 years, 200 quarters, 600 months, or approximately 2500 weeks. Including $T=2500$ also lets us check our Monte Carlo results against known large-sample results.

The autoregressive DGP in (\ref{dgp_ar1}) matches the design in \cite{lazarus2018har}. We explore $\rho\in\{ 0, .3, .5, .7, .9, .95, .99\}$, which spans the relevant range for economics. All $\rho$ values are positive, as economic time series are generally positively serially correlated, and they range from white noise to the very strong serial correlation often of relevance in macroeconomic series. Including the white noise case ($\rho=0$) allows us to check our Monte Carlo results  against known results for the iid case.

In the simulations for the triangular VAR DGP in (\ref{dgp_tri_var}), we consider the following values for the matrix $\Psi$:
\[
\Psi_{1}=%
\begin{pmatrix}
	0.4 & 0.7\\
	0 & 0.5
\end{pmatrix}
 \quad\quad\Psi_{1}^{*}=%
\begin{pmatrix}
	0.4 & 0.7\\
	0 & 0.6
\end{pmatrix}.
\]
$\Psi_{1}^*$ has a larger leading eigenvalue than $\Psi_{1}$  (0.6 versus 0.5) and hence exhibits stronger autoregressive features. 
 
For the unrestricted VAR DGP in (\ref{dgp_unr_var}), we consider the following values:
\[
\Psi_{2}=%
\begin{pmatrix}
	0.4 & 0.7\\
	0.3 & 0.5
\end{pmatrix}
 \quad\quad\Psi_{2}^{*}=%
\begin{pmatrix}
	0.4 & 0.7\\
	0.3 & 0.6
\end{pmatrix}
\]
As before, $\Psi_{2}^{*}$ was selected to be similar to $\Psi_{2}$ but with a larger leading eigenvalue (0.97 for $\Psi_{2}^{*}$ and 0.91 for $\Psi_{2}$).

For the dynamic regression DGP in (\ref{dgp_dyn_reg}), we consider various parameter values for the the coefficient on $y_{t-1}$, namely  $\rho\in\{0, 0.5, 0.7, 0.95\}$. When $\rho=0.5$, the common factor restriction introduced in footnote \ref{cf} holds, in which case we expect FGLS and FGLS-D to perform well.

For all DGPs in our simulations we perform 10,000 Monte Carlo replications. We simulate exact realizations of $x$ and $u$ by drawing $x_{0}$ and $u_{0}$ from their stationary distribution at each Monte Carlo replication, and we use common random numbers whenever appropriate.

\subsection{Operational Considerations}
\label{operational}

Next, we detail operational matters relating to our implementation of the various estimators we use in our simulations.

\subsubsection{OLS-HAC}
\label{op_OLS}

OLS-HAC estimation proceeds from the approach previously outlined in section \ref{est1}; namely
\[
T^{1/2}(\hat{\beta}_{OLS}-\beta)\rightarrow N(0,V),
\]
where $V=Q^{-1}\Omega Q^{-1}$ and
\[
\Omega=\sum_{\tau=-\infty}^{\infty}\Gamma(\tau)
\]
where $\Gamma(\tau)=cov(x_{t}u_{t},x_{t-\tau}u_{t-\tau}),$ and $\tau=0,\pm1,...$

The key object in $V$ is $\Omega$, the spectrum of $xu$ at frequency zero. The OLS-HAC approach uses
\[
\widehat{V}=Q^{-1}\widehat{\Omega}Q^{-1},
\]
where $\widehat{\Omega}$ is a consistent estimator of $\Omega$ and hence $\widehat{V}$ delivers a consistent estimator of $V$.

A large literature on consistent estimation of $\Omega$ can be traced back to at least \cite{hansen1980forward}. The most popular approach is due to \cite{newey1987simple}, who propose lag-window estimation with linearly-decreasing (Bartlett) lag window:
\begin{equation}
	\label{Main}\widehat{\Omega}=\left(  \frac{1}{T}\sum_{t=1}^{T}(x_{t}%
	x_{t}^{\prime})\hat{u}_{t}^{2}+\sum_{\tau=1}^{h}\left(  1-\frac{\tau}%
	{h+1}\right)  (\boldsymbol{\widehat{\Gamma}}_{\tau}%
	+\boldsymbol{\ \widehat{\Gamma}}_{-\tau})\right)  ,
\end{equation}
where
\[
\boldsymbol{\widehat{\Gamma}}_{\tau}=\frac{1}{T}\sum_{t=1}^{T}\hat{u}_{t}%
x_{t}x_{t-\tau}^{\prime}\hat{u}_{t-\tau},
\]
the $\hat{u}_{t}$ are OLS regression residuals, and $T$ is sample size. Indeed, many leading HAC estimators are of the form (\ref{Main}), distinguished only by their choice of truncation lag $h$.

We will explore several leading truncation lag choices, including:

\begin{enumerate}
	\item NW: Newey-West (\ref{Main}) with $h=\lceil(T/100)^{2/9}\rceil$. This $h$ choice is a standard textbook recommendation (e.g.,\cite{wooldridge2015introductory}).
	
	\item NW-A: Newey-West (\ref{Main}) with $h = \lceil0.75T^{1/3}\rceil$. This $h$ choice is also standard, arising when a formula in \cite{andrews1991heteroskedasticity} is specialized to the case of a first-order autoregression with coefficient 0.25.
	
	\item NW-LLSW: Newey-West (\ref{Main}) with $h=\lceil1.3T^{1/2}\rceil$, as proposed by \cite{lazarus2018har}. Its use of $T^{1/2}$ rather than $T^{2/9}$ or $T^{1/3}$ as in NW or NW-A, respectively, produces higher truncation lags. For example, if $T=200$, then NW selects $h=5$ but NW-LLSW selects $h=19$.
	
	\item NW-KV: Newey-West (\ref{Main}) with $h=T$, as proposed by \cite{kiefer2002heteroskedasticity}, which builds on \citet{kiefer2000simple}. Setting $h=T$ is of course the maximum possible truncation lag.
\end{enumerate}

We will also explore the \cite{muller2014hac} HAC estimator (we denote it by M), which is not in the Newey-West family. Instead, it is an orthogonal series estimator, that uses a type-II discrete cosine transform to produce an equally-weighted average of projections on cosines. The M estimator is:
\[
\widehat{\Omega}=\frac{1}{\nu} \sum_{j=1}^{\nu}\widehat{\Lambda}%
_{j}\widehat{\Lambda}_{j}^{\prime},
\]
where
\[
\widehat{\Lambda}_{j}= \sqrt{\frac{2}{T}}\sum_{t=1}^{T} (x_{t}\hat{u}_{t}%
)\cos\left(  \pi j \left(  \frac{t-1/2}{T}\right)  \right)
\]
\noindent The M truncation parameter, $\nu$, is the total number of cosines included in the average projection. \cite{lazarus2018har} suggest setting $\nu=\lfloor0.4T^{2/3} \rfloor$, producing the M-LLSW estimator.

\subsubsection{FGLS and FGLS-D}
\label{op_FGLS}

If the data follow the DGP in (\ref{dgp}), namely $y_{t}=x_{t}\beta+u_{t}$, and there exists a known lag operator polynomial (filter) $\Phi(L)$ that reduces $u_{t}$ to white noise $\varepsilon_{t}$ (i.e., $\Phi(L)u_{t}=\varepsilon_{t}$), then GLS estimation of $\beta$ is appropriate, and it  amounts to running an OLS regression on transformed data. Specifically, one regresses $\tilde{y}_{t}$ on $\tilde{x}_{t}$, where $\tilde{y}_{t}=\Phi(L)y_{t}$ and $\tilde{x}_{t}=\Phi(L)x_{t}$. 

In practice, however, $\Phi(L)$ is unknown and needs to be approximated. The FGLS estimator uses  $\phi(L)=1-\phi_{1}L-\phi_{2}L^{2}-...-\phi_{p}L^{p}$ and proceeds as follows:

\begin{enumerate}
	\item \label{one} Run an OLS regression of $y_{t}$ on $x_{t}$, and obtain the residuals $\hat{u}_{t}$. 
	\item \label{two} Fit an AR$(p)$ model to $\hat{u}_{t}$ (in particular, run an OLS regression of  $\hat{u}_{t}$ on $\hat{u}_{t-1},...,\hat{u}_{t-p}$, with $p$ selected by AIC or BIC), and obtain the coefficients $\hat{\phi}_{1}, ...,\hat{\phi}_{p}$. 
	\item Construct the transformed data,
\begin{align*}
	\tilde{x}_{t} &  =x_{t}-\hat{\phi}_{1} x_{t-1}- ...- \hat{\phi}_{p} x_{t-p}\\
	\tilde{y}_{t} &  =y_{t}-\hat{\phi}_{1} y_{t-1}- ...- \hat{\phi}_{p} y_{t-p}.
\end{align*}
\item Run an OLS regression of $\tilde{y}_{t}$ on $\tilde{x}_{t}$ to obtain the FGLS estimator of $\beta$.
\end{enumerate}

The FGLS-D estimator relies on a different first-stage procedure, replacing the  regressions in steps \ref{one} and \ref{two} above with a single  DURBIN regression, proceeding as follows:
	
\begin{enumerate}
	\item Run the  OLS DURBIN regression  (with $p$ selected by AIC or BIC),
\begin{align*}
	y_{t}  &  =\sum_{j=1}^{p}\varphi_{j}y_{t-j}+ \sum_{i=1}^{k} \beta_{i} x_{i,t}
	+\sum_{j=1}^{p} \sum_{i=1}^{k} \gamma_{i,j}x_{i,t-j}+\varepsilon_{t}.%
\end{align*}
\item Use the estimated coefficients on the lags of $y_{t}$, $\hat{\varphi}_{1},...,\hat{\varphi}_{p}$, to construct the transformed data,
\begin{align*}
	\tilde{x}_{t}  &  =x_{t}-\hat{\varphi}_{1} x_{t-1}- ...- \hat{\varphi}_{p}
	x_{t-p}\\
	\tilde{y}_{t}  &  =y_{t}-\hat{\varphi}_{1} y_{t-1}- ...- \hat{\varphi}_{p}
	y_{t-p}.
\end{align*}
\item Run the OLS regression of $\tilde{y}_{t}$ on $\tilde{x}_{t}$ to obtain the FGLS-D estimator of $\beta$. 

\end{enumerate}

\subsubsection{DURBIN}
\label{op_Durbin}

As previously noted, the DURBIN regression augments regression (\ref{dgp}) with lags of $y$ and $x$ to capture dynamics, very much in the spirit of an arbitrary equation in a vector autoregression, as suggested by \cite{durbin1970testing}.\footnote{Also related is the important recent work of \cite{Plagborg}, who study lag-augmented local projection estimators of impulse-response functions in vector autoregressions.} The $p^{\mathrm{th}}$-order DURBIN regression is
\begin{equation}
	y_{t}=\sum_{j=1}^{p}\phi_{j}y_{t-j}+\sum_{i=1}^{k}\beta_{i}x_{i,t}+\sum
	_{j=1}^{p}\sum_{i=1}^{k}\gamma_{i,j}x_{i,t-j}+\varepsilon_{t}, \label{dynreg}%
\end{equation}
which has $p+k+kp$ parameters.  

If $u_{t}$ in equation \eqref{dgp} is a finite-ordered AR$(p)$ process with $p$ known, then DURBIN holds exactly. In particular, we have\footnote{Note that DURBIN does \textit{not} impose the common factor restriction embedded in (\ref{comfac}), namely that $\gamma_{i,j}=\beta_{i}\phi_{j} ~ \forall i,j$, in which case DURBIN  coincides with FGLS. See \cite{Sargan1964} and \cite{HendryMizon1978}.}
\begin{align}
	y_{t}  &  =\sum_{j=1}^{p}\phi_{j}y_{t-j}+\sum_{i=1}^{k}\beta_{i}x_{i,t}%
	+\sum_{j=1}^{p}\sum_{i=1}^{k}\beta_{i}\phi_{j}x_{i,t-j}+\varepsilon
	_{t}\label{comfac}\\
	&  =\sum_{j=1}^{p}\phi_{j}y_{t-j}+\sum_{i=1}^{k}\beta_{i}x_{i,t}+\sum
	_{j=1}^{p}\sum_{i=1}^{k}\gamma_{i,j}x_{i,t-j}+\varepsilon_{t}.\nonumber
\end{align}
Hence the usual asymptotic inference is immediately available:
\begin{equation}
	T^{1/2}(\widehat{\vartheta}_{OLS}-\vartheta)\rightarrow N(0,Q^{-1}), \label{inf1}%
\end{equation}
where $\vartheta_{OLS}$ is the vector of DURBIN parameters,
\begin{equation}
	Q=plim\left(  T^{-1}\sum_{t=1}^{T}z_{t}z_{t}^{\prime}\right)  , \label{inf2}%
\end{equation}
and $z_{t}^{\prime}=\left(  y_{t-1},...,y_{t-p},~x_{1,t},...,x_{k,t}%
,~x_{1,t-1},...,x_{k,t-1},~...,~x_{1,t-p},...,x_{k,t-p}\right)  $.

In the more compelling case where $p$ is \textit{un}known and must be selected (implemented in our Monte Carlo below), the DURBIN regression (\ref{dynreg}) is approximate rather than exact. However, the limiting distribution (\ref{inf1}) remains valid if $p$ is selected suitably \citep{grenander1981abstract, hannan2012statistical}, as achieved by standard criteria with well-known optimality properties.\footnote{OLS-HAC regression, in contrast, typically relies on one or another of various \textquotedblleft rules of thumb" for bandwidth (truncation, $h$ or $\nu$) selection. ``Automatic" bandwidth selection has, however, been considered in Andrews-Newey-West environments by \cite{andrews1991heteroskedasticity}, \cite{AndrewsMonahan1992}, and \cite{NeweyWest1994}, among others.} In particular, if a $p_{max}$ is known such that $p\leq p_{max}$, then a consistent selection criterion (in the model selection sense) like BIC is a natural choice. Alternatively, in the absence of a $p_{max}$, an efficient selection criterion (in the model selection sense) like AIC is a natural choice.\footnote{In the Gaussian case, we have $\text{BIC}=T\log(\mathrm{SSE})+\log(T)(p+k+kp)$ and $\mathrm{AIC}=T\log(\mathrm{SSE})+ 2(p + k + kp)$, where SSE is the DURBIN regression sum of squared errors.}

\subsection{Estimation Accuracy}
\label{estimation}

We first examine the accuracy of our four estimators (OLS, FGLS, FGLS-D, and DURBIN) under our four DGPs ($NDY + BD$, $NDY + GEXOG$, $NDY + EBD$, $BD$). The key object of interest is $\mathrm{RE_{est}}$, the efficiency of DURBIN relative to OLS, FGLS or FGLS-D. For example:
\[
\mathrm{RE_{est}(OLS) = \frac{MSE(\text{OLS})}{MSE(DURBIN)}. }%
\]
We also show MSE and bias.\footnote{Note that {all} OLS-HAC estimators simply use the OLS estimator of $\beta$. Particular HAC estimators will have particular effects on the \textit{standard errors} of $\hat{\beta}$, but not on $\hat{\beta}$ itself, which always remains just $\hat{\beta}_{OLS}$.}

\begin{table}[p]
	\caption{Bias, MSE, and  Relative Efficiency \\ Estimators: OLS, FGLS, FGLS-D, DURBIN \\ DGP: Autoregressive Disturbances,  {$NDY + BD$}}%
	\label{tbl_MSE_AR}%
	\vspace{-0.5cm} \renewcommand{\arraystretch}{1}
	\par
	\begin{center}
		{\scriptsize
			\begin{tabular}
				[c]{llrrrrrrr}\hline\hline
				\multicolumn{9}{c}{\textbf{T=50}}\\\hline
				&  & $\rho=0$ & $\rho=0.3$ & $\rho=0.5$ & $\rho=0.7$ & $\rho=0.9$ &
				$\rho=0.95$ & $\rho=0.99$\\\hline
				\multirow{4}{*}{Bias} & OLS & 0.0006 & -0.0021 & -0.0008 & 0.0025 &
				0.0093 & 0.0098 & -0.0270\\
				& FGLS & 0.0006 & -0.0015 & -0.0002 & 0.0000 & 0.0030 & -0.0002 & -0.0068\\
				& FGLS-D & 0.0007 & -0.0016 & -0.0002 & 0.0006 & 0.0021 & -0.0002 & 0.0014\\
				& DURBIN & 0.0006 & -0.0010 & 0.0006 & 0.0001 & 0.0034 & -0.0006 &
				0.0006\\\hline
				\multirow{4}{*}{MSE} & OLS & 0.0112 & 0.0180 & 0.0289 & 0.0599 & 0.3013 &
				1.3368 & 80.0070\\
				& FGLS & 0.0121 & 0.0174 & 0.0216 & 0.0237 & 0.0251 & 0.0304 & 0.2953\\
				& FGLS-D & 0.0114 & 0.0179 & 0.0222 & 0.0231 & 0.0207 & 0.0198 & 0.0183\\
				& DURBIN & 0.0131 & 0.0201 & 0.0237 & 0.0229 & 0.0226 & 0.0234 &
				0.0227\\\hline
				\multirow{3}{*}{RE$_\text{est}$} & OLS & 0.8597 & 0.8952 & 1.2189 &
				2.6140 & 13.3211 & 57.1114 & 3530.5932\\
				& FGLS & 0.9261 & 0.8666 & 0.9116 & 1.0356 & 1.1115 & 1.2986 & 13.0298\\
				& FGLS-D & 0.8685 & 0.8869 & 0.9370 & 1.0063 & 0.9142 & 0.8476 &
				0.8092\\\hline
				&  &  &  &  &  &  &  & \\
				\multicolumn{9}{c}{\textbf{T=200}}\\\hline
				&  & $\rho=0$ & $\rho=0.3$ & $\rho=0.5$ & $\rho=0.7$ & $\rho=0.9$ &
				$\rho=0.95$ & $\rho=0.99$\\\hline
				\multirow{4}{*}{Bias} & OLS & -0.0003 & -0.0006 & -0.0007 & -0.0012 &
				0.0029 & -0.0082 & -0.0441\\
				& FGLS & -0.0003 & -0.0008 & -0.0004 & -0.0009 & -0.0005 & -0.0008 & 0.0001\\
				& FGLS-D & -0.0003 & -0.0008 & -0.0004 & -0.0008 & -0.0005 & -0.0005 &
				0.0000\\
				& DURBIN & -0.0002 & -0.0009 & -0.0003 & -0.0009 & -0.0002 & -0.0004 &
				-0.0003\\\hline
				\multirow{4}{*}{MSE} & OLS & 0.0026 & 0.0043 & 0.0072 & 0.0147 & 0.0635 &
				0.1884 & 8.9415\\
				& FGLS & 0.0027 & 0.0039 & 0.0048 & 0.0051 & 0.0048 & 0.0046 & 0.0048\\
				& FGLS-D & 0.0026 & 0.0040 & 0.0048 & 0.0051 & 0.0048 & 0.0045 & 0.0044\\
				& DURBIN & 0.0027 & 0.0051 & 0.0052 & 0.0051 & 0.0052 & 0.0052 &
				0.0051\\\hline
				\multirow{3}{*}{RE$_\text{est}$} & OLS & 0.9611 & 0.8526 & 1.3750 &
				2.8763 & 12.3138 & 36.3636 & 1753.3931\\
				& FGLS & 0.9761 & 0.7730 & 0.9319 & 1.0083 & 0.9296 & 0.8909 & 0.9385\\
				& FGLS-D & 0.9614 & 0.7801 & 0.9306 & 1.0024 & 0.9221 & 0.8767 &
				0.8567\\\hline
				&  &  &  &  &  &  &  & \\
				\multicolumn{9}{c}{\textbf{T=600}}\\\hline
				&  & $\rho=0$ & $\rho=0.3$ & $\rho=0.5$ & $\rho=0.7$ & $\rho=0.9$ &
				$\rho=0.95$ & $\rho=0.99$\\\hline
				\multirow{4}{*}{Bias} & OLS & 0.0001 & 0.0001 & -0.0008 & 0.0009 &
				0.0005 & 0.0007 & 0.0119\\
				& FGLS & 0.0001 & 0.0000 & -0.0006 & 0.0001 & 0.0002 & 0.0002 & -0.0006\\
				& FGLS-D & 0.0001 & 0.0000 & -0.0006 & 0.0001 & 0.0002 & 0.0002 & -0.0006\\
				& DURBIN & 0.0001 & 0.0001 & -0.0005 & 0.0001 & 0.0003 & 0.0004 &
				-0.0006\\\hline
				\multirow{4}{*}{MSE} & OLS & 0.0009 & 0.0015 & 0.0024 & 0.0048 & 0.0203 &
				0.0493 & 1.1932\\
				& FGLS & 0.0009 & 0.0013 & 0.0016 & 0.0017 & 0.0015 & 0.0015 & 0.0015\\
				& FGLS-D & 0.0009 & 0.0013 & 0.0016 & 0.0017 & 0.0015 & 0.0015 & 0.0014\\
				& DURBIN & 0.0009 & 0.0017 & 0.0017 & 0.0017 & 0.0016 & 0.0017 &
				0.0017\\\hline
				\multirow{3}{*}{RE$_\text{est}$} & OLS & 0.9900 & 0.8551 & 1.3794 &
				2.8902 & 12.4793 & 29.5992 & 705.7406\\
				& FGLS & 0.9959 & 0.7625 & 0.9287 & 1.0027 & 0.9257 & 0.8918 & 0.8731\\
				& FGLS-D & 0.9899 & 0.7622 & 0.9278 & 1.0026 & 0.9250 & 0.8912 &
				0.8464\\\hline
				&  &  &  &  &  &  &  & \\
				\multicolumn{9}{c}{\textbf{T=2500}}\\\hline
				&  & $\rho=0$ & $\rho=0.3$ & $\rho=0.5$ & $\rho=0.7$ & $\rho=0.9$ &
				$\rho=0.95$ & $\rho=0.99$\\\hline
				\multirow{4}{*}{Bias} & OLS & 0.0000 & 0.0002 & -0.0001 & 0.0001 &
				-0.0008 & 0.0008 & -0.0010\\
				& FGLS & 0.0000 & 0.0002 & 0.0000 & 0.0001 & 0.0002 & 0.0002 & -0.0001\\
				& FGLS-D & 0.0000 & 0.0002 & 0.0000 & 0.0001 & 0.0002 & 0.0002 & -0.0001\\
				& DURBIN & 0.0000 & 0.0004 & 0.0001 & 0.0001 & 0.0001 & 0.0003 &
				0.0000\\\hline
				\multirow{4}{*}{MSE} & OLS & 0.0002 & 0.0003 & 0.0006 & 0.0011 & 0.0048 &
				0.0106 & 0.1116\\
				& FGLS & 0.0002 & 0.0003 & 0.0004 & 0.0004 & 0.0004 & 0.0004 & 0.0004\\
				& FGLS-D & 0.0002 & 0.0003 & 0.0004 & 0.0004 & 0.0004 & 0.0004 & 0.0004\\
				& DURBIN & 0.0002 & 0.0004 & 0.0004 & 0.0004 & 0.0004 & 0.0004 &
				0.0004\\\hline
				\multirow{3}{*}{RE$_\text{est}$} & OLS & 0.9946 & 0.8436 & 1.4542 &
				2.8534 & 12.0802 & 26.5763 & 272.6119\\
				& FGLS & 0.9954 & 0.7551 & 0.9355 & 0.9998 & 0.9212 & 0.8891 & 0.8598\\
				& FGLS-D & 0.9946 & 0.7549 & 0.9353 & 1.0000 & 0.9213 & 0.8887 &
				0.8569\\\hline\hline
			\end{tabular}
		}
	\end{center}
	\par
	\begin{spacing}{1} \footnotesize
		Notes:  All shocks are $N(0,1)$ white noise. We select  FGLS, FGLS-D, and DURBIN lag orders using BIC.  $\rm RE_{est}$  denotes the relative estimation efficiency of DURBIN.   We perform 10000 Monte Carlo replications, drawing $x_0$ and $u_0$ from their stationary distributions and using common random numbers whenever possible. See text for details.
	\end{spacing}
\end{table}

\begin{figure}[t]
	\caption{Efficiency of DURBIN Relative to OLS  \\DGP: Autoregressive Disturbances, $NDY + BD$}
	\label{fig_REest}
	\begin{center}
		\includegraphics[trim=0 320 0 43,
		clip,width=0.8\linewidth]{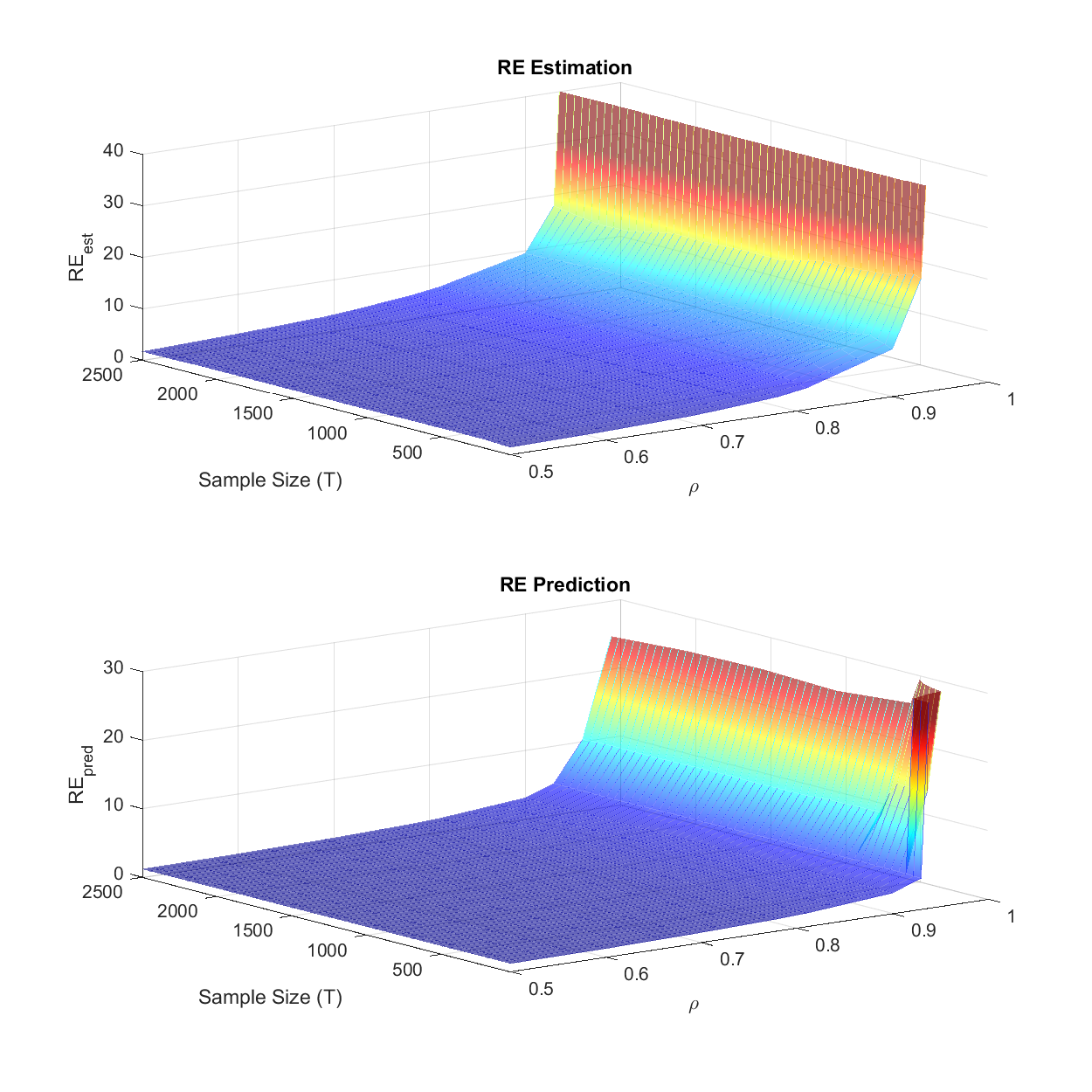}
	\end{center}
	\par
	\begin{spacing}{1} \footnotesize
		Notes:  All shocks are $N(0,1)$ white noise. We select DURBIN lag order using  BIC. We perform 10000 Monte Carlo replications, drawing $x_0$ and $u_0$ from their stationary distributions and using common random numbers whenever possible.  We do not plot values for $\rho=0.99$, due to their extreme magnitude as shown in Table \ref{tbl_MSE_AR}.  See text for details.
	\end{spacing}
\end{figure}

\paragraph{Autoregressive Disturbances DGP ($NDY + BD$).}

Results appear in Table \ref{tbl_MSE_AR}. Let us begin directly with the $\mathrm{RE_{est}}$ results for DURBIN relative to OLS. For any fixed sample size $T$, $\mathrm{RE_{est}}$ is increasing in serial correlation strength $\rho$. Consider, for example, a leading case like $T=200$ corresponding, to fifty years of quarterly data. For $\rho=0$, $\mathrm{RE_{est}}$ is close to 1, as it should be since there is no serial correlation. $\mathrm{RE_{est}}$ grows quickly as $\rho$ increases, however, reaching 2.9 when $\rho=0.7$ and 36.3 when $\rho= 0.95$.

In contrast, for any fixed serial correlation strength $\rho$, $\mathrm{RE_{est}}$ stabilizes quickly in sample size $T$ and remains approximately constant. Consider, for example, a realistic case like $\rho=0.9$. $\mathrm{RE_{est}}$ remains at approximately $\mathrm{RE_{est} =12}$ for all sample sizes $T\in\{50, 200, 600, 2500\}$. Hence $\mathrm{RE_{est}}$ is clearly driven by serial correlation strength and not by sample size.

In Figure \ref{fig_REest} we provide a  visual representation of the $\mathrm{RE_{est}}$ of DURBIN relative to OLS presented in Table \ref{tbl_MSE_AR}. It reveals clearly that $\mathrm{RE_{est}}$ is driven entirely by the degree of serial correlation and not by sample size.

Now consider separately the MSEs for OLS and DURBIN that underlie $\mathrm{RE_{est}}$. For any fixed sample size $T$,  the MSE of OLS is strongly increasing in serial correlation strength $\rho$ (because the OLS estimator ignores serial correlation), whereas the MSE from DURBIN is invariant to serial correlation strength (because the DURBIN estimator controls for serial correlation). That is why the $\mathrm{RE_{est}}$ ratio is also strongly increasing in $\rho$, as documented earlier. In contrast, for any fixed serial correlation strength $\rho$, the MSEs for \textit{both} OLS and DURBIN decrease with sample size $T$ (as they must, since both OLS and DURBIN are consistent), but they decrease proportionately, so that the $\mathrm{RE_{est}}$ ratio is invariant to $T$, as documented earlier.

Next, let us examine the bias and variance components that underlie the MSEs. First consider bias. Both the OLS and DURBIN estimators are theoretically unbiased for any serial correlation strength and sample size, and the Monte Carlo confirms the theory: the estimated biases are always negligible and invariant to $\rho$.\footnote{Moreover the estimated biases decrease with $T$, as expected, by consistency.} Moreover, given the scale of the bias, the patterns mentioned above for MSE will correspond to patterns in variance: OLS variance increases sharply with serial correlation strength (because OLS ignores serial correlation), whereas DURBIN variance does not (because DURBIN controls for serial correlation), and both variances decrease with sample size (by consistency), but they do so proportionately. That is, the MSE patterns between OLS and DURBIN, and hence the corresponding $\mathrm{RE_{est}}$ patterns, are driven entirely by variance.

\paragraph{Triangular and Unrestricted VAR DGPs ($NDY + GEXOG$,  $NDY + EBD$).}

Results appear in Table \ref{tbl_MSE_VAR}.  $\Psi_{1}$ and $\Psi_{1}^{*}$ correspond to different parameterizations  of the $NDY + GEXOG$ DGP, and $\Psi_{2}$ and $\Psi_{2}^{*}$ correspond to different parameterizations  of the $NDY + EBD$ DGP.\footnote{Recall that  $\Psi_{1}^{*}$ has a larger leading eigenvalue than does $\Psi_{1}$, and $\Psi_{2}^{*}$ 
	has a larger leading eigenvalue than $\Psi_{2}$.} For all sample sizes, OLS and FGLS exhibit large bias and MSE, which is expected since they are indeed inconsistent under both $NDY + GEXOG$ and $NDY + EBD$. As a result, the RE$_{\text{est}}$'s for DURBIN relative to OLS and FGLS in Table \ref{tbl_MSE_VAR} are very large: DURBIN dominates both.

\paragraph{Dynamic Regression DGP ($EBD$).}

Results appear in Table \ref{tbl_MSE_lags}. In the $EBD$ case, OLS, FGLS and FGLS-D are in general inconsistent, whereas DURBIN remains consistent. This is reflected in the large biases and MSEs of the other estimators compared to DURBIN, and hence the high efficiency of DURBIN relative to OLS and FGLS.

A notable exception is when $\rho=0.5$, in which case the common factor restriction holds, so that it is possible to write the dynamic regression as a single-regressor equation (with just $x_{t}$) and a disturbance with AR(1) serial correlation. Put differently, in this case the DGP in (\ref{dgp_dyn_reg}) can be rewritten in the form of (\ref{dgp_ar1}), so that FGLS and FGLS-D are consistent and efficient and should have lower MSE than Durbin. Table \ref{tbl_MSE_lags} shows that this is the case for all sample sizes. This result highlights the role that the common factor restriction plays; if it holds, it guarantees that all dynamics enter through the disturbance term, so that FGLS and FGLS-D dominate DURBIN, but if it does not hold (and there is no reason why it should  hold), DURBIN dominates.

\begin{table}[p]
	\caption{Bias, MSE, and Relative Efficiency \\ Estimators: OLS, FGLS, FGLS-D, DURBIN \\ DGPs: (1) Triangular VAR, $NDY + GEXOG$, (2) Unrestricted VAR, $NDY + EBD$}%
	\label{tbl_MSE_VAR}%
	\vspace{-0.5cm} \renewcommand{\arraystretch}{1}
	\par
	\begin{center} 
		{\scriptsize
			\begin{tabular}
				[c]{llrrrr}\hline\hline
				\multicolumn{6}{c}{\textbf{T=50}}\\\hline
				&  & $\Psi_{1}$ & $\Psi_{1}^{*}$ & $\Psi_{2}$ & $\Psi_{2}^{*}$\\\hline
				\multirow{4}{*}{Bias} & OLS & 0.2376 & 0.3054 & 0.5286 & 0.6828\\
				& FGLS & 0.1083 & 0.1484 & 0.4622 & 0.6367\\
				& FGLS-D & 0.0708 & 0.0520 & 0.1889 & 0.2902\\
				& DURBIN & 0.0637 & 0.0386 & 0.0268 & 0.0074\\\hline
				\multirow{4}{*}{MSE} & OLS & 0.0705 & 0.1082 & 0.2989 & 0.4834\\
				& FGLS & 0.0286 & 0.0400 & 0.2494 & 0.4385\\
				& FGLS-D & 0.0384 & 0.0352 & 0.0980 & 0.1672\\
				& DURBIN & 0.0408 & 0.0374 & 0.0409 & 0.0300\\\hline
				\multirow{3}{*}{RE$_{\text{est}}$} & OLS & 1.7284 & 2.8912 & 7.3025 &
				16.1337\\
				& FGLS & 0.7021 & 1.0693 & 6.0917 & 14.6353\\
				& FGLS-D & 0.9430 & 0.9393 & 2.3938 & 5.5796\\\hline
				&  &  &  &  & \\
				\multicolumn{6}{c}{\textbf{T=200}}\\\hline
				&  & $\Psi_{1}$ & $\Psi_{1}^{*}$ & $\Psi_{2}$ & $\Psi_{2}^{*}$\\\hline
				\multirow{4}{*}{Bias} & OLS & 0.2445 & 0.3174 & 0.5621 & 0.7252\\
				& FGLS & 0.1001 & 0.1413 & 0.5037 & 0.6986\\
				& FGLS-D & 0.0033 & 0.0034 & 0.1947 & 0.3410\\
				& DURBIN & 0.0009 & -0.0007 & -0.0003 & 0.0002\\\hline
				\multirow{4}{*}{MSE} & OLS & 0.0632 & 0.1045 & 0.3206 & 0.5291\\
				& FGLS & 0.0141 & 0.0246 & 0.2644 & 0.4944\\
				& FGLS-D & 0.0052 & 0.0051 & 0.0557 & 0.1451\\
				& DURBIN & 0.0052 & 0.0051 & 0.0052 & 0.0052\\\hline
				\multirow{3}{*}{RE$_{\text{est}}$} & OLS & 12.0344 & 20.6612 & 62.1667 &
				101.5652\\
				& FGLS & 2.6880 & 4.8570 & 51.2578 & 94.9083\\
				& FGLS-D & 0.9968 & 1.0006 & 10.8053 & 27.8557\\\hline
				&  &  &  &  & \\
				\multicolumn{6}{c}{\textbf{T=600}}\\\hline
				&  & $\Psi_{1}$ & $\Psi_{1}^{*}$ & $\Psi_{2}$ & $\Psi_{2}^{*}$\\\hline
				\multirow{4}{*}{Bias} & OLS & 0.2460 & 0.3218 & 0.5721 & 0.7378\\
				& FGLS & 0.0980 & 0.1409 & 0.5045 & 0.7176\\
				& FGLS-D & 0.0000 & 0.0012 & 0.2007 & 0.3667\\
				& DURBIN & -0.0008 & -0.0002 & 0.0001 & 0.0008\\\hline
				\multirow{4}{*}{MSE} & OLS & 0.0617 & 0.1048 & 0.3288 & 0.5454\\
				& FGLS & 0.0110 & 0.0214 & 0.2592 & 0.5169\\
				& FGLS-D & 0.0017 & 0.0017 & 0.0464 & 0.1454\\
				& DURBIN & 0.0017 & 0.0017 & 0.0017 & 0.0017\\\hline
				\multirow{3}{*}{RE$_{\text{est}}$} & OLS & 37.0807 & 61.9455 & 199.1306 &
				329.6430\\
				& FGLS & 6.6214 & 12.6578 & 156.9929 & 312.4485\\
				& FGLS-D & 1.0163 & 1.0083 & 28.1260 & 87.8873\\\hline
				&  &  &  &  & \\
				\multicolumn{6}{c}{\textbf{T=2500}}\\\hline
				&  & $\Psi_{1}$ & $\Psi_{1}^{*}$ & $\Psi_{2}$ & $\Psi_{2}^{*}$\\\hline
				\multirow{4}{*}{Bias} & OLS & 0.2472 & 0.3220 & 0.5756 & 0.7438\\
				& FGLS & 0.0984 & 0.1400 & 0.4654 & 0.7213\\
				& FGLS-D & 0.0004 & 0.0001 & 0.2034 & 0.3763\\
				& DURBIN & 0.0002 & -0.0003 & 0.0001 & -0.0001\\\hline
				\multirow{4}{*}{MSE} & OLS & 0.0614 & 0.1040 & 0.3316 & 0.5534\\
				& FGLS & 0.0100 & 0.0200 & 0.2179 & 0.5209\\
				& FGLS-D & 0.0004 & 0.0004 & 0.0430 & 0.1444\\
				& DURBIN & 0.0004 & 0.0004 & 0.0004 & 0.0004\\\hline
				\multirow{3}{*}{RE$_{\text{est}}$} & OLS & 153.9376 & 258.1795 &
				830.3128 & 1370.2607\\
				& FGLS & 25.1163 & 49.6206 & 545.5996 & 1289.7121\\
				& FGLS-D & 1.0261 & 1.0147 & 107.5807 & 357.5905\\\hline\hline
			\end{tabular}
		}
	\end{center}
	\par
	\begin{spacing}{1} \footnotesize
Notes:  All shocks are $N(0,1)$ white noise. We select  FGLS, FGLS-D, and DURBIN lag orders using BIC.  $\rm RE_{est}$  denotes the relative estimation efficiency of DURBIN.   We perform 10000 Monte Carlo replications, drawing $x_0$ and $u_0$ from their stationary distributions and using common random numbers whenever possible. See text for details.
		\end{spacing}
\end{table}

\begin{table}[p]
	\caption{Bias, MSE, and Relative Efficiency \\ Estimators: OLS, FGLS, FGLS-D, DURBIN \\ DGP: Dynamic Regression, {$EBD$}}%
	\label{tbl_MSE_lags}%
	\vspace{-0.5cm} \renewcommand{\arraystretch}{1}
	\par
	\begin{center}
		{\scriptsize
			\begin{tabular}
				[c]{llrrrrr}%
				\hline\hline
				\multicolumn{7}{c}{\textbf{T=50}}\\\hline
				&  & $\rho=0$ & $\rho=0.5$ & $\rho=0.7$ & $\rho=0.9$ & $\rho=0.95$\\\hline
				\multirow{4}{*}{Bias} & OLS & -0.3366 & -0.0003 & 0.2531 & 0.6886 &
				0.8700\\
				& FGLS & -0.3369 & -0.0002 & 0.0329 & -0.1000 & -0.1536\\
				& FGLS-D & -0.3336 & -0.0004 & 0.0170 & -0.1317 & -0.1891\\
				& DURBIN & -0.0484 & -0.0001 & -0.0009 & 0.0000 & -0.0004\\\hline
				\multirow{4}{*}{MSE} & OLS & 0.1275 & 0.0292 & 0.1305 & 0.9036 & 1.9497\\
				& FGLS & 0.1297 & 0.0215 & 0.0303 & 0.0423 & 0.0597\\
				& FGLS-D & 0.1275 & 0.0220 & 0.0267 & 0.0400 & 0.0570\\
				& DURBIN & 0.0435 & 0.0234 & 0.0224 & 0.0231 & 0.0234\\\hline
				\multirow{3}{*}{RE$_\text{est}$} & OLS & 2.9300 & 1.2443 & 5.8337 &
				39.1776 & 83.4520\\
				& FGLS & 2.9827 & 0.9161 & 1.3561 & 1.8348 & 2.5563\\
				& FGLS-D & 2.9310 & 0.9401 & 1.1944 & 1.7358 & 2.4410\\\hline
				&  &  &  &  &  & \\
				\multicolumn{7}{c}{\textbf{T=200}}\\\hline
				&  & $\rho=0$ & $\rho=0.5$ & $\rho=0.7$ & $\rho=0.9$ & $\rho=0.95$\\\hline
				\multirow{4}{*}{Bias} & OLS & -0.3462 & -0.0004 & 0.2678 & 0.7371 &
				0.9235\\
				& FGLS & -0.3464 & -0.0004 & 0.0082 & -0.1369 & -0.1895\\
				& FGLS-D & -0.3457 & -0.0004 & 0.0049 & -0.1422 & -0.1946\\
				& DURBIN & 0.0000 & -0.0003 & 0.0004 & 0.0005 & 0.0005\\\hline
				\multirow{4}{*}{MSE} & OLS & 0.1231 & 0.0071 & 0.0879 & 0.6648 & 1.1970\\
				& FGLS & 0.1236 & 0.0047 & 0.0061 & 0.0242 & 0.0408\\
				& FGLS-D & 0.1231 & 0.0047 & 0.0060 & 0.0254 & 0.0425\\
				& DURBIN & 0.0051 & 0.0050 & 0.0051 & 0.0053 & 0.0051\\\hline
				\multirow{3}{*}{RE$_\text{est}$} & OLS & 24.1127 & 1.4081 & 17.2231 &
				124.8315 & 232.6470\\
				& FGLS & 24.1985 & 0.9384 & 1.2010 & 4.5375 & 7.9253\\
				& FGLS-D & 24.0998 & 0.9342 & 1.1725 & 4.7644 & 8.2576\\\hline
				&  &  &  &  &  & \\
				\multicolumn{7}{c}{\textbf{T=600}}\\\hline
				&  & $\rho=0$ & $\rho=0.5$ & $\rho=0.7$ & $\rho=0.9$ & $\rho=0.95$\\\hline
				\multirow{4}{*}{Bias} & OLS & -0.3489 & 0.0005 & 0.2734 & 0.7502 &
				0.9393\\
				& FGLS & -0.3488 & 0.0002 & 0.0036 & -0.1431 & -0.1938\\
				& FGLS-D & -0.3486 & 0.0002 & 0.0026 & -0.1447 & -0.1952\\
				& DURBIN & -0.0002 & 0.0002 & 0.0009 & -0.0001 & 0.0010\\\hline
				\multirow{4}{*}{MSE} & OLS & 0.1228 & 0.0024 & 0.0803 & 0.6050 & 0.9995\\
				& FGLS & 0.1229 & 0.0015 & 0.0020 & 0.0221 & 0.0391\\
				& FGLS-D & 0.1227 & 0.0015 & 0.0020 & 0.0226 & 0.0396\\
				& DURBIN & 0.0017 & 0.0017 & 0.0017 & 0.0017 & 0.0017\\\hline
				\multirow{3}{*}{RE$_\text{est}$} & OLS & 73.7851 & 1.4430 & 46.0534 &
				364.0187 & 596.3733\\
				& FGLS & 73.8221 & 0.9363 & 1.1720 & 13.3167 & 23.3308\\
				& FGLS-D & 73.7053 & 0.9356 & 1.1602 & 13.5727 & 23.6306\\\hline
				&  &  &  &  &  & \\
				\multicolumn{7}{c}{\textbf{T=2500}}\\\hline
				&  & $\rho=0$ & $\rho=0.5$ & $\rho=0.7$ & $\rho=0.9$ & $\rho=0.95$\\\hline
				\multirow{4}{*}{Bias} & OLS & -0.3495 & 0.0001 & 0.2740 & 0.7544 &
				0.9374\\
				& FGLS & -0.3494 & 0.0002 & 0.0007 & -0.1453 & -0.1959\\
				& FGLS-D & -0.3494 & 0.0002 & 0.0005 & -0.1456 & -0.1962\\
				& DURBIN & 0.0003 & 0.0002 & 0.0000 & -0.0002 & 0.0002\\\hline
				\multirow{4}{*}{MSE} & OLS & 0.1224 & 0.0006 & 0.0764 & 0.5791 & 0.9073\\
				& FGLS & 0.1224 & 0.0004 & 0.0005 & 0.0215 & 0.0387\\
				& FGLS-D & 0.1223 & 0.0004 & 0.0004 & 0.0216 & 0.0388\\
				& DURBIN & 0.0004 & 0.0004 & 0.0004 & 0.0004 & 0.0004\\\hline
				\multirow{3}{*}{RE$_\text{est}$} & OLS & 311.2664 & 1.4398 & 193.3029 &
				1445.5630 & 2249.2027\\
				& FGLS & 311.1810 & 0.9367 & 1.1432 & 53.6830 & 96.0556\\
				& FGLS-D & 311.1251 & 0.9365 & 1.1370 & 53.8808 & 96.2890\\\hline\hline
			\end{tabular}
		}
	\end{center}
	\par
	\begin{spacing}{1} \footnotesize
Notes:  All shocks are $N(0,1)$ white noise. We select  FGLS, FGLS-D, and DURBIN lag orders using BIC.  $\rm RE_{est}$  denotes the relative estimation efficiency of DURBIN.   We perform 10000 Monte Carlo replications, drawing $x_0$ and $u_0$ from their stationary distributions and using common random numbers whenever possible. See text for details.
	\end{spacing}
\end{table}

\subsection{Prediction Accuracy}
\label{forecast}

One of the primary uses of regression and dynamic regression is for ex ante prediction. There is substantial previous literature related to the task of prediction. In particular, \cite{Baillie1979} has considered the situation of predictions from the regression model with $AR(p)$ errors and the properties of prediction from static regressions and also with optimal multi-step predictions in the sense of minimum MSE\ predictions. \cite{Baillie1979} also derived results on the efficiency of these predictors with and without estimated parameters. One conclusion concerns the importance of including the full effects of dynamics from the $AR(p)$ regression model in the predictor. In this case, the complete structural dynamic predictor generally has substantial asymptotic and small sample efficiency gains over predictors from static regressions. Similar effects and properties are found in more complicated dynamic models such as the DGP considered in section \ref{general} of this paper.

We now consider one-step-ahead predictions relying on the OLS and DURBIN estimation strategies. The results reflect that an explicit modeling of autocorrelation can be used for improved prediction. OLS estimators neglect this and therefore produce suboptimal predictions. To see this, first consider the case of a DGP with autoregressive disturbances and known parameter $\beta=1$.\footnote{We start with the case of known parameter $\beta$, as it can easily be solved analytically.} Specifically consider the DGP given by
\begin{align*}
	y_{t}  &  =x_{t}+u_{t}\\
	x_{t}  &  =\rho x_{t-1}+\epsilon_{x,t}\\
	u_{t}  &  =\rho u_{t-1}+\epsilon_{u,t},
\end{align*}
with all shocks $N(0,1)$ and orthogonal at all leads and lags. For this DGP, the optimal prediction accounting for serial correlation in $u$ is
\begin{align}
	\label{fullopt}%
	\begin{split}
		y_{t+1,t}^{opt}  &  =x_{t+1,t}+u_{t+1,t}\\
		&  =\rho x_{t}+\rho u_{t},
	\end{split}
\end{align}
and the corresponding prediction error is $e_{t+1}^{opt}=\varepsilon_{x,t+1}{+}\varepsilon_{u,t+1}$, with variance $\sigma_{opt}^{2}{=}2$.

The suboptimal prediction, failing to account for serial correlation in $u$, is just the first term in (\ref{fullopt}),
\[
y^{subopt}_{t+1,t} = \rho x_{t},
\]
with corresponding prediction error $e^{subopt}_{t+1} = \varepsilon_{x,t+1}{+}u_{t+1}$, and variance $\sigma^{2}_{subopt}{=}1 {+}\frac{1}{1{-}\rho^{2}}$.

Both predictions are unbiased, so the prediction efficiency of DURBIN relative to OLS ($\mathrm{RE_{pred}}$) is just the relative variance, which is
\begin{equation}
	\label{repred}\mathrm{RE_{pred} = \frac{\sigma^{2}_{subopt}}{\sigma^{2}_{opt}}
		= \frac{1}{2 } + \frac{1}{2(1 - \rho^{2})}.}%
\end{equation}
$\mathrm{RE_{pred}}$ is bounded below by 1, which occurs when $\rho{=}0$, and $\mathrm{RE_{pred} {\rightarrow} \infty}$ monotonically as $\rho{\rightarrow}1$.

Now we consider the case of estimated parameters, which is more complicated. In Table \ref{tbl_RE_pred} we show $\mathrm{RE_{pred}}$ estimated by Monte Carlo, accounting for parameter estimation uncertainty. For all but the most extreme cases (e.g., $T=50$ with $\rho=0.99$) the Monte Carlo results are almost identical to the analytic result (\ref{repred}) that ignores parameter estimation uncertainty.\footnote{This is because the effects of parameter estimation uncertainty on MSPE vanish quickly (like $1/T$ rather than $1/\sqrt{T}$), as is well known. Hence the earlier-documented poor estimation efficiency of OLS relative to DURBIN, although a large problem for some purposes, is not an important problem for prediction.} Hence $\mathrm{RE_{pred}}$ depends strongly on $\rho$ but not on $T$. More precisely, for any $T$ we of course obtain $\mathrm{RE_{pred}=1}$ in the white noise case ($\rho=0$), but then $\mathrm{RE_{pred}}$ grows quickly in $\rho$, and for any $\rho$, $\mathrm{RE_{pred}}$ stabilizes extremely quickly in $T$ and is basically constant.

\begin{table}[t]
	\caption{Prediction Efficiency of DURBIN Relative to OLS\\ DGP: Autoregressive Disturbances, $NDY + EBD$}%
	\label{tbl_RE_pred}%
	\vspace{-0.5cm} \renewcommand{\arraystretch}{1}
	\par
	\begin{center}%
		\begin{tabular}
			[c]{llrrrrrrr}\hline\hline
			\multicolumn{9}{c}{Relative Prediction Efficiency ($\mathrm{RE_{pred}}$%
				)}\\\hline
			T &  & $\rho=0$ & $\rho=0.3$ & $\rho=0.5$ & $\rho=0.7$ & $\rho=0.9$ &
			$\rho=0.95$ & $\rho=0.99$\\\hline
			50 &  & 0.989 & 1.042 & 1.160 & 1.452 & 3.033 & 5.865 & 391.908\\
			200 &  & 0.997 & 1.051 & 1.168 & 1.476 & 3.121 & 5.698 & 47.361\\
			600 &  & 1.000 & 1.047 & 1.152 & 1.505 & 3.214 & 5.605 & 25.569\\
			2500 &  & 1.000 & 1.049 & 1.163 & 1.469 & 3.101 & 5.656 & 25.648\\\hline\hline
		\end{tabular}
	\end{center}
	\par
	\begin{spacing}{1} \footnotesize
		Notes:  All shocks are $N(0,1)$ white noise.  $\rm RE_{pred}$  is the relative  predictive  efficiency of DURBIN,  $\rm RE_{pred}  {=} MSPE(OLS)/MSPE(DURBIN)$, where MSPE is 1-step-ahead mean squared prediction error. We select the DURBIN lag order using BIC.  We perform 10000 Monte Carlo replications, drawing $x_0$ and $u_0$ from their stationary distributions and using common random numbers whenever possible. See text for details.
	\end{spacing}
\end{table}

\subsection{Inference}
\label{inference}

Now we consider the finite-sample properties of hypothesis tests associated with the various estimation procedures. We first consider test sizes, after which we consider rejection frequencies. In all tables in this section we consider the following estimators: OLS with unadjusted standard errors, five OLS-HAC estimators (NW, NW-A, NW-LLSW, NW-KV, and M-LLSW), FGLS, FGLS-D, and two implementations of DURBIN, one using BIC for lag order selection and the other using AIC. Additionally, we have included two Hausman tests; the first null hypothesis is that FGLS is efficient relative to OLS, and the second is that FGLS-D is efficient relative to DURBIN.

\begin{table}[p]
	\caption{Empirical Size of Nominal 5\% t-test of $H_{0}:\beta=1$\\ 
		DGP: Autoregressive Disturbances, $NDY + BD$}
	\label{tbl_size_AR}%
	\vspace{-0.5cm} \renewcommand{\arraystretch}{1}
	\par
	\begin{center}
		{\scriptsize
			\begin{tabular}
				[c]{llrrrrrrr}\hline\hline
				\multicolumn{9}{c}{\textbf{T=50}}\\\hline
				& Truncation & $\rho=0$ & $\rho=0.3$ & $\rho=0.5$ & $\rho=0.7$ & $\rho=0.9$ &
				$\rho=0.95$ & $\rho=0.99$\\\hline
				OLS & $-$ & 0.051 & 0.106 & 0.167 & 0.245 & 0.346 & 0.380 & 0.407\\
				NW & $h=\lceil4(T/100)^{2/9} \rceil$ & 0.066 & 0.088 & 0.113 & 0.141 & 0.200 &
				0.227 & 0.263\\
				NW-A & $h=\lceil0.75 T^{1/3} \rceil$ & 0.064 & 0.093 & 0.121 & 0.160 & 0.230 &
				0.264 & 0.294\\
				NW-LLSW & $h=\lceil1.3 T^{1/2} \rceil$ & 0.064 & 0.078 & 0.091 & 0.110 &
				0.123 & 0.137 & 0.196\\
				NW-KV & $h=T$ & 0.061 & 0.075 & 0.081 & 0.097 & 0.091 & 0.090 & 0.155\\
				M-LLSW & $\nu=\lfloor4(T/100)^{2/9} \rfloor$ & 0.065 & 0.069 & 0.077 & 0.087 &
				0.086 & 0.102 & 0.167\\
				FGLS & BIC & 0.066 & 0.076 & 0.082 & 0.076 & 0.076 & 0.084 & 0.054\\
				FGLS-D & BIC & 0.054 & 0.095 & 0.090 & 0.069 & 0.057 & 0.059 & 0.052\\
				DURBIN & BIC & 0.060 & 0.099 & 0.082 & 0.058 & 0.053 & 0.058 & 0.051\\
				DURBIN & AIC & 0.086 & 0.093 & 0.088 & 0.078 & 0.076 & 0.080 & 0.076\\\hline
				Hausman 1 & OLS vs FGLS &  & 0.738 & 0.632 & 0.446 & 0.253 & 0.239 & 0.274\\
				Hausman 2 & DURBIN vs FGLS-D &  & 0.052 & 0.091 & 0.121 & 0.119 & 0.118 &
				0.100\\\hline
				\multicolumn{9}{c}{\textbf{T=200}}\\\hline
				& Truncation & $\rho=0$ & $\rho=0.3$ & $\rho=0.5$ & $\rho=0.7$ & $\rho=0.9$ &
				$\rho=0.95$ & $\rho=0.99$\\\hline
				OLS & $-$ & 0.051 & 0.110 & 0.174 & 0.252 & 0.352 & 0.386 & 0.413\\
				NW & $h=\lceil4(T/100)^{2/9} \rceil$ & 0.059 & 0.067 & 0.085 & 0.107 & 0.142 &
				0.157 & 0.174\\
				NW-A & $h=\lceil0.75 T^{1/3} \rceil$ & 0.059 & 0.067 & 0.085 & 0.107 & 0.142 &
				0.157 & 0.174\\
				NW-LLSW & $h=\lceil1.3 T^{1/2} \rceil$ & 0.059 & 0.060 & 0.068 & 0.072 &
				0.080 & 0.076 & 0.084\\
				NW-KV & $h=T$ & 0.053 & 0.056 & 0.061 & 0.060 & 0.061 & 0.049 & 0.032\\
				M-LLSW & $\nu=\lfloor4(T/100)^{2/9} \rfloor$ & 0.060 & 0.059 & 0.063 & 0.063 &
				0.066 & 0.063 & 0.076\\
				FGLS & BIC & 0.055 & 0.054 & 0.056 & 0.055 & 0.052 & 0.053 & 0.048\\
				FGLS-D & BIC & 0.051 & 0.061 & 0.054 & 0.054 & 0.053 & 0.051 & 0.050\\
				DURBIN & BIC & 0.053 & 0.064 & 0.049 & 0.052 & 0.051 & 0.053 & 0.049\\
				DURBIN & AIC & 0.066 & 0.055 & 0.053 & 0.056 & 0.056 & 0.057 & 0.052\\\hline
				Hausman 1 & OLS vs FGLS &  & 0.622 & 0.455 & 0.212 & 0.126 & 0.115 & 0.123\\
				Hausman 2 & DURBIN vs FGLS-D &  & 0.048 & 0.067 & 0.093 & 0.074 & 0.066 &
				0.062\\\hline
				\multicolumn{9}{c}{\textbf{T=600}}\\\hline
				& Truncation & $\rho=0$ & $\rho=0.3$ & $\rho=0.5$ & $\rho=0.7$ & $\rho=0.9$ &
				$\rho=0.95$ & $\rho=0.99$\\\hline
				OLS & $-$ & 0.049 & 0.117 & 0.175 & 0.249 & 0.351 & 0.379 & 0.408\\
				NW & $h=\lceil4(T/100)^{2/9} \rceil$ & 0.049 & 0.065 & 0.070 & 0.086 & 0.114 &
				0.120 & 0.134\\
				NW-A & $h=\lceil0.75 T^{1/3} \rceil$ & 0.049 & 0.064 & 0.068 & 0.082 & 0.104 &
				0.109 & 0.121\\
				NW-LLSW & $h=\lceil1.3 T^{1/2} \rceil$ & 0.051 & 0.057 & 0.057 & 0.057 &
				0.065 & 0.058 & 0.050\\
				NW-KV & $h=T$ & 0.049 & 0.050 & 0.053 & 0.053 & 0.050 & 0.042 & 0.017\\
				M-LLSW & $\nu=\lfloor4(T/100)^{2/9} \rfloor$ & 0.051 & 0.054 & 0.054 & 0.055 &
				0.057 & 0.051 & 0.052\\
				FGLS & BIC & 0.050 & 0.055 & 0.051 & 0.049 & 0.047 & 0.051 & 0.051\\
				FGLS-D & BIC & 0.049 & 0.055 & 0.050 & 0.049 & 0.047 & 0.051 & 0.051\\
				DURBIN & BIC & 0.049 & 0.055 & 0.049 & 0.049 & 0.045 & 0.048 & 0.049\\
				DURBIN & AIC & 0.063 & 0.056 & 0.050 & 0.050 & 0.046 & 0.048 & 0.051\\\hline
				Hausman 1 & OLS vs FGLS &  & 0.533 & 0.296 & 0.115 & 0.088 & 0.077 & 0.069\\
				Hausman 2 & DURBIN vs FGLS-D &  & 0.052 & 0.052 & 0.089 & 0.059 & 0.051 &
				0.054\\\hline
				\multicolumn{9}{c}{\textbf{T=2500}}\\\hline
				& Truncation & $\rho=0$ & $\rho=0.3$ & $\rho=0.5$ & $\rho=0.7$ & $\rho=0.9$ &
				$\rho=0.95$ & $\rho=0.99$\\\hline
				OLS & $-$ & 0.050 & 0.110 & 0.179 & 0.246 & 0.352 & 0.376 & 0.405\\
				NW & $h=\lceil4(T/100)^{2/9} \rceil$ & 0.050 & 0.053 & 0.063 & 0.068 & 0.088 &
				0.093 & 0.093\\
				NW-A & $h=\lceil0.75 T^{1/3} \rceil$ & 0.050 & 0.052 & 0.060 & 0.065 & 0.081 &
				0.085 & 0.082\\
				NW-LLSW & $h=\lceil1.3 T^{1/2} \rceil$ & 0.051 & 0.048 & 0.052 & 0.052 &
				0.057 & 0.056 & 0.042\\
				NW-KV & $h=T$ & 0.050 & 0.047 & 0.050 & 0.049 & 0.050 & 0.046 & 0.027\\
				M-LLSW & $\nu=\lfloor4(T/100)^{2/9} \rfloor$ & 0.051 & 0.047 & 0.052 & 0.051 &
				0.053 & 0.051 & 0.042\\
				FGLS & BIC & 0.050 & 0.048 & 0.051 & 0.050 & 0.048 & 0.050 & 0.052\\
				FGLS-D & BIC & 0.049 & 0.048 & 0.050 & 0.050 & 0.048 & 0.050 & 0.053\\
				DURBIN & BIC & 0.050 & 0.050 & 0.049 & 0.050 & 0.048 & 0.053 & 0.049\\
				DURBIN & AIC & 0.064 & 0.050 & 0.049 & 0.050 & 0.047 & 0.052 & 0.050\\\hline
				Hausman 1 & OLS vs FGLS &  & 0.429 & 0.155 & 0.072 & 0.068 & 0.066 & 0.049\\
				Hausman 2 & DURBIN vs FGLS-D &  & 0.051 & 0.050 & 0.086 & 0.052 & 0.052 &
				0.052\\\hline\hline
			\end{tabular}
		}
	\end{center}
	\par
	\begin{spacing}{1} \footnotesize
			Notes:  All shocks are $N(0,1)$ white noise.   We perform 10000 Monte Carlo replications, drawing $x_0$ and $u_0$ from their stationary distributions and using common random numbers whenever possible. See text for details.
	\end{spacing}
\end{table}

\begin{table}[p]
	\caption{Empirical Size of Nominal 5\% t-test of $H_{0}:\beta=1$\\ DGPs: (1) Triangular VAR, $NDY + GEXOG$, (2)  Unrestricted VAR,  $NDY + EBD$ }%
	\label{tbl_size_VAR}%
	\vspace{-0.5cm} \renewcommand{\arraystretch}{1}
	\par
	\begin{center}
		{\scriptsize
			\begin{tabular}
				[c]{llrrrr}\hline\hline
				\multicolumn{6}{c}{\textbf{T=50}}\\\hline
				& Truncation & $\Psi_{1}$ & $\Psi_{1}^{*}$ & $\Psi_{2}$ & $\Psi_{2}^{*}%
				$\\\hline
				OLS & $-$ & 0.613 & 0.787 & 0.969 & 0.992\\
				NW & $h=\lceil4(T/100)^{2/9} \rceil$ & 0.558 & 0.730 & 0.954 & 0.989\\
				NW-A & $h=\lceil0.75 T^{1/3} \rceil$ & 0.568 & 0.743 & 0.959 & 0.990\\
				NW-LLSW & $h=\lceil1.3 T^{1/2} \rceil$ & 0.507 & 0.676 & 0.928 & 0.980\\
				NW-KV & $h=T$ & 0.438 & 0.586 & 0.874 & 0.958\\
				M-LLSW & $\nu=\lfloor4(T/100)^{2/9} \rfloor$ & 0.447 & 0.607 & 0.899 & 0.970\\
				FGLS & BIC & 0.221 & 0.322 & 0.883 & 0.959\\
				FGLS-D & BIC & 0.309 & 0.248 & 0.505 & 0.627\\
				DURBIN & BIC & 0.276 & 0.200 & 0.124 & 0.070\\
				DURBIN & AIC & 0.133 & 0.099 & 0.087 & 0.078\\\hline
				Hausman 1 & OLS vs FGLS & 0.880 & 0.899 & 0.693 & 0.740\\
				Hausman 2 & DURBIN vs FGLS-D & 0.006 & 0.009 & 0.427 & 0.640\\\hline
				\multicolumn{6}{c}{\textbf{T=200}}\\\hline
				& Truncation & $\Psi_{1}$ & $\Psi_{1}^{*}$ & $\Psi_{2}$ & $\Psi_{2}^{*}%
				$\\\hline
				OLS & $-$ & 0.990 & 0.999 & 1.000 & 1.000\\
				NW & $h=\lceil4(T/100)^{2/9} \rceil$ & 0.983 & 0.999 & 1.000 & 1.000\\
				NW-A & $h=\lceil0.75 T^{1/3} \rceil$ & 0.983 & 0.999 & 1.000 & 1.000\\
				NW-LLSW & $h=\lceil1.3 T^{1/2} \rceil$ & 0.974 & 0.998 & 1.000 & 1.000\\
				NW-KV & $h=T$ & 0.872 & 0.959 & 0.999 & 1.000\\
				M-LLSW & $\nu=\lfloor4(T/100)^{2/9} \rfloor$ & 0.969 & 0.996 & 1.000 & 1.000\\
				FGLS & BIC & 0.453 & 0.688 & 0.999 & 1.000\\
				FGLS-D & BIC & 0.125 & 0.123 & 0.735 & 0.903\\
				DURBIN & BIC & 0.047 & 0.047 & 0.048 & 0.052\\
				DURBIN & AIC & 0.047 & 0.051 & 0.054 & 0.054\\\hline
				Hausman 1 & OLS vs FGLS & 0.994 & 0.998 & 0.516 & 0.457\\
				Hausman 2 & DURBIN vs FGLS-D & 0.000 & 0.000 & 0.910 & 0.978\\\hline
				\multicolumn{6}{c}{\textbf{T=600}}\\\hline
				& Truncation & $\Psi_{1}$ & $\Psi_{1}^{*}$ & $\Psi_{2}$ & $\Psi_{2}^{*}%
				$\\\hline
				OLS & $-$ & 1.000 & 1.000 & 1.000 & 1.000\\
				NW & $h=\lceil4(T/100)^{2/9} \rceil$ & 1.000 & 1.000 & 1.000 & 1.000\\
				NW-A & $h=\lceil0.75 T^{1/3} \rceil$ & 1.000 & 1.000 & 1.000 & 1.000\\
				NW-LLSW & $h=\lceil1.3 T^{1/2} \rceil$ & 1.000 & 1.000 & 1.000 & 1.000\\
				NW-KV & $h=T$ & 0.997 & 1.000 & 1.000 & 1.000\\
				M-LLSW & $\nu=\lfloor4(T/100)^{2/9} \rfloor$ & 1.000 & 1.000 & 1.000 & 1.000\\
				FGLS & BIC & 0.831 & 0.978 & 1.000 & 1.000\\
				FGLS-D & BIC & 0.122 & 0.128 & 0.956 & 0.998\\
				DURBIN & BIC & 0.048 & 0.051 & 0.051 & 0.048\\
				DURBIN & AIC & 0.049 & 0.053 & 0.052 & 0.047\\\hline
				Hausman 1 & OLS vs FGLS & 1.000 & 1.000 & 0.669 & 0.273\\
				Hausman 2 & DURBIN vs FGLS-D & 0.000 & 0.000 & 1.000 & 1.000\\\hline
				\multicolumn{6}{c}{\textbf{T=2500}}\\\hline
				& Truncation & $\Psi_{1}$ & $\Psi_{1}^{*}$ & $\Psi_{2}$ & $\Psi_{2}^{*}%
				$\\\hline
				OLS & $-$ & 1.000 & 1.000 & 1.000 & 1.000\\
				NW & $h=\lceil4(T/100)^{2/9} \rceil$ & 1.000 & 1.000 & 1.000 & 1.000\\
				NW-A & $h=\lceil0.75 T^{1/3} \rceil$ & 1.000 & 1.000 & 1.000 & 1.000\\
				NW-LLSW & $h=\lceil1.3 T^{1/2} \rceil$ & 1.000 & 1.000 & 1.000 & 1.000\\
				NW-KV & $h=T$ & 1.000 & 1.000 & 1.000 & 1.000\\
				M-LLSW & $\nu=\lfloor4(T/100)^{2/9} \rfloor$ & 1.000 & 1.000 & 1.000 & 1.000\\
				FGLS & BIC & 1.000 & 1.000 & 1.000 & 1.000\\
				FGLS-D & BIC & 0.123 & 0.129 & 1.000 & 1.000\\
				DURBIN & BIC & 0.048 & 0.050 & 0.050 & 0.050\\
				DURBIN & AIC & 0.048 & 0.049 & 0.050 & 0.050\\\hline
				Hausman 1 & OLS vs FGLS & 1.000 & 1.000 & 0.999 & 0.465\\
				Hausman 2 & DURBIN vs FGLS-D & 0.000 & 0.000 & 1.000 & 1.000\\\hline\hline
			\end{tabular}
		}
	\end{center}
	\par
	\begin{spacing}{1} \footnotesize
					Notes:  All shocks are $N(0,1)$ white noise.   We perform 10000 Monte Carlo replications, drawing $x_0$ and $u_0$ from their stationary distributions and using common random numbers whenever possible. See text for details.
	\end{spacing}
\end{table}

\begin{table}[p]
	\caption{Empirical Size of Nominal 5\% t-test of $H_{0}:\beta=1$\\DGP: Dynamic
		Regression, $EBD$}%
	\label{tbl_size_lags}%
	\renewcommand{\arraystretch}{1} \vspace{-0.5cm}
	\par
	\begin{center}
		{\scriptsize
			\begin{tabular}
				[c]{llrrrrr}%
				\hline\hline
				\multicolumn{7}{c}{\textbf{T=50}}\\\hline
				Test & Truncation & $\rho=0$ & $\rho=0.5$ & $\rho=0.7$ & $\rho=0.9$ &
				$\rho=0.95$\\\hline
				OLS & $-$ & 0.785 & 0.162 & 0.441 & 0.504 & 0.406\\
				NW & $h=\lceil4(T/100)^{2/9} \rceil$ & 0.779 & 0.105 & 0.302 & 0.330 & 0.247\\
				NW-A & $h=\lceil0.75 T^{1/3} \rceil$ & 0.783 & 0.117 & 0.331 & 0.372 & 0.282\\
				NW-LLSW & $h=\lceil1.3 T^{1/2} \rceil$ & 0.742 & 0.087 & 0.237 & 0.228 &
				0.159\\
				NW-KV & $h=T$ & 0.662 & 0.082 & 0.200 & 0.184 & 0.122\\
				M-LLSW & $\nu=\lfloor4(T/100)^{2/9} \rfloor$ & 0.687 & 0.075 & 0.194 & 0.170 &
				0.124\\
				FGLS & BIC & 0.775 & 0.077 & 0.097 & 0.113 & 0.183\\
				FGLS-D & BIC & 0.768 & 0.085 & 0.078 & 0.124 & 0.198\\
				DURBIN & BIC & 0.205 & 0.078 & 0.052 & 0.055 & 0.059\\
				DURBIN & AIC & 0.098 & 0.082 & 0.071 & 0.078 & 0.078\\\hline
				Hausman 1 & OLS vs FGLS & 0.782 & 0.624 & 0.583 & 0.491 & 0.332\\
				Hausman 2 & DURBIN vs FGLS-D & 0.770 & 0.085 & 0.565 & 0.930 & 0.955\\\hline
				\multicolumn{7}{c}{\textbf{T=200}}\\\hline
				Test & Truncation & $\rho=0$ & $\rho=0.5$ & $\rho=0.7$ & $\rho=0.9$ &
				$\rho=0.95$\\\hline
				OLS & $-$ & 1.000 & 0.169 & 0.820 & 0.903 & 0.779\\
				NW & $h=\lceil4(T/100)^{2/9} \rceil$ & 1.000 & 0.081 & 0.656 & 0.738 & 0.516\\
				NW-A & $h=\lceil0.75 T^{1/3} \rceil$ & 1.000 & 0.081 & 0.656 & 0.738 & 0.516\\
				NW-LLSW & $h=\lceil1.3 T^{1/2} \rceil$ & 0.999 & 0.063 & 0.560 & 0.620 &
				0.352\\
				NW-KV & $h=T$ & 0.979 & 0.058 & 0.421 & 0.447 & 0.248\\
				M-LLSW & $\nu=\lfloor4(T/100)^{2/9} \rfloor$ & 0.999 & 0.060 & 0.536 & 0.583 &
				0.319\\
				FGLS & BIC & 1.000 & 0.053 & 0.066 & 0.415 & 0.683\\
				FGLS-D & BIC & 1.000 & 0.051 & 0.062 & 0.441 & 0.716\\
				DURBIN & BIC & 0.049 & 0.047 & 0.050 & 0.055 & 0.052\\
				DURBIN & AIC & 0.051 & 0.051 & 0.054 & 0.058 & 0.058\\\hline
				Hausman 1 & OLS vs FGLS & 0.714 & 0.443 & 0.827 & 0.899 & 0.665\\
				Hausman 2 & DURBIN vs FGLS-D & 1.000 & 0.061 & 0.974 & 1.000 & 1.000\\\hline
				\multicolumn{7}{c}{\textbf{T=600}}\\\hline
				Test & Truncation & $\rho=0$ & $\rho=0.5$ & $\rho=0.7$ & $\rho=0.9$ &
				$\rho=0.95$\\\hline
				OLS & $-$ & 1.000 & 0.174 & 0.995 & 0.999 & 0.989\\
				NW & $h=\lceil4(T/100)^{2/9} \rceil$ & 1.000 & 0.075 & 0.971 & 0.991 & 0.928\\
				NW-A & $h=\lceil0.75 T^{1/3} \rceil$ & 1.000 & 0.072 & 0.970 & 0.991 & 0.920\\
				NW-LLSW & $h=\lceil1.3 T^{1/2} \rceil$ & 1.000 & 0.058 & 0.950 & 0.985 &
				0.882\\
				NW-KV & $h=T$ & 1.000 & 0.050 & 0.801 & 0.843 & 0.646\\
				M-LLSW & $\nu=\lfloor4(T/100)^{2/9} \rfloor$ & 1.000 & 0.057 & 0.946 & 0.984 &
				0.864\\
				FGLS & BIC & 1.000 & 0.049 & 0.066 & 0.912 & 0.995\\
				FGLS-D & BIC & 1.000 & 0.049 & 0.065 & 0.922 & 0.997\\
				DURBIN & BIC & 0.050 & 0.047 & 0.053 & 0.050 & 0.050\\
				DURBIN & AIC & 0.051 & 0.048 & 0.055 & 0.050 & 0.050\\\hline
				Hausman 1 & OLS vs FGLS & 0.675 & 0.293 & 0.997 & 1.000 & 0.990\\
				Hausman 2 & DURBIN vs FGLS-D & 1.000 & 0.053 & 1.000 & 1.000 & 1.000\\\hline
				\multicolumn{7}{c}{\textbf{T=2500}}\\\hline
				Test & Truncation & $\rho=0$ & $\rho=0.5$ & $\rho=0.7$ & $\rho=0.9$ &
				$\rho=0.95$\\\hline
				OLS & $-$ & 1.000 & 0.173 & 1.000 & 1.000 & 1.000\\
				NW & $h=\lceil4(T/100)^{2/9} \rceil$ & 1.000 & 0.062 & 1.000 & 1.000 & 1.000\\
				NW-A & $h=\lceil0.75 T^{1/3} \rceil$ & 1.000 & 0.060 & 1.000 & 1.000 & 1.000\\
				NW-LLSW & $h=\lceil1.3 T^{1/2} \rceil$ & 1.000 & 0.053 & 1.000 & 1.000 &
				1.000\\
				NW-KV & $h=T$ & 1.000 & 0.050 & 0.997 & 0.998 & 0.981\\
				M-LLSW & $\nu=\lfloor4(T/100)^{2/9} \rfloor$ & 1.000 & 0.053 & 1.000 & 1.000 &
				1.000\\
				FGLS & BIC & 1.000 & 0.051 & 0.056 & 1.000 & 1.000\\
				FGLS-D & BIC & 1.000 & 0.051 & 0.053 & 1.000 & 1.000\\
				DURBIN & BIC & 0.049 & 0.048 & 0.049 & 0.052 & 0.050\\
				DURBIN & AIC & 0.049 & 0.049 & 0.050 & 0.052 & 0.050\\\hline
				Hausman 1 & OLS vs FGLS & 0.636 & 0.154 & 1.000 & 1.000 & 1.000\\
				Hausman 2 & DURBIN vs FGLS-D & 1.000 & 0.051 & 1.000 & 1.000 &
				1.000\\\hline\hline
			\end{tabular}
		}
	\end{center}
	\par
	\begin{spacing}{1} \footnotesize
			Notes:  All shocks are $N(0,1)$ white noise.   We perform 10000 Monte Carlo replications, drawing $x_0$ and $u_0$ from their stationary distributions and using common random numbers whenever possible. See text for details.
	\end{spacing}
\end{table}

\subsubsection{Size}
\label{size}

Table 	\ref{tbl_size_AR} contains results for the autoregressive disturbances DGP, $NDY + BD$. First, tests based on OLS are incorrectly sized for all $(\rho,T)$ combinations, except when $\rho=0$, and the size distortions become huge as $\rho$ grows. Second, the various NW HAC corrections reduce but do not eliminate the size distortion. In particular, distortion generally remains in the economically crucial region of $\rho\in[0.5, 0.99]$, depending on the sample size and the precise NW version used. NW and NW-A are worst, NW-LLSW are better, and NW-KV is the best. The M-LLSW HAC correction is different in that it exhibits an approximately correct size across $(\rho,T)$ combinations. Finally, tests based on FGLS, FGLS-D and DURBIN, in contrast, are correctly sized for all $(\rho,T)$ combinations, even with extremely strong autocorrelation. This holds regardless of whether DURBIN lag order selection is done with BIC or AIC.

Table 	\ref{tbl_size_VAR} contains results for the two VAR DGPs, $NDY + GEXOG$ and $NDY + EBD$. 
In the {\textit{NDY}+ \textit{GEXOG}} environment, OLS and FGLS are inconsistent, which produces large size distortions. In contrast, DURBIN and FGLS-D are consistent; they should outperform OLS and FGLS, and they do. DURBIN and FLGS-D should perform similarly, and they do. In the {\textit{NDY} +\textit{EBD}} environment, OLS, FGLS, \textit{and} FGLS-D are inconsistent, and all have large size distortions. DURBIN, however, remains consistent and performs admirably.

Finally, Table \ref{tbl_size_lags} contains results for the dynamic regression DGP, $EBD$. In this  environment DURBIN should perform well, and it does, whereas all other test sizes are distorted, except at or near the very special common-factor case of $\rho=0.5$.

\begin{figure}[pt]
	\caption{Empirical Rejection Frequencies of Nominal 5\% t-Tests of 
		$\mathrm{H_{0}{:}~\beta{=}1}$, $T=200$}%
	\label{fig_powerT_AR}
	\begin{center}
		\includegraphics[trim=10 10 0 30, clip,
		scale=.79	]{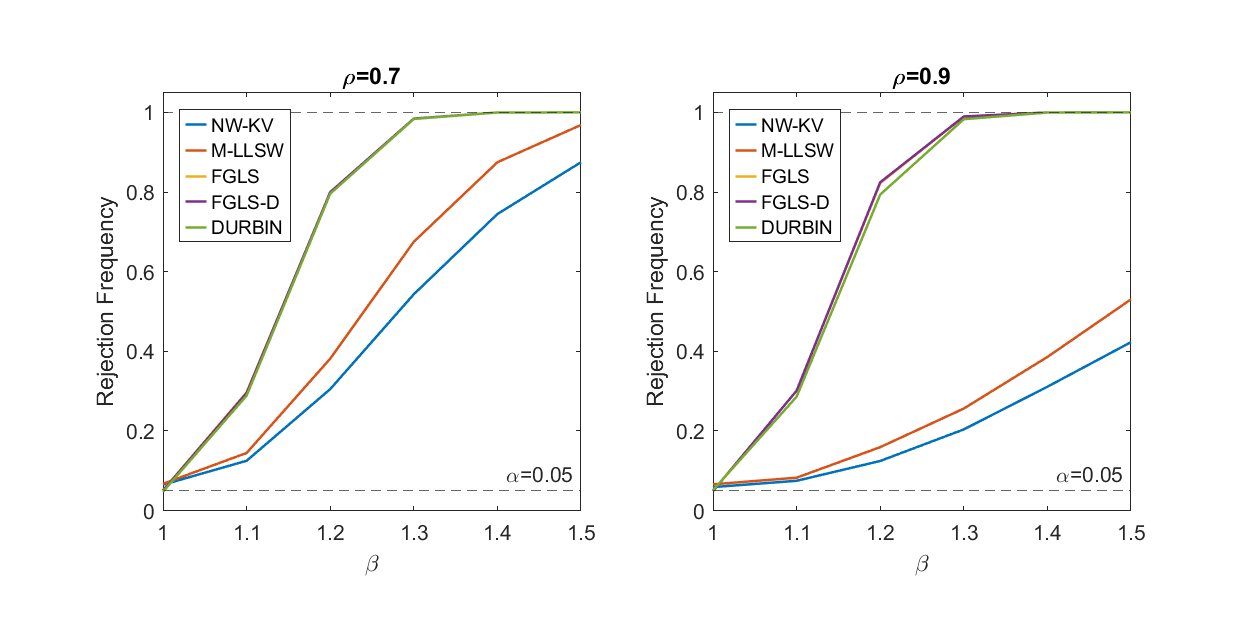}
		\includegraphics[trim=10 10 0 25, clip,
		scale=.79,	]{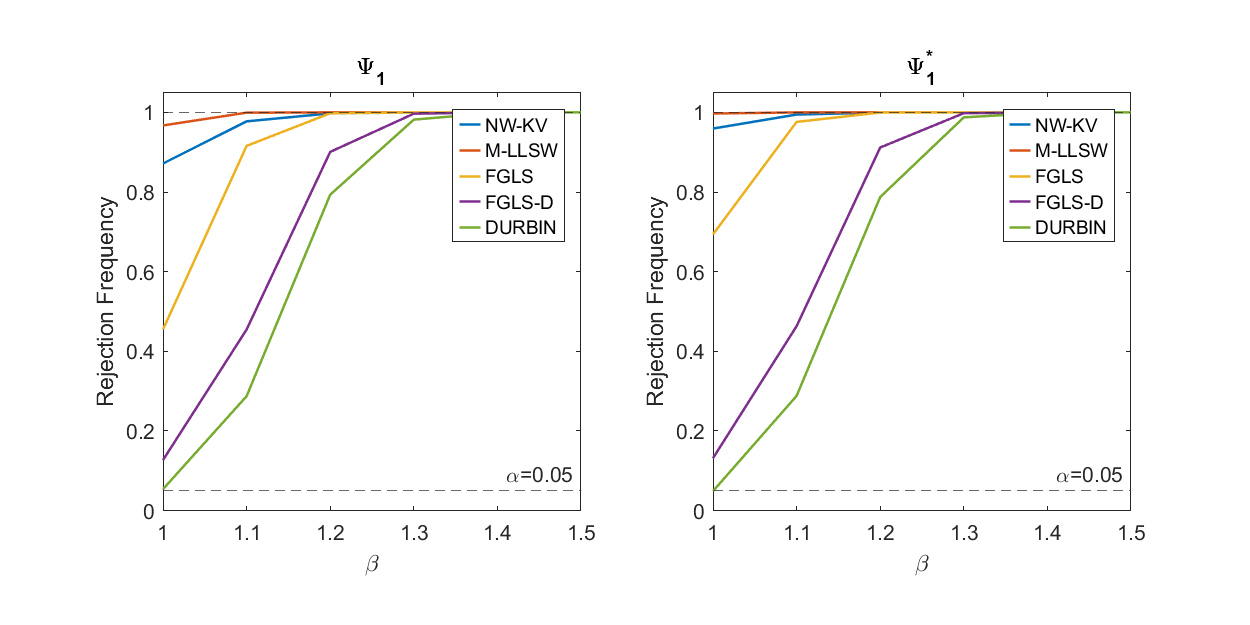}
		\includegraphics[trim=10 10 0 25, clip,
		scale=.79,	]{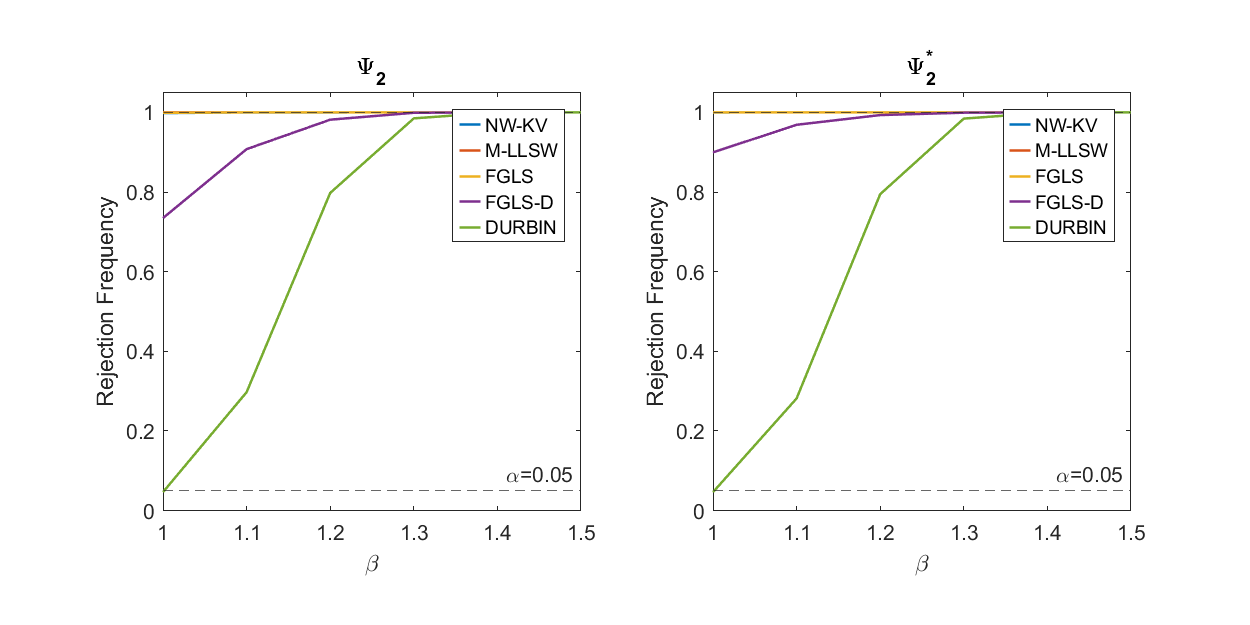}
	\end{center}
	\par
	\begin{spacing}{1} \footnotesize
		Notes: DGPs: $NDY + BD$ (top row),  $NDY + GEXOG$  (middle row), and  $NDY + EBD$ (bottom row).  See text for details.	\end{spacing}
\end{figure}

\subsubsection{Power}
\label{power}

Only tests that are correctly sized are of real interest, because only correctly-sized tests produce trustworthy and interpretable rejections. As we have shown, DURBIN satisfies that requirement, whereas OLS-HAC regression does not. One could simply stop there, but it is of interest to compare rejection frequencies in a few laboratory environments where the DGP is known. We do so in Figure \ref{fig_powerT_AR} for three of our DGPs with $T=200$ and various persistence parameters, comparing OLS-HAC (Kiefer-Vogelsang, LLSW), FGLS, and FGLS-D.

In the top row Figure \ref{fig_powerT_AR}  we show rejection frequencies for the autoregressive disturbances environment, $NDY + BD$. All estimators are consistent, and all tests have correct size when $\beta=1$, i.e., when the true parameter equals its value under the null hypothesis. Moving away from the null however, it is clear that OLS-HAC power is inferior to that of DURBIN, because OLS is inefficient relative to DURBIN. Moreover, the inferior power performance of OLS-HAC increases with disturbance persistence ($\rho$), precisely because the relative inefficiency of OLS increases with persistence. Finally, DURBIN, FGLS and FGLS-D have virtually identical power curves.

In the middle row Figure \ref{fig_powerT_AR}  we show rejection frequencies for the  triangular VAR case,   $NDY + GEXOG$.  OLS-HAC and FGLS are so badly mis-sized that it is not worth discussing them, whereas FGLS-D is asymptotically correctly sized but is still over-sized for $T=200$. Only DURBIN is trustworthy. Moving from the middle-left to middle-right panel (higher persistence), the superiority of Durbin is amplified.

In the bottom row Figure \ref{fig_powerT_AR}  we show rejection frequencies for the unrestricted VAR case,  $NDY + EBD$. FGLS-D fails even asymptotically, so it is not surprising that its finite-sample performance is much worse than in the middle-row triangular VAR   $NDY + GEXOG$ case.   DURBIN, however, remains trustworthy. Moving from the bottom-left  to bottom-right panel (higher persistence), the superiority of DURBIN is amplified, just as in the triangular case.

\section{Concluding Remarks and Directions for Future Research}
\label{concl}

We have considered issues surrounding the time-series application of OLS regression with HAC standard errors. Although the OLS-HAC methodology is often sensible in cross-section regression situations, we argued that it is not generally an effective procedure in time-series regressions. Such regressions usually possess persistent autocorrelation, which causes OLS-HAC regressions to be highly sub-optimal for parameter estimation (in terms of efficiency), inference (in terms of both test size and power), and prediction.

We showed that the OLS-HAC problems are largely avoided by the use of a simple dynamic regression procedure, DURBIN. We demonstrated the significant advantages of DURBIN with detailed simulations covering a range of practical environments and issues.  Effectively, DURBIN is a powerful tool for pre-whitened HAC estimation, in the tradition of \cite{AndrewsMonahan1992} -- indeed \textit{such} a good pre-whitening tool that there's rarely any need for subsequent HAC estimation.

On the other hand, DURBIN is of course not a panacea. For example, DURBIN  may struggle in small samples when dynamics have a strong moving-average component. Our Monte Carlo makes clear, however,  that for all but the most  extreme environments,   DURBIN with lag order selected using standard information criteria performs consistently well. Indeed that is the key message of our paper.
	 
In future work, one could generalize the DURBIN regression in various ways. For example, one could allow different lag lengths for $y$ and the $x_{i}$'s. One could also allow for heteroskedasticity, which we suppressed in this paper so as to focus exclusively on autocorrelation, for example by allowing for $GARCH$ disturbances in the DURBIN regression.

\clearpage

\appendix
\appendixpage
\addappheadtotoc

\section{Additional Monte Carlo: AR Disturbances}
\label{AR_BIC}

\renewcommand{\theequation}{A\arabic{equation}} \setcounter{equation}{0}
\renewcommand{\thefigure}{A\arabic{figure}} \setcounter{figure}{0}
\renewcommand{\thetable}{A\arabic{table}} \setcounter{table}{0}

\begin{table}[h]
	\caption{Selected Lags by test \\ Estimators: FGLS, DURBIN AIC, DURBIN BIC \\ DGP: Autoregressive Disturbances,  {$NDY + BD$}}%
	\label{tbl_lag_AR}%
	\vspace{-0.5cm} 
	\renewcommand{\arraystretch}{1} 
	\par
	\begin{center}
		{\footnotesize
			\begin{tabular}
				[c]{cllrrrrrrr}\hline\hline
				$T$ &  &  & $\rho=0$ & $\rho=0.3$ & $\rho=0.5$ & $\rho=0.7$ & $\rho=0.9$ &
				$\rho=0.95$ & $\rho=0.99$\\\hline
				\multirow{6}{*}{50} & \multirow{3}{*}{Median} & FGLS BIC & 1 & 1 & 1 & 1 & 1 &
				1 & 1\\
				&  & DURBIN BIC & 0 & 0 & 1 & 1 & 1 & 1 & 1\\
				&  & DURBIN AIC & 0 & 1 & 1 & 1 & 1 & 1 & 1\\\cline{2-10}
				& \multirow{3}{*}{Mean} & FGLS BIC & 1.2 & 1.2 & 1.3 & 1.2 & 1.3 & 1.3 & 1.5\\
				&  & DURBIN BIC & 0.1 & 0.4 & 0.9 & 1.1 & 1.1 & 1.1 & 1.1\\
				&  & DURBIN AIC & 2.0 & 2.5 & 2.9 & 3.0 & 3.0 & 3.1 & 3.1\\\hline
				\multirow{6}{*}{200} & \multirow{3}{*}{Median} & FGLS BIC & 1 & 1 & 1 & 1 &
				1 & 1 & 1\\
				&  & DURBIN BIC & 0 & 1 & 1 & 1 & 1 & 1 & 1\\
				&  & DURBIN AIC & 0 & 1 & 1 & 1 & 1 & 1 & 1\\\cline{2-10}
				& \multirow{3}{*}{Mean} & FGLS BIC & 1.1 & 1.1 & 1.1 & 1.1 & 1.1 & 1.1 & 1.5\\
				&  & DURBIN BIC & 0.0 & 0.9 & 1.0 & 1.0 & 1.0 & 1.0 & 1.0\\
				&  & DURBIN AIC & 1.1 & 2.1 & 2.2 & 2.2 & 2.2 & 2.2 & 2.2\\\hline
				\multirow{6}{*}{600} & \multirow{3}{*}{Median} & FGLS BIC & 1 & 1 & 1 & 1 &
				1 & 1 & 1\\
				&  & DURBIN BIC & 0 & 1 & 1 & 1 & 1 & 1 & 1\\
				&  & DURBIN AIC & 0 & 1 & 1 & 1 & 1 & 1 & 1\\\cline{2-10}
				& \multirow{3}{*}{Mean} & FGLS BIC & 1.0 & 1.0 & 1.0 & 1.0 & 1.0 & 1.0 & 1.5\\
				&  & DURBIN BIC & 0.0 & 1.0 & 1.0 & 1.0 & 1.0 & 1.0 & 1.0\\
				&  & DURBIN AIC & 0.7 & 1.8 & 1.7 & 1.7 & 1.8 & 1.8 & 1.7\\\hline
				\multirow{6}{*}{2500} & \multirow{3}{*}{Median} & FGLS BIC & 1 & 1 & 1 & 1 &
				1 & 1 & 1\\
				&  & DURBIN BIC & 0 & 1 & 1 & 1 & 1 & 1 & 1\\
				&  & DURBIN AIC & 0 & 1 & 1 & 1 & 1 & 1 & 1\\\cline{2-10}
				& \multirow{3}{*}{Mean} & FGLS BIC & 1.0 & 1.0 & 1.0 & 1.0 & 1.0 & 1.0 & 1.1\\
				&  & DURBIN BIC & 0.0 & 1.0 & 1.0 & 1.0 & 1.0 & 1.0 & 1.0\\
				&  & DURBIN AIC & 0.7 & 1.7 & 1.7 & 1.7 & 1.7 & 1.7 & 1.7\\\hline\hline
			\end{tabular}
		}
	\end{center}
	\par
	\begin{spacing}{1} \footnotesize
		Notes: All shocks are $N(0,1)$ white noise.   We perform 10000 Monte Carlo replications, drawing $x_0$ and $u_0$ from their stationary distributions and using common random numbers whenever possible. See text for details.
	\end{spacing}
\end{table}

\clearpage

\section{Additional Monte Carlo: MA Disturbances}
\label{MA_BIC}

\renewcommand{\theequation}{B\arabic{equation}} \setcounter{equation}{0}
\renewcommand{\thefigure}{B\arabic{figure}} \setcounter{figure}{0}
\renewcommand{\thetable}{B\arabic{table}} \setcounter{table}{0}

\begin{table}[h]
	\caption{Selected Lags by test \\ Estimators: FGLS, DURBIN AIC, DURBIN BIC \\ DGP: Moving Average Disturbances}%
	\label{tbl_lag_MA}%
	\vspace{-0.5cm} 
	\renewcommand{\arraystretch}{1} 
	\par
	\begin{center}
		{\footnotesize
			\begin{tabular}
				[c]{cllrrrrrrr}\hline\hline
				$T$ &  &  & $\theta=0$ & $\theta=0.3$ & $\theta=0.5$ & $\theta=0.7$ &
				$\theta=0.9$ & $\theta=0.95$ & $\theta=0.99$\\\hline
				\multirow{6}{*}{50} & \multirow{3}{*}{Median} & FGLS & 1 & 1 & 1 & 2 & 2 & 3 &
				3\\
				&  & DURBIN BIC & 0 & 0 & 1 & 1 & 2 & 2 & 2\\
				&  & DURBIN AIC & 0 & 1 & 2 & 3 & 5 & 5 & 5\\\cline{2-10}
				& \multirow{3}{*}{Mean} & FGLS & 1.3 & 1.3 & 1.6 & 2.2 & 2.9 & 3.0 & 3.1\\
				&  & DURBIN BIC & 0.1 & 0.3 & 0.8 & 1.4 & 1.9 & 2.0 & 2.0\\
				&  & DURBIN AIC & 2.0 & 2.6 & 3.4 & 4.2 & 5.2 & 5.6 & 5.6\\\hline
				\multirow{6}{*}{200} & \multirow{3}{*}{Median} & FGLS & 1 & 1 & 2 & 3 & 5 &
				5 & 5\\
				&  & DURBIN BIC & 0 & 1 & 1 & 2 & 3 & 3 & 3\\
				&  & DURBIN AIC & 0 & 1 & 2 & 4 & 8 & 9 & 10\\\cline{2-10}
				& \multirow{3}{*}{Mean} & FGLS & 1.1 & 1.2 & 1.9 & 3.1 & 4.9 & 5.2 & 5.4\\
				&  & DURBIN BIC & 0.0 & 0.8 & 1.4 & 2.3 & 3.3 & 3.5 & 3.5\\
				&  & DURBIN AIC & 1.1 & 2.4 & 3.6 & 5.4 & 9.1 & 10.7 & 11.5\\\hline
				\multirow{6}{*}{600} & \multirow{3}{*}{Median} & FGLS & 1 & 1 & 2 & 4 & 7 &
				8 & 9\\
				&  & DURBIN BIC & 0 & 1 & 2 & 3 & 5 & 6 & 6\\
				&  & DURBIN AIC & 0 & 2 & 3 & 6 & 12 & 14 & 16\\\cline{2-10}
				& \multirow{3}{*}{Mean} & FGLS & 1.0 & 1.4 & 2.5 & 4.3 & 7.6 & 8.5 & 8.9\\
				&  & DURBIN BIC & 0.0 & 1.1 & 2.1 & 3.4 & 5.4 & 5.9 & 6.1\\
				&  & DURBIN AIC & 0.7 & 2.4 & 3.8 & 6.2 & 12.3 & 15.2 & 17.0\\\hline
				\multirow{6}{*}{2500} & \multirow{3}{*}{Median} & FGLS & 1 & 2 & 3 & 6 & 12 &
				15 & 17\\
				&  & DURBIN BIC & 0 & 2 & 3 & 5 & 9 & 11 & 12\\
				&  & DURBIN AIC & 0 & 2 & 4 & 8 & 17 & 24 & 28\\\cline{2-10}
				& \multirow{3}{*}{Mean} & FGLS & 1.0 & 2.0 & 3.4 & 5.9 & 12.2 & 15.3 & 17.3\\
				&  & DURBIN BIC & 0.0 & 1.6 & 2.9 & 4.9 & 9.5 & 11.2 & 12.0\\
				&  & DURBIN AIC & 0.7 & 3.0 & 4.8 & 8.1 & 18.1 & 24.1 & 27.2\\\hline\hline
			\end{tabular}
		}
	\end{center}
	\par
	\begin{spacing}{1} \footnotesize
		Notes: The data-generating process is $y_{t}= x_{t}+u_{t}$, $x_{t}=0.7 x_{t-1}+\protect\epsilon _{x,t}$, $u_{t}= \theta  \epsilon_{u,t-1}+\protect\epsilon _{u,t}$,  $t=1, ..., T$.  All shocks are $N(0,1)$ white noise.   We perform 10000 Monte Carlo replications, drawing $x_0$ and $u_0$ from their stationary distributions and using common random numbers whenever possible. See text for details.
	\end{spacing}
\end{table}

\begin{table}[h]
	\caption{Bias, MSE, and Relative Efficiency \\ Estimators: OLS, FGLS, FGLS-D, DURBIN \\ DGP: Moving Average Disturbances}%
	\label{tbl_MSE_MA}%
	\vspace{-0.5cm} 
	\renewcommand{\arraystretch}{1} 
	\par
	\begin{center}
		{\scriptsize
			\begin{tabular}
				[c]{llrrrrrrr}\hline\hline
				\multicolumn{9}{c}{\textbf{T=50}}\\\hline
				&  & $\theta=0$ & $\theta=0.3$ & $\theta=0.5$ & $\theta=0.7$ & $\theta=0.9$ &
				$\theta=0.95$ & $\theta=0.99$\\
				\hline
				\multirow{4}{*}{Bias} & OLS & 0.0004 & -0.0013 & 0.0008 & 0.0018 &
				0.0024 & -0.0002 & -0.0034\\
				& FGLS & 0.0004 & -0.0010 & 0.0004 & 0.0018 & 0.0027 & -0.0008 & -0.0014\\
				& FGLS-D & 0.0004 & -0.0010 & 0.0013 & 0.0009 & 0.0028 & -0.0022 & -0.0022\\
				& DURBIN & 0.0004 & -0.0002 & 0.0028 & 0.0011 & 0.0032 & -0.0020 &
				-0.0028\\\hline
				\multirow{4}{*}{MSE} & OLS & 0.0112 & 0.0165 & 0.0208 & 0.0279 & 0.0329 &
				0.0344 & 0.0360\\
				& FGLS & 0.0121 & 0.0167 & 0.0195 & 0.0224 & 0.0231 & 0.0237 & 0.0246\\
				& FGLS-D & 0.0114 & 0.0168 & 0.0203 & 0.0235 & 0.0248 & 0.0256 & 0.0267\\
				& DURBIN & 0.0131 & 0.0192 & 0.0237 & 0.0275 & 0.0314 & 0.0334 &
				0.0350\\\hline
				\multirow{3}{*}{RE$_\text{est}$} & OLS & 0.8570 & 0.8612 & 0.8795 &
				1.0140 & 1.0485 & 1.0300 & 1.0288\\
				& FGLS & 0.9250 & 0.8730 & 0.8225 & 0.8167 & 0.7355 & 0.7084 & 0.7032\\
				& FGLS-D & 0.8668 & 0.8755 & 0.8573 & 0.8542 & 0.7900 & 0.7654 &
				0.7638\\\hline
				&  &  &  &  &  &  &  & \\
				\multicolumn{9}{c}{\textbf{T=200}}\\\hline
				&  & $\theta=0$ & $\theta=0.3$ & $\theta=0.5$ & $\theta=0.7$ & $\theta=0.9$ &
				$\theta=0.95$ & $\theta=0.99$\\\hline
				\multirow{4}{*}{Bias} & OLS & -0.0008 & -0.0005 & -0.0005 & 0.0001 &
				-0.0006 & -0.0015 & -0.0004\\
				& FGLS & -0.0008 & -0.0003 & -0.0004 & 0.0004 & -0.0005 & 0.0001 & -0.0004\\
				& FGLS-D & -0.0008 & -0.0004 & -0.0003 & 0.0003 & -0.0004 & -0.0001 &
				-0.0002\\
				& DURBIN & -0.0007 & 0.0000 & -0.0001 & 0.0000 & -0.0005 & 0.0000 &
				-0.0005\\\hline
				\multirow{4}{*}{MSE} & OLS & 0.0026 & 0.0039 & 0.0049 & 0.0064 & 0.0076 &
				0.0083 & 0.0088\\
				& FGLS & 0.0026 & 0.0037 & 0.0043 & 0.0044 & 0.0040 & 0.0039 & 0.0040\\
				& FGLS-D & 0.0026 & 0.0037 & 0.0043 & 0.0045 & 0.0044 & 0.0045 & 0.0047\\
				& DURBIN & 0.0027 & 0.0050 & 0.0053 & 0.0056 & 0.0061 & 0.0063 &
				0.0066\\\hline
				\multirow{3}{*}{RE$_\text{est}$} & OLS & 0.9666 & 0.7802 & 0.9311 &
				1.1610 & 1.2500 & 1.3162 & 1.3237\\
				& FGLS & 0.9824 & 0.7383 & 0.8057 & 0.7966 & 0.6489 & 0.6166 & 0.6115\\
				& FGLS-D & 0.9673 & 0.7484 & 0.8185 & 0.8160 & 0.7254 & 0.7078 &
				0.7102\\\hline
				&  &  &  &  &  &  &  & \\
				\multicolumn{9}{c}{\textbf{T=600}}\\\hline
				&  & $\theta=0$ & $\theta=0.3$ & $\theta=0.5$ & $\theta=0.7$ & $\theta=0.9$ &
				$\theta=0.95$ & $\theta=0.99$\\\hline
				\multirow{4}{*}{Bias} & OLS & 0.0004 & 0.0001 & 0.0003 & -0.0004 &
				0.0005 & 0.0001 & -0.0011\\
				& FGLS & 0.0004 & 0.0000 & 0.0006 & -0.0001 & 0.0002 & 0.0005 & -0.0006\\
				& FGLS-D & 0.0004 & 0.0000 & 0.0005 & -0.0002 & 0.0001 & 0.0005 & -0.0007\\
				& DURBIN & 0.0003 & 0.0001 & 0.0007 & 0.0000 & 0.0005 & 0.0005 &
				-0.0015\\\hline
				\multirow{4}{*}{MSE} & OLS & 0.0009 & 0.0013 & 0.0017 & 0.0021 & 0.0026 &
				0.0027 & 0.0029\\
				& FGLS & 0.0009 & 0.0012 & 0.0014 & 0.0014 & 0.0010 & 0.0009 & 0.0009\\
				& FGLS-D & 0.0009 & 0.0012 & 0.0014 & 0.0014 & 0.0012 & 0.0011 & 0.0011\\
				& DURBIN & 0.0009 & 0.0017 & 0.0017 & 0.0017 & 0.0018 & 0.0019 &
				0.0020\\\hline
				\multirow{3}{*}{RE$_\text{est}$} & OLS & 0.9762 & 0.7668 & 0.9680 &
				1.2116 & 1.4440 & 1.4420 & 1.4588\\
				& FGLS & 0.9805 & 0.7113 & 0.7987 & 0.7756 & 0.5660 & 0.4812 & 0.4708\\
				& FGLS-D & 0.9761 & 0.7112 & 0.8060 & 0.7905 & 0.6386 & 0.5750 &
				0.5733\\\hline
				&  &  &  &  &  &  &  & \\
				\multicolumn{9}{c}{\textbf{T=2500}}\\\hline
				&  & $\theta=0$ & $\theta=0.3$ & $\theta=0.5$ & $\theta=0.7$ & $\theta=0.9$ &
				$\theta=0.95$ & $\theta=0.99$\\\hline
				\multirow{4}{*}{Bias} & OLS & 0.0001 & 0.0003 & 0.0000 & 0.0001 &
				-0.0001 & 0.0002 & 0.0004\\
				& FGLS & 0.0001 & 0.0002 & 0.0000 & 0.0002 & -0.0001 & 0.0001 & 0.0001\\
				& FGLS-D & 0.0001 & 0.0002 & 0.0000 & 0.0002 & -0.0001 & 0.0002 & 0.0001\\
				& DURBIN & 0.0001 & 0.0002 & 0.0000 & 0.0000 & 0.0000 & 0.0003 &
				0.0001\\\hline
				\multirow{4}{*}{MSE} & OLS & 0.0002 & 0.0003 & 0.0004 & 0.0005 & 0.0006 &
				0.0007 & 0.0007\\
				& FGLS & 0.0002 & 0.0003 & 0.0003 & 0.0003 & 0.0002 & 0.0002 & 0.0001\\
				& FGLS-D & 0.0002 & 0.0003 & 0.0003 & 0.0003 & 0.0002 & 0.0002 & 0.0002\\
				& DURBIN & 0.0002 & 0.0004 & 0.0004 & 0.0004 & 0.0004 & 0.0004 &
				0.0004\\\hline
				\multirow{3}{*}{RE$_\text{est}$} & OLS & 0.9948 & 0.7708 & 0.9920 &
				1.2350 & 1.5383 & 1.5773 & 1.5744\\
				& FGLS & 0.9952 & 0.7127 & 0.8024 & 0.7610 & 0.4866 & 0.3710 & 0.3012\\
				& FGLS-D & 0.9948 & 0.7146 & 0.8034 & 0.7692 & 0.5248 & 0.4369 &
				0.3862\\\hline\hline
			\end{tabular}
		}
	\end{center}
	\par
	\begin{spacing}{1} \scriptsize
		Notes: The data-generating process is $y_{t}= x_{t}+u_{t}$, $x_{t}=0.7 x_{t-1} + \epsilon _{x,t}$, $u_{t}= \theta  \epsilon_{u,t-1} + \epsilon _{u,t}$,  $t=1, ..., T$.  All shocks are $N(0,1)$ white noise. We select  FGLS, FGLS-D, and DURBIN lag orders using BIC.  $\rm RE_{est}$  denotes the relative estimation efficiency of DURBIN.   We perform 10000 Monte Carlo replications, drawing $x_0$ and $u_0$ from their stationary distributions and using common random numbers whenever possible. See text for details.
	\end{spacing}
\end{table}

\thispagestyle{empty}
\begin{table}[h]
	\caption{Empirical Size of Nominal 5\% t-test of $H_{0}:\beta=1$\\	DGP: Moving Average Disturbances}%
	\label{tbl_size_MA}%
	\vspace{-0.5cm} 
	\renewcommand{\arraystretch}{1} 
	\par
	\begin{center}
		{\scriptsize
			\begin{tabular}
				[c]{llrrrrrrr}\hline\hline
				\multicolumn{9}{c}{\textbf{T=50}}\\\hline
				& Truncation & $\theta=0$ & $\theta=0.3$ & $\theta=0.5$ & $\theta=0.7$ &
				$\theta=0.9$ & $\theta=0.95$ & $\theta=0.99$\\\hline
				OLS & $-$ & 0.051 & 0.097 & 0.115 & 0.132 & 0.129 & 0.129 & 0.130\\
				NW & $h=\lceil4(T/100)^{2/9} \rceil$ & 0.066 & 0.082 & 0.085 & 0.094 & 0.092 &
				0.088 & 0.093\\
				NW-A & $h=\lceil0.75 T^{1/3} \rceil$ & 0.064 & 0.084 & 0.088 & 0.097 & 0.096 &
				0.093 & 0.095\\
				NW-LLSW & $h=\lceil1.3 T^{1/2} \rceil$ & 0.064 & 0.074 & 0.080 & 0.083 &
				0.082 & 0.082 & 0.086\\
				NW-KV & $h=T$ & 0.061 & 0.073 & 0.077 & 0.079 & 0.075 & 0.075 & 0.076\\
				M-LLSW & $\nu=\lfloor4(T/100)^{2/9} \rfloor$ & 0.065 & 0.072 & 0.072 & 0.073 &
				0.076 & 0.072 & 0.077\\
				FGLS & BIC & 0.067 & 0.076 & 0.082 & 0.097 & 0.101 & 0.099 & 0.104\\
				FGLS-D & BIC & 0.054 & 0.092 & 0.095 & 0.093 & 0.092 & 0.092 & 0.094\\
				DURBIN & BIC & 0.060 & 0.096 & 0.094 & 0.075 & 0.075 & 0.074 & 0.078\\
				DURBIN & AIC & 0.086 & 0.096 & 0.087 & 0.080 & 0.085 & 0.085 & 0.088\\\hline
				Hausman 1 & OLS vs FGLS &  & 0.753 & 0.680 & 0.574 & 0.477 & 0.466 & 0.449\\
				Hausman 2 & DURBIN vs FGLS-D &  & 0.047 & 0.073 & 0.088 & 0.097 & 0.099 &
				0.095\\\hline
				\multicolumn{9}{c}{\textbf{T=200}}\\\hline
				& Truncation & $\theta=0$ & $\theta=0.3$ & $\theta=0.5$ & $\theta=0.7$ &
				$\theta=0.9$ & $\theta=0.95$ & $\theta=0.99$\\\hline
				OLS & $-$ & 0.050 & 0.090 & 0.110 & 0.132 & 0.123 & 0.133 & 0.134\\
				NW & $h=\lceil4(T/100)^{2/9} \rceil$ & 0.054 & 0.060 & 0.065 & 0.070 & 0.064 &
				0.070 & 0.066\\
				NW-A & $h=\lceil0.75 T^{1/3} \rceil$ & 0.054 & 0.060 & 0.065 & 0.070 & 0.064 &
				0.070 & 0.066\\
				NW-LLSW & $h=\lceil1.3 T^{1/2} \rceil$ & 0.053 & 0.057 & 0.055 & 0.059 &
				0.056 & 0.066 & 0.062\\
				NW-KV & $h=T$ & 0.049 & 0.055 & 0.056 & 0.059 & 0.053 & 0.057 & 0.060\\
				M-LLSW & $\nu=\lfloor4(T/100)^{2/9} \rfloor$ & 0.056 & 0.057 & 0.056 & 0.059 &
				0.053 & 0.062 & 0.061\\
				FGLS & BIC & 0.052 & 0.053 & 0.059 & 0.063 & 0.056 & 0.057 & 0.059\\
				FGLS-D & BIC & 0.050 & 0.059 & 0.057 & 0.061 & 0.053 & 0.055 & 0.056\\
				DURBIN & BIC & 0.052 & 0.062 & 0.050 & 0.051 & 0.055 & 0.055 & 0.055\\
				DURBIN & AIC & 0.067 & 0.053 & 0.054 & 0.056 & 0.059 & 0.063 & 0.058\\\hline
				Hausman 1 & OLS vs FGLS &  & 0.666 & 0.545 & 0.385 & 0.224 & 0.208 & 0.199\\
				Hausman 2 & DURBIN vs FGLS-D &  & 0.049 & 0.055 & 0.059 & 0.059 & 0.064 &
				0.060\\\hline
				\multicolumn{9}{c}{\textbf{T=600}}\\\hline
				& Truncation & $\theta=0$ & $\theta=0.3$ & $\theta=0.5$ & $\theta=0.7$ &
				$\theta=0.9$ & $\theta=0.95$ & $\theta=0.99$\\\hline
				OLS & $-$ & 0.052 & 0.094 & 0.117 & 0.125 & 0.132 & 0.130 & 0.134\\
				NW & $h=\lceil4(T/100)^{2/9} \rceil$ & 0.054 & 0.057 & 0.063 & 0.066 & 0.060 &
				0.060 & 0.066\\
				NW-A & $h=\lceil0.75 T^{1/3} \rceil$ & 0.054 & 0.056 & 0.062 & 0.065 & 0.060 &
				0.058 & 0.065\\
				NW-LLSW & $h=\lceil1.3 T^{1/2} \rceil$ & 0.055 & 0.055 & 0.056 & 0.059 &
				0.055 & 0.054 & 0.057\\
				NW-KV & $h=T$ & 0.052 & 0.050 & 0.047 & 0.054 & 0.052 & 0.051 & 0.049\\
				M-LLSW & $\nu=\lfloor4(T/100)^{2/9} \rfloor$ & 0.055 & 0.053 & 0.055 & 0.057 &
				0.054 & 0.052 & 0.055\\
				FGLS & BIC & 0.053 & 0.047 & 0.058 & 0.053 & 0.047 & 0.040 & 0.043\\
				FGLS-D & BIC & 0.052 & 0.046 & 0.058 & 0.050 & 0.047 & 0.039 & 0.041\\
				DURBIN & BIC & 0.053 & 0.051 & 0.055 & 0.052 & 0.049 & 0.050 & 0.052\\
				DURBIN & AIC & 0.067 & 0.051 & 0.054 & 0.056 & 0.048 & 0.050 & 0.052\\\hline
				Hausman 1 & OLS vs FGLS &  & 0.588 & 0.421 & 0.242 & 0.101 & 0.095 & 0.088\\
				Hausman 2 & DURBIN vs FGLS-D &  & 0.050 & 0.051 & 0.049 & 0.048 & 0.054 &
				0.054\\\hline
				\multicolumn{9}{c}{\textbf{T=2500}}\\\hline
				& Truncation & $\theta=0$ & $\theta=0.3$ & $\theta=0.5$ & $\theta=0.7$ &
				$\theta=0.9$ & $\theta=0.95$ & $\theta=0.99$\\\hline
				OLS & $-$ & 0.046 & 0.098 & 0.115 & 0.132 & 0.133 & 0.131 & 0.127\\
				NW & $h=\lceil4(T/100)^{2/9} \rceil$ & 0.047 & 0.057 & 0.057 & 0.057 & 0.058 &
				0.059 & 0.056\\
				NW-A & $h=\lceil0.75 T^{1/3} \rceil$ & 0.046 & 0.056 & 0.057 & 0.055 & 0.058 &
				0.058 & 0.056\\
				NW-LLSW & $h=\lceil1.3 T^{1/2} \rceil$ & 0.047 & 0.053 & 0.054 & 0.052 &
				0.055 & 0.052 & 0.050\\
				NW-KV & $h=T$ & 0.050 & 0.049 & 0.046 & 0.050 & 0.050 & 0.047 & 0.048\\
				M-LLSW & $\nu=\lfloor4(T/100)^{2/9} \rfloor$ & 0.046 & 0.055 & 0.053 & 0.051 &
				0.054 & 0.051 & 0.050\\
				FGLS & BIC & 0.046 & 0.055 & 0.051 & 0.053 & 0.047 & 0.045 & 0.034\\
				FGLS-D & BIC & 0.046 & 0.053 & 0.051 & 0.053 & 0.044 & 0.039 & 0.035\\
				DURBIN & BIC & 0.046 & 0.049 & 0.050 & 0.050 & 0.048 & 0.051 & 0.050\\
				DURBIN & AIC & 0.058 & 0.049 & 0.050 & 0.049 & 0.049 & 0.051 & 0.048\\\hline
				Hausman 1 & OLS vs FGLS &  & 0.473 & 0.291 & 0.119 & 0.067 & 0.057 & 0.054\\
				Hausman 2 & DURBIN vs FGLS-D &  & 0.052 & 0.050 & 0.052 & 0.053 & 0.050 &
				0.053\\\hline\hline
			\end{tabular}
		}
	\end{center}
	\par
	\begin{spacing}{1} \footnotesize
		Notes: The data-generating process is $y_{t}= x_{t}+u_{t}$, $x_{t}=0.7 x_{t-1}+\protect\epsilon _{x,t}$, $u_{t}= \theta  \epsilon_{u,t-1}+\protect\epsilon _{u,t}$,  $t=1, ..., T$.  All shocks are $N(0,1)$ white noise.   We perform 10000 Monte Carlo replications, drawing $x_0$ and $u_0$ from their stationary distributions and using common random numbers whenever possible. See text for details.
	\end{spacing}
\end{table}

\begin{figure}[h]
	\caption{Empirical Rejection frequencies of Nominal 5\% t-Test of $\mathrm{H_{0}{:}~\beta{=}1}$ \\
		DGP: Moving Average Disturbances, $T=200$}
	\label{tbl_powerT_MA}
	\vspace{-0.5cm} 
	\renewcommand{\arraystretch}{1} 
	\begin{center}
		\includegraphics[trim=0 30 0 30,
		clip,	width=1.1\linewidth]{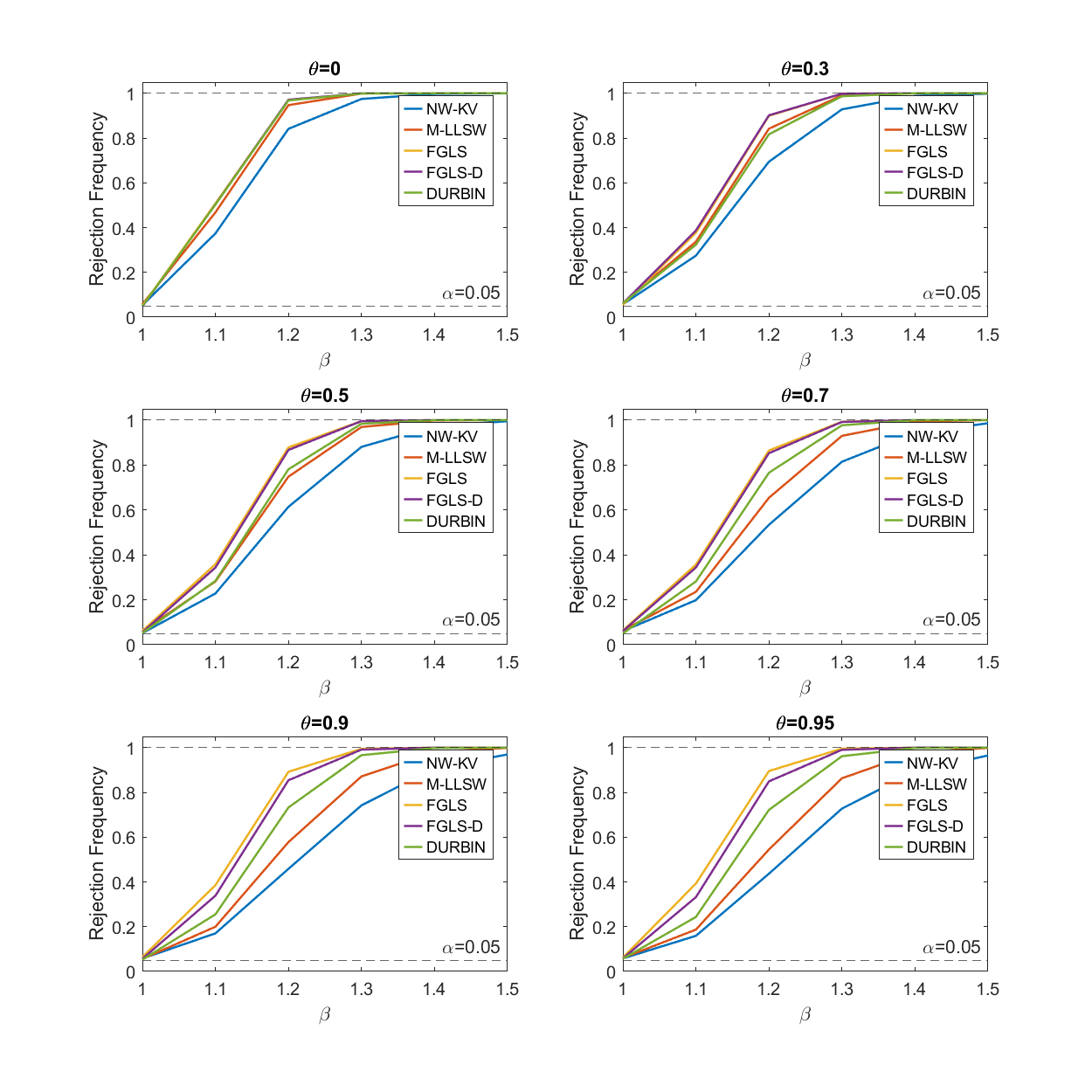}
	\end{center}
	\par
	\begin{spacing}{1} \footnotesize
		Notes:  The data-generating process is $y_{t}=\beta x_{t}+u_{t}$, $x_{t}=0.7 x_{t-1}+\protect\epsilon _{x,t}$, $u_{t}= \rho  \epsilon_{u,t-1}+\protect\epsilon_{u,t}$,  $t=1, ..., 200$.  All shocks are $N(0,1)$ white noise.   We perform 10000 Monte Carlo replications, drawing $x_0$ and $u_0$ from their stationary distributions and using common random numbers whenever possible. See text for details.
	\end{spacing}
\end{figure}

\begin{figure}[h]
	\caption{Empirical Rejection frequencies of Nominal 5\% t-Test of $\mathrm{H_{0}{:} ~\beta{=}1}$\\
		DGP: Moving Average Disturbances, $\theta=0.7$}%
	\label{tbl_powerRho_MA}
	\vspace{-0.5cm} 
	\renewcommand{\arraystretch}{1} 
	\begin{center}
		\includegraphics[trim=0 30 0 30,
		clip,	width=1.1\linewidth]{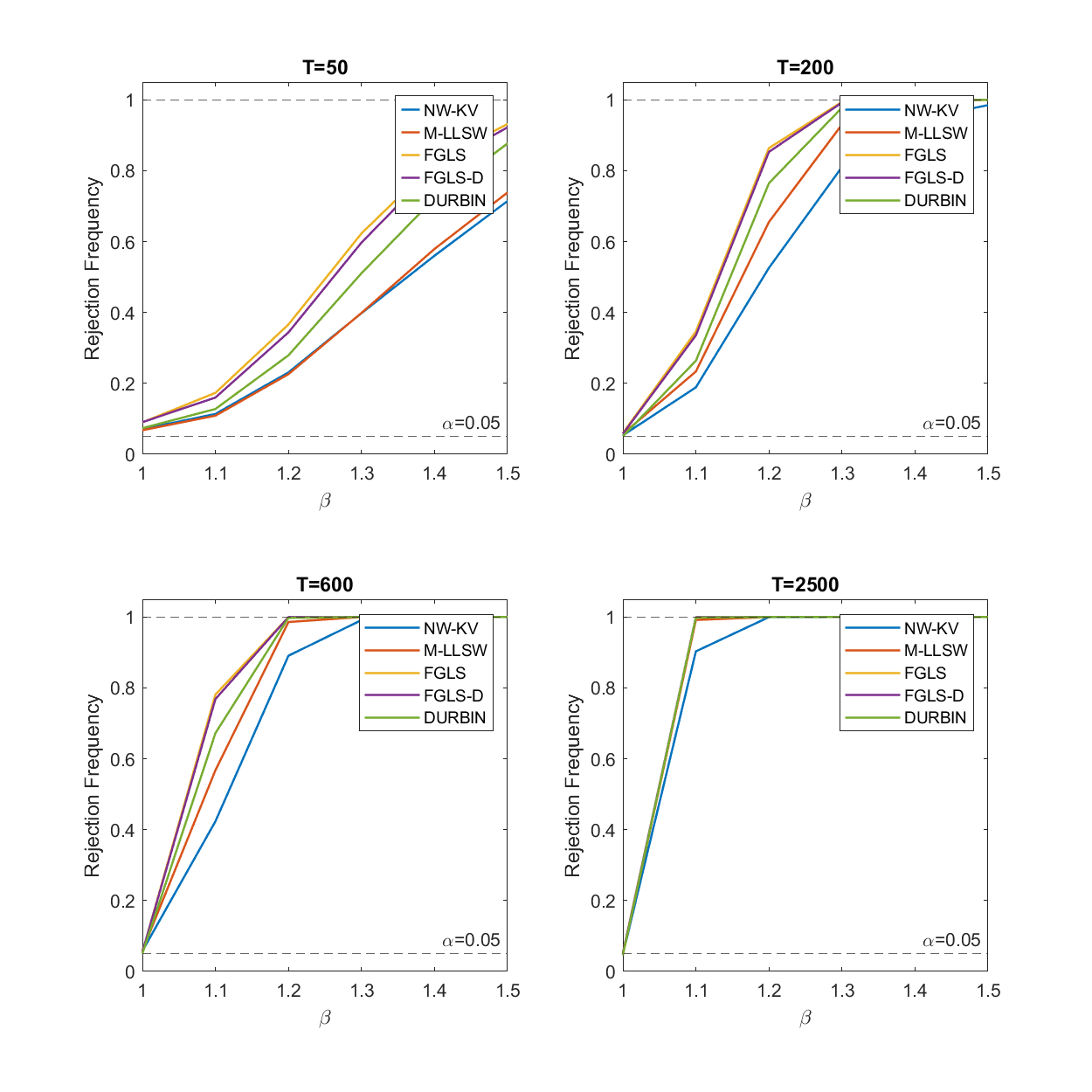}
	\end{center}
	\par
	\begin{spacing}{1} \footnotesize
		Notes:  The data-generating process is $y_{t}=\beta x_{t}+u_{t}$, $x_{t}=0.7 x_{t-1}+\protect\epsilon _{x,t}$, $u_{t}= 0.7  \epsilon_{u,t-1}+\protect\epsilon_{u,t}$,  $t=1, ..., T$.  All shocks are $N(0,1)$ white noise.   We perform 10000 Monte Carlo replications, drawing $x_0$ and $u_0$ from their stationary distributions and using common random numbers whenever possible. See text for details.
	\end{spacing}
\end{figure}

\clearpage

\section{Additional Monte Carlo: ARMA Disturbances}
\label{ARMA_BIC}

\renewcommand{\theequation}{C\arabic{equation}} \setcounter{equation}{0}
\renewcommand{\thefigure}{C\arabic{figure}} \setcounter{figure}{0}
\renewcommand{\thetable}{C\arabic{table}} \setcounter{table}{0}

\begin{table}[h]
	\caption{Selected Lags by test \\ Estimators: FGLS, DURBIN AIC, DURBIN BIC \\ DGP: ARMA Disturbances}%
	\label{tbl_lag_ARMA}%
	\vspace{-0.5cm} 
	\renewcommand{\arraystretch}{1} 
	\par
	\begin{center}
		{\footnotesize
			\begin{tabular}
				[c]{cllrrrrrrr}\hline\hline
				$T$ &  &  & $\theta=0$ & $\theta=0.3$ & $\theta=0.5$ & $\theta=0.7$ &
				$\theta=0.9$ & $\theta=0.95$ & $\theta=0.99$\\\hline
				\multirow{6}{*}{50} & \multirow{3}{*}{Median} & FGLS & 1 & 1 & 2 & 2 & 2 & 2 &
				1\\
				&  & DURBIN BIC & 1 & 1 & 2 & 2 & 2 & 3 & 3\\
				&  & DURBIN AIC & 1 & 2 & 3 & 4 & 6 & 6 & 6\\\cline{2-10}
				& \multirow{3}{*}{Mean} & FGLS & 1.2 & 1.6 & 2.1 & 2.6 & 2.9 & 2.7 & 2.0\\
				&  & DURBIN BIC & 1.1 & 1.3 & 1.7 & 2.3 & 2.9 & 3.0 & 3.1\\
				&  & DURBIN AIC & 3.0 & 3.5 & 4.1 & 5.0 & 6.0 & 6.2 & 6.3\\\hline
				\multirow{6}{*}{200} & \multirow{3}{*}{Median} & FGLS & 1 & 2 & 3 & 4 & 5 &
				5 & 4\\
				&  & DURBIN BIC & 1 & 2 & 2 & 3 & 4 & 4 & 4\\
				&  & DURBIN AIC & 1 & 2 & 3 & 5 & 9 & 10 & 11\\\cline{2-10}
				& \multirow{3}{*}{Mean} & FGLS & 1.1 & 2.0 & 2.6 & 3.7 & 4.9 & 4.9 & 3.9\\
				&  & DURBIN BIC & 1.0 & 1.6 & 2.3 & 3.1 & 4.2 & 4.4 & 4.5\\
				&  & DURBIN AIC & 2.2 & 3.4 & 4.4 & 6.3 & 10.2 & 11.6 & 12.5\\\hline
				\multirow{6}{*}{600} & \multirow{3}{*}{Median} & FGLS & 1 & 2 & 3 & 5 & 7 &
				8 & 7\\
				&  & DURBIN BIC & 1 & 2 & 3 & 4 & 6 & 7 & 7\\
				&  & DURBIN AIC & 1 & 3 & 4 & 6 & 12 & 15 & 17\\\cline{2-10}
				& \multirow{3}{*}{Mean} & FGLS & 1.0 & 2.2 & 3.3 & 4.9 & 7.5 & 8.0 & 7.1\\
				&  & DURBIN BIC & 1.0 & 2.0 & 2.9 & 4.2 & 6.3 & 6.8 & 7.0\\
				&  & DURBIN AIC & 1.7 & 3.3 & 4.7 & 7.0 & 13.1 & 16.1 & 18.0\\\hline
				\multirow{6}{*}{2500} & \multirow{3}{*}{Median} & FGLS & 1 & 3 & 4 & 7 & 12 &
				14 & 15\\
				&  & DURBIN BIC & 1 & 2 & 4 & 6 & 10 & 12 & 13\\
				&  & DURBIN AIC & 1 & 3 & 5 & 8 & 18 & 25 & 28\\\cline{2-10}
				& \multirow{3}{*}{Mean} & FGLS & 1.0 & 2.8 & 4.2 & 6.7 & 12.3 & 14.6 & 14.9\\
				&  & DURBIN BIC & 1.0 & 2.3 & 3.7 & 5.8 & 10.3 & 12.0 & 13.0\\
				&  & DURBIN AIC & 1.7 & 3.8 & 5.6 & 8.9 & 18.9 & 24.7 & 27.5\\\hline\hline
			\end{tabular}
		}
	\end{center}
	\par
	\begin{spacing}{1} \footnotesize
		Notes: The data-generating process is $y_{t}= x_{t}+u_{t}$, $x_{t}=0.7 x_{t-1}+\protect\epsilon _{x,t}$, $u_{t}=0.7  u_{t-1}+ \theta  \epsilon_{u,t-1}+\protect\epsilon _{u,t}$,  $t=1, ..., T$.  All shocks are $N(0,1)$ white noise.   We perform 10000 Monte Carlo replications, drawing $x_0$ and $u_0$ from their stationary distributions and using common random numbers whenever possible. See text for details.
	\end{spacing}
\end{table}

\begin{table}[h]
	\caption{Bias, MSE, and Relative Efficiency \\ Estimators: OLS, FGLS, FGLS-D, DURBIN \\ DGP: ARMA Disturbances}%
	\label{tbl_MSE_ARMA}%
	\vspace{-0.5cm} 
	\renewcommand{\arraystretch}{1} 
	\par
	\begin{center}
		{\scriptsize
			\begin{tabular}
				[c]{llrrrrrrr}\hline\hline
				\multicolumn{9}{c}{\textbf{T=50}}\\\hline
				&  & $\theta=0$ & $\theta=0.3$ & $\theta=0.5$ & $\theta=0.7$ & $\theta=0.9$ &
				$\theta=0.95$ & $\theta=0.99$\\\hline
				\multirow{4}{*}{Bias} & OLS & -0.0026 & -0.0017 & -0.0022 & 0.0006 &
				0.0023 & -0.0039 & -0.0234\\
				& FGLS & -0.0002 & 0.0009 & 0.0009 & 0.0018 & 0.0005 & -0.0046 & 0.0023\\
				& FGLS-D & 0.0001 & 0.0006 & 0.0008 & 0.0030 & -0.0006 & -0.0024 & 0.0021\\
				& DURBIN & 0.0003 & 0.0004 & 0.0002 & 0.0031 & 0.0001 & -0.0053 &
				0.0014\\\hline
				\multirow{4}{*}{MSE} & OLS & 0.0559 & 0.0945 & 0.1271 & 0.1698 & 0.3421 &
				0.8138 & 15.4933\\
				& FGLS & 0.0242 & 0.0250 & 0.0244 & 0.0232 & 0.0276 & 0.0358 & 0.1380\\
				& FGLS-D & 0.0235 & 0.0231 & 0.0219 & 0.0198 & 0.0191 & 0.0199 & 0.0200\\
				& DURBIN & 0.0234 & 0.0245 & 0.0268 & 0.0292 & 0.0349 & 0.0369 &
				0.0376\\\hline
				\multirow{3}{*}{RE$_\text{est}$} & OLS & 2.3893 & 3.8547 & 4.7469 &
				5.8198 & 9.8056 & 22.0441 & 411.8199\\
				& FGLS & 1.0369 & 1.0207 & 0.9108 & 0.7964 & 0.7900 & 0.9687 & 3.6672\\
				& FGLS-D & 1.0059 & 0.9441 & 0.8198 & 0.6778 & 0.5486 & 0.5383 &
				0.5310\\\hline
				&  &  &  &  &  &  &  & \\
				\multicolumn{9}{c}{\textbf{T=200}}\\\hline
				&  & $\theta=0$ & $\theta=0.3$ & $\theta=0.5$ & $\theta=0.7$ & $\theta=0.9$ &
				$\theta=0.95$ & $\theta=0.99$\\\hline
				\multirow{4}{*}{Bias} & OLS & 0.0002 & 0.0016 & -0.0029 & -0.0022 &
				-0.0004 & -0.0032 & -0.0024\\
				& FGLS & -0.0006 & 0.0009 & -0.0010 & -0.0007 & -0.0001 & -0.0003 & -0.0005\\
				& FGLS-D & -0.0006 & 0.0008 & -0.0009 & -0.0003 & 0.0000 & 0.0001 & -0.0004\\
				& DURBIN & -0.0006 & 0.0007 & -0.0008 & -0.0004 & -0.0006 & 0.0010 &
				-0.0006\\\hline
				\multirow{4}{*}{MSE} & OLS & 0.0146 & 0.0238 & 0.0318 & 0.0420 & 0.0591 &
				0.0932 & 1.0482\\
				& FGLS & 0.0051 & 0.0048 & 0.0042 & 0.0034 & 0.0026 & 0.0029 & 0.0046\\
				& FGLS-D & 0.0051 & 0.0048 & 0.0042 & 0.0034 & 0.0026 & 0.0027 & 0.0027\\
				& DURBIN & 0.0051 & 0.0052 & 0.0053 & 0.0057 & 0.0060 & 0.0066 &
				0.0068\\\hline
				\multirow{3}{*}{RE$_\text{est}$} & OLS & 2.8789 & 4.5810 & 5.9923 &
				7.3801 & 9.7832 & 14.1853 & 155.2254\\
				& FGLS & 1.0071 & 0.9277 & 0.7980 & 0.5970 & 0.4372 & 0.4432 & 0.6771\\
				& FGLS-D & 1.0026 & 0.9303 & 0.7937 & 0.5921 & 0.4311 & 0.4053 &
				0.3961\\\hline
				&  &  &  &  &  &  &  & \\
				\multicolumn{9}{c}{\textbf{T=600}}\\\hline
				&  & $\theta=0$ & $\theta=0.3$ & $\theta=0.5$ & $\theta=0.7$ & $\theta=0.9$ &
				$\theta=0.95$ & $\theta=0.99$\\\hline
				\multirow{4}{*}{Bias} & OLS & -0.0007 & 0.0007 & -0.0012 & -0.0007 &
				-0.0005 & 0.0004 & -0.0019\\
				& FGLS & 0.0007 & 0.0000 & -0.0001 & 0.0001 & 0.0000 & 0.0003 & -0.0003\\
				& FGLS-D & 0.0007 & 0.0000 & 0.0000 & 0.0002 & -0.0001 & 0.0004 & -0.0003\\
				& DURBIN & 0.0007 & -0.0001 & 0.0002 & -0.0001 & 0.0002 & 0.0007 &
				-0.0006\\\hline
				\multirow{4}{*}{MSE} & OLS & 0.0049 & 0.0082 & 0.0106 & 0.0135 & 0.0178 &
				0.0227 & 0.1273\\
				& FGLS & 0.0016 & 0.0015 & 0.0013 & 0.0009 & 0.0006 & 0.0005 & 0.0007\\
				& FGLS-D & 0.0016 & 0.0015 & 0.0013 & 0.0010 & 0.0006 & 0.0006 & 0.0005\\
				& DURBIN & 0.0016 & 0.0017 & 0.0017 & 0.0017 & 0.0019 & 0.0019 &
				0.0020\\\hline
				\multirow{3}{*}{RE$_\text{est}$} & OLS & 2.9909 & 4.8906 & 6.1570 &
				7.8409 & 9.6024 & 11.9589 & 63.9303\\
				& FGLS & 0.9998 & 0.9163 & 0.7628 & 0.5497 & 0.3109 & 0.2779 & 0.3544\\
				& FGLS-D & 0.9988 & 0.9152 & 0.7705 & 0.5623 & 0.3314 & 0.2911 &
				0.2746\\\hline
				&  &  &  &  &  &  &  & \\
				\multicolumn{9}{c}{\textbf{T=2500}}\\\hline
				&  & $\theta=0$ & $\theta=0.3$ & $\theta=0.5$ & $\theta=0.7$ & $\theta=0.9$ &
				$\theta=0.95$ & $\theta=0.99$\\\hline
				\multirow{4}{*}{Bias} & OLS & 0.0002 & 0.0006 & 0.0004 & 0.0002 &
				-0.0005 & -0.0003 & 0.0020\\
				& FGLS & -0.0003 & 0.0003 & 0.0000 & 0.0002 & 0.0001 & 0.0000 & 0.0000\\
				& FGLS-D & -0.0003 & 0.0003 & 0.0000 & 0.0002 & 0.0001 & 0.0000 & 0.0000\\
				& DURBIN & -0.0003 & 0.0004 & 0.0000 & 0.0001 & 0.0002 & 0.0002 &
				0.0002\\\hline
				\multirow{4}{*}{MSE} & OLS & 0.0011 & 0.0019 & 0.0026 & 0.0033 & 0.0041 &
				0.0044 & 0.0111\\
				& FGLS & 0.0004 & 0.0004 & 0.0003 & 0.0002 & 0.0001 & 0.0001 & 0.0001\\
				& FGLS-D & 0.0004 & 0.0004 & 0.0003 & 0.0002 & 0.0001 & 0.0001 & 0.0001\\
				& DURBIN & 0.0004 & 0.0004 & 0.0004 & 0.0004 & 0.0004 & 0.0004 &
				0.0004\\\hline
				\multirow{3}{*}{RE$_\text{est}$} & OLS & 2.8997 & 4.6975 & 6.4698 &
				7.9824 & 9.8683 & 10.4866 & 25.2690\\
				& FGLS & 0.9991 & 0.9153 & 0.7524 & 0.5116 & 0.2320 & 0.1695 & 0.1637\\
				& FGLS-D & 0.9989 & 0.9183 & 0.7556 & 0.5188 & 0.2460 & 0.1876 &
				0.1657\\\hline\hline
			\end{tabular}
		}
	\end{center}
	\par
	\begin{spacing}{1} \scriptsize
		Notes: The data-generating process is $y_{t}= x_{t}+u_{t}$, $x_{t}=0.7 x_{t-1} + \epsilon _{x,t}$, $u_{t}=0.7  u_{t-1}+ \theta  \epsilon_{u,t-1} + \epsilon _{u,t}$,  $t=1, ..., T$. All shocks are $N(0,1)$ white noise. We select  FGLS, FGLS-D, and DURBIN lag orders using BIC.  $\rm RE_{est}$  denotes the relative estimation efficiency of DURBIN. We perform 10000 Monte Carlo replications, drawing $x_0$ and $u_0$ from their stationary distributions and using common random numbers whenever possible. 
	\end{spacing}
\end{table}

\begin{table}[h]
	\caption{Empirical Size of Nominal 5\% t-test of $H_{0}:\beta=1$\\	DGP: ARMA Disturbances}%
	\label{tbl_size_ARMA}%
	\vspace{-0.5cm} 
	\renewcommand{\arraystretch}{1} 
	\par
	\begin{center}
		{\scriptsize
			\begin{tabular}
				[c]{llrrrrrrr}\hline\hline
				\multicolumn{9}{c}{\textbf{T=50}}\\\hline
				& Truncation & $\theta=0$ & $\theta=0.3$ & $\theta=0.5$ & $\theta=0.7$ &
				$\theta=0.9$ & $\theta=0.95$ & $\theta=0.99$\\\hline
				OLS & $-$ & 0.237 & 0.263 & 0.277 & 0.279 & 0.291 & 0.299 & 0.325\\
				NW & $h=\lceil4(T/100)^{2/9} \rceil$ & 0.136 & 0.143 & 0.151 & 0.150 & 0.136 &
				0.119 & 0.045\\
				NW-A & $h=\lceil0.75 T^{1/3} \rceil$ & 0.154 & 0.165 & 0.172 & 0.172 & 0.164 &
				0.148 & 0.065\\
				NW-LLSW & $h=\lceil1.3 T^{1/2} \rceil$ & 0.103 & 0.107 & 0.109 & 0.111 &
				0.090 & 0.071 & 0.021\\
				NW-KV & $h=T$ & 0.088 & 0.096 & 0.097 & 0.093 & 0.073 & 0.052 & 0.013\\
				M-LLSW & $\nu=\lfloor4(T/100)^{2/9} \rfloor$ & 0.081 & 0.084 & 0.088 & 0.084 &
				0.061 & 0.045 & 0.014\\
				FGLS & BIC & 0.076 & 0.084 & 0.086 & 0.081 & 0.081 & 0.081 & 0.104\\
				FGLS-D & BIC & 0.068 & 0.068 & 0.068 & 0.065 & 0.062 & 0.060 & 0.061\\
				DURBIN & BIC & 0.055 & 0.057 & 0.063 & 0.062 & 0.074 & 0.073 & 0.073\\
				DURBIN & AIC & 0.074 & 0.078 & 0.082 & 0.075 & 0.088 & 0.085 & 0.081\\\hline
				Hausman 1 & OLS vs FGLS &  & 0.312 & 0.250 & 0.213 & 0.164 & 0.120 & 0.038\\
				Hausman 2 & DURBIN vs FGLS-D &  & 0.129 & 0.124 & 0.105 & 0.104 & 0.104 &
				0.107\\\hline
				\multicolumn{9}{c}{\textbf{T=200}}\\\hline
				& Truncation & $\theta=0$ & $\theta=0.3$ & $\theta=0.5$ & $\theta=0.7$ &
				$\theta=0.9$ & $\theta=0.95$ & $\theta=0.99$\\\hline
				OLS & $-$ & 0.248 & 0.269 & 0.274 & 0.285 & 0.283 & 0.296 & 0.304\\
				NW & $h=\lceil4(T/100)^{2/9} \rceil$ & 0.106 & 0.105 & 0.112 & 0.114 & 0.102 &
				0.101 & 0.042\\
				NW-A & $h=\lceil0.75 T^{1/3} \rceil$ & 0.106 & 0.105 & 0.112 & 0.114 & 0.102 &
				0.101 & 0.042\\
				NW-LLSW & $h=\lceil1.3 T^{1/2} \rceil$ & 0.074 & 0.074 & 0.077 & 0.078 &
				0.068 & 0.063 & 0.022\\
				NW-KV & $h=T$ & 0.065 & 0.062 & 0.063 & 0.061 & 0.056 & 0.045 & 0.016\\
				M-LLSW & $\nu=\lfloor4(T/100)^{2/9} \rfloor$ & 0.066 & 0.066 & 0.068 & 0.067 &
				0.053 & 0.045 & 0.017\\
				FGLS & BIC & 0.051 & 0.053 & 0.051 & 0.049 & 0.038 & 0.040 & 0.039\\
				FGLS-D & BIC & 0.051 & 0.051 & 0.048 & 0.045 & 0.037 & 0.040 & 0.039\\
				DURBIN & BIC & 0.049 & 0.048 & 0.050 & 0.056 & 0.052 & 0.056 & 0.055\\
				DURBIN & AIC & 0.053 & 0.053 & 0.055 & 0.057 & 0.058 & 0.060 & 0.061\\\hline
				Hausman 1 & OLS vs FGLS &  & 0.140 & 0.123 & 0.108 & 0.095 & 0.087 & 0.034\\
				Hausman 2 & DURBIN vs FGLS-D &  & 0.076 & 0.052 & 0.058 & 0.063 & 0.060 &
				0.060\\\hline
				\multicolumn{9}{c}{\textbf{T=600}}\\\hline
				& Truncation & $\theta=0$ & $\theta=0.3$ & $\theta=0.5$ & $\theta=0.7$ &
				$\theta=0.9$ & $\theta=0.95$ & $\theta=0.99$\\\hline
				OLS & $-$ & 0.257 & 0.278 & 0.276 & 0.276 & 0.281 & 0.298 & 0.298\\
				NW & $h=\lceil4(T/100)^{2/9} \rceil$ & 0.086 & 0.090 & 0.088 & 0.087 & 0.089 &
				0.091 & 0.052\\
				NW-A & $h=\lceil0.75 T^{1/3} \rceil$ & 0.081 & 0.084 & 0.083 & 0.082 & 0.082 &
				0.086 & 0.047\\
				NW-LLSW & $h=\lceil1.3 T^{1/2} \rceil$ & 0.058 & 0.064 & 0.061 & 0.061 &
				0.058 & 0.060 & 0.030\\
				NW-KV & $h=T$ & 0.054 & 0.054 & 0.051 & 0.053 & 0.048 & 0.045 & 0.019\\
				M-LLSW & $\nu=\lfloor4(T/100)^{2/9} \rfloor$ & 0.054 & 0.058 & 0.056 & 0.058 &
				0.051 & 0.047 & 0.020\\
				FGLS & BIC & 0.046 & 0.050 & 0.049 & 0.045 & 0.038 & 0.033 & 0.024\\
				FGLS-D & BIC & 0.046 & 0.049 & 0.047 & 0.044 & 0.038 & 0.034 & 0.031\\
				DURBIN & BIC & 0.046 & 0.050 & 0.050 & 0.050 & 0.050 & 0.050 & 0.053\\
				DURBIN & AIC & 0.047 & 0.051 & 0.049 & 0.051 & 0.052 & 0.049 & 0.054\\\hline
				Hausman 1 & OLS vs FGLS &  & 0.089 & 0.082 & 0.074 & 0.071 & 0.073 & 0.037\\
				Hausman 2 & DURBIN vs FGLS-D &  & 0.052 & 0.050 & 0.052 & 0.054 & 0.056 &
				0.058\\\hline
				\multicolumn{9}{c}{\textbf{T=2500}}\\\hline
				& Truncation & $\theta=0$ & $\theta=0.3$ & $\theta=0.5$ & $\theta=0.7$ &
				$\theta=0.9$ & $\theta=0.95$ & $\theta=0.99$\\\hline
				OLS & $-$ & 0.247 & 0.273 & 0.284 & 0.284 & 0.282 & 0.278 & 0.300\\
				NW & $h=\lceil4(T/100)^{2/9} \rceil$ & 0.067 & 0.067 & 0.076 & 0.072 & 0.074 &
				0.068 & 0.057\\
				NW-A & $h=\lceil0.75 T^{1/3} \rceil$ & 0.063 & 0.064 & 0.071 & 0.068 & 0.069 &
				0.064 & 0.053\\
				NW-LLSW & $h=\lceil1.3 T^{1/2} \rceil$ & 0.051 & 0.051 & 0.056 & 0.056 &
				0.053 & 0.050 & 0.043\\
				NW-KV & $h=T$ & 0.045 & 0.050 & 0.053 & 0.052 & 0.050 & 0.045 & 0.033\\
				M-LLSW & $\nu=\lfloor4(T/100)^{2/9} \rfloor$ & 0.048 & 0.049 & 0.054 & 0.053 &
				0.052 & 0.045 & 0.029\\
				FGLS & BIC & 0.047 & 0.050 & 0.050 & 0.047 & 0.038 & 0.029 & 0.026\\
				FGLS-D & BIC & 0.047 & 0.050 & 0.049 & 0.047 & 0.039 & 0.033 & 0.032\\
				DURBIN & BIC & 0.047 & 0.050 & 0.051 & 0.052 & 0.048 & 0.051 & 0.051\\
				DURBIN & AIC & 0.047 & 0.051 & 0.052 & 0.053 & 0.049 & 0.050 & 0.050\\\hline
				Hausman 1 & OLS vs FGLS &  & 0.063 & 0.067 & 0.065 & 0.060 & 0.056 & 0.047\\
				Hausman 2 & DURBIN vs FGLS-D &  & 0.051 & 0.050 & 0.053 & 0.051 & 0.054 &
				0.055\\\hline\hline
			\end{tabular}
		}
	\end{center}
	\par
	\begin{spacing}{1} \footnotesize
		Notes: The data-generating process is $y_{t}= x_{t}+u_{t}$, $x_{t}=0.7 x_{t-1}+\protect\epsilon _{x,t}$, $u_{t}= 0.7  u_{t-1}+ \theta  \epsilon_{u,t-1}+\protect\epsilon _{u,t}$,  $t=1, ..., T$.  All shocks are $N(0,1)$ white noise.   We perform 10000 Monte Carlo replications, drawing $x_0$ and $u_0$ from their stationary distributions and using common random numbers whenever possible. See text for details.
	\end{spacing}
\end{table}

\begin{figure}[h]
	\caption{Empirical Rejection Frequencies of Nominal 5\% t-Test of $\mathrm{H_{0}{:}~\beta{=}1}$,\\
		DGP: ARMA Disturbances, $T=200$}
	\label{tbl_powerT_ARMA}
	\vspace{-0.5cm} 
	\renewcommand{\arraystretch}{1} 
	\begin{center}
		\includegraphics[trim=0 30 0 30,
		clip,	width=1.1\linewidth]{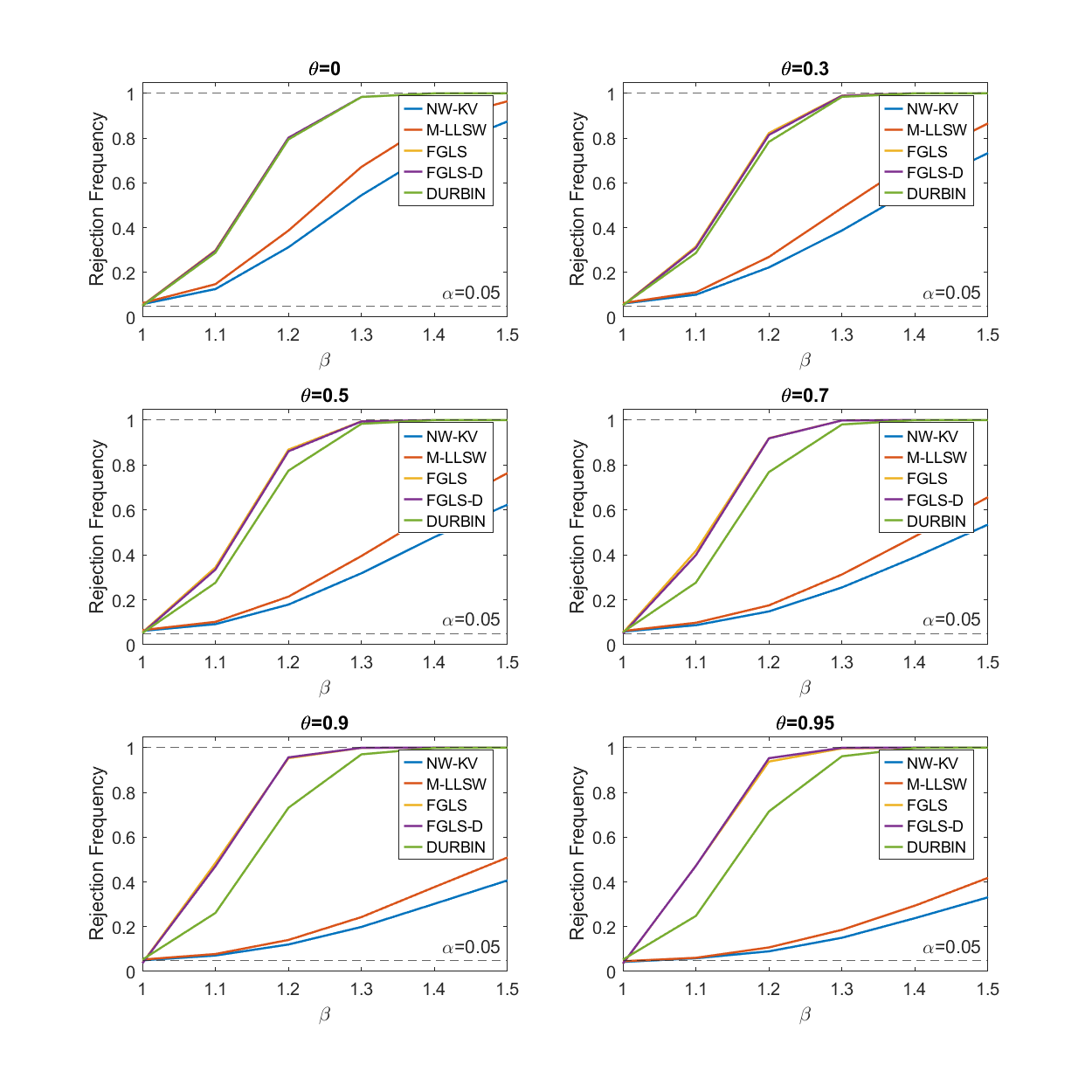}
	\end{center}
	\par
	\begin{spacing}{1} \footnotesize
		Notes:  The data-generating process is $y_{t}=\beta x_{t}+u_{t}$, $x_{t}=0.7 x_{t-1}+\protect\epsilon _{x,t}$, $u_{t}=0.7  u_{t-1}+ \theta  \epsilon_{u,t-1}+\protect\epsilon_{u,t}$,  $t=1, ..., 200$.  All shocks are $N(0,1)$ white noise.   We perform 10000 Monte Carlo replications, drawing $x_0$ and $u_0$ from their stationary distributions and using common random numbers whenever possible. See text for details.
	\end{spacing}
\end{figure}
\begin{figure}[h]
	\caption{Empirical Rejection Frequencies of Nominal 5\% t-Test of $\mathrm{H_{0}{:} ~\beta{=}1}$,\\
		DGP: ARMA Disturbances, $\theta=0.5$}%
	\label{tbl_powerRho_ARMA}
	\vspace{-0.5cm} 
	\renewcommand{\arraystretch}{1} 
	\begin{center}
		\includegraphics[trim=0 30 0 30,
		clip,	width=1.1\linewidth]{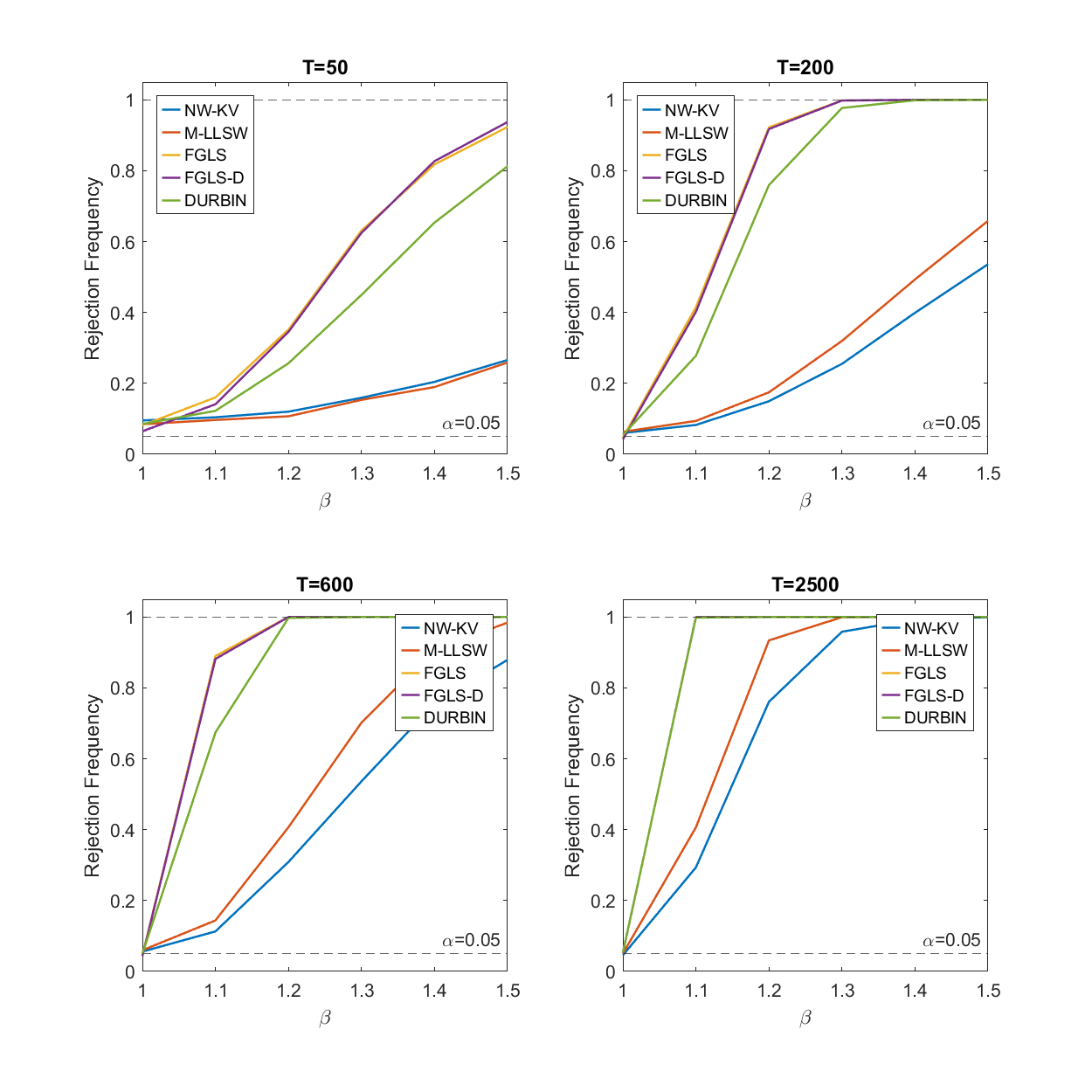}
	\end{center}
	\par
	\begin{spacing}{1} \footnotesize
		Notes:  The data-generating process is $y_{t}=\beta x_{t}+u_{t}$, $x_{t}=0.7 x_{t-1}+\protect\epsilon _{x,t}$, $u_{t}= 0.7  u_{t-1}+ 0.5  \epsilon_{u,t-1}+\protect\epsilon_{u,t}$,  $t=1, ..., T$.  All shocks are $N(0,1)$ white noise.   We perform 10000 Monte Carlo replications, drawing $x_0$ and $u_0$ from their stationary distributions and using common random numbers whenever possible. See text for details.
	\end{spacing}
\end{figure}

\clearpage

\addcontentsline{toc}{section}{References}
\bibliographystyle{Diebold}
\bibliography{Bibliography}

@article{RW2017,
  title={Resurrecting Weighted Least Squares},
  author={Romano, J.P. and Wolf, M.},
	journal = {Journal of Econometrics},
  volume={197},
  pages={1--19},
  year={2017}
}

@article{dav2002,
  title={Establishing Conditions for the Functional Central Limit Theorem in Nonlinear and Semiparametric Time Series Processes},
  author={Davidson, J.},
  journal = {Journal of Econometrics},
  volume={106},
  pages={243--269},
  year={2002}
}

@article{Baillie1979,
  title={The Asymptotic Mean Squared Error of Multistep Prediction from the Regression Model with autoregressive Errors},
  author={Baillie, R.T.},
  journal = {Journal of the American Statistical Association},
  volume={74},
  pages={175-184},
  year={1979}
}

@article{White1980,
  title={A Heteroskedasticity-Consistent Covariance Matrix Estimator and a Direct Test for Heteroskedasticity},
  author={White, H.},
  journal={Econometrica},
  volume={48},
  pages={817-838},
  year={1980}
}

@article{Plagborg,
author = {Montiel Olea, J.L. and Plagborg-Møller, M.},
title = {Local Projection Inference Is Simpler and More Robust Than You Think},
journal = {Econometrica},
volume = {89},
number = {4},
pages = {1789-1823},
doi = {https://doi.org/10.3982/ECTA18756},
url = {https://onlinelibrary.wiley.com/doi/abs/10.3982/ECTA18756},
eprint = {https://onlinelibrary.wiley.com/doi/pdf/10.3982/ECTA18756},
year = {2021}
}

@article{glsp,
  title={Feasible GLS for Time Series Regression},
  author={Perron, P. and González-Coya, E.},
  note={Manuscript, Department of Economics, Boston University},
  year={2022},
}

@article{amemiya1973generalized,
  title={Generalized Least Squares with an Estimated Autocovariance Matrix},
  author={Amemiya, T.},
  journal={Econometrica},
  volume={41},
  number={4},
  pages={723--732},
  year={1973},
  publisher={JSTOR}
}

@book{grenander1981abstract,
  title={Abstract Inference},
  author={Grenander, U.},
  year={1981},
  publisher={Wiley, New York}
}

@book{hannan2012statistical,
  title={The Statistical Theory Of Linear Systems},
  author={Hannan, E.J. and Deistler, M.},
  year={1988},
  publisher={Wiley}
}

@article{hansen1980forward,
  title={Forward Exchange Rates as Optimal Predictors of Future Spot Rates: An Econometric Analysis},
  author={Hansen, L.P. and Hodrick, R.J.},
  journal={Journal of Political Economy},
  volume={88},
  number={5},
  pages={829--853},
  year={1980},
  publisher={The University of Chicago Press}
}

@article{kapetanios2016semiparametric,
  title={Semiparametric Sieve-Type Generalized Least Squares Inference},
  author={Kapetanios, G. and Psaradakis, Z.},
  journal={Econometric Reviews},
  volume={35},
  number={6},
  pages={951--985},
  year={2016},
  publisher={Taylor \& Francis}
}

@article{kiefer2002heteroskedasticity,
  title={Heteroskedasticity-Autocorrelation Robust Standard Errors using the Bartlett Kernel without Truncation},
  author={Kiefer, N.M. and Vogelsang, T.J.},
  journal={Econometrica},
  volume={70},
  number={5},
  pages={2093--2095},
  year={2002},
  publisher={JSTOR}
}

@article{lazarus2018har,
  title={HAR Inference: Recommendations For Practice},
  author={Lazarus, E. and Lewis, D.J. and Stock, J.H. and Watson, M.W.},
  journal={Journal of Business and Economic Statistics},
  volume={36},
  number={4},
  pages={541--559},
  year={2018},
  publisher={Taylor \& Francis}
}

@article{muller2014hac,
  title={HAC Corrections for Strongly Autocorrelated Time Series},
  author={M{\"u}ller, U.K.},
  journal={Journal of Business and Economic Statistics},
  volume={32},
  number={3},
  pages={311--322},
  year={2014},
  publisher={Taylor \& Francis}
}

@article{newey1987simple,
  title={A Simple, Positive Semi-Definite, Heteroskedasticity and Autocorrelation Consistent Covariance Matrix},
  author={Newey, W.K. and West, K.D.},
  journal={Econometrica},
  volume={55},
  number={3},
  pages={703--708},
  year={1987},
  publisher={Econometric Society}
}

@article{kiefer2000simple,
  title={Simple Robust Testing of Regression Hypotheses},
  author={Kiefer, N.M. and Vogelsang, T.J. and Bunzel, H.},
  journal={Econometrica},
  volume={68},
  number={3},
  pages={695--714},
  year={2000},
  publisher={JSTOR}
}

@article{andrews1991heteroskedasticity,
  title={Heteroskedasticity and Autocorrelation Consistent Covariance Matrix Estimation},
  author={Andrews, D.W.K.},
  journal={Econometrica},
  pages={817--858},
  year={1991},
  publisher={JSTOR}
}

@book{wooldridge2015introductory,
  title={Introductory Econometrics: A Modern Approach},
  author={Wooldridge, J.M.},
  year={2015},
  publisher={Cengage Learning}
}

@book{angrist2008mostly,
  title={Mostly Harmless Econometrics: An Empiricist's Companion},
  author={Angrist, J.D. and Pischke, J.-S.},
  year={2008},
  publisher={Princeton University Press}
}

@article{muller2022spatial,
  title={Spatial Correlation Robust Inference},
  author={M{\"u}ller, U.K. and Watson, M.W.},
  journal={Econometrica},
  volume={90},
  number={6},
  pages={2901--2935},
  year={2022},
  publisher={Wiley Online Library}
}

@article{HendryMizon1978,
  title={Serial Correlation as a Convenient Simplification, Not a Nuisance: A Comment on a Study of the Demand for Money by the Bank of England},
  author={Hendry, D.F. and Mizon, G.E.},
  journal={The Economic Journal},
  volume={88},
  number={351},
  pages={549--563},
  year={1978},
  publisher={JSTOR}
}

@article{Sargan1964,
  title={Wages and Prices in the United Kingdom: A Study in Econometric Methodology},
  author={Sargan, J.D.},
  note={in P.E. Hart, G. Mills and J.K. Whitaker (eds.), \textit{Econometric Analysis for National Economic Planning}, Butterworths, New York, 25-54},
  year={1964}
}

@article{neweyWest1994,
  title={Automatic Lag Selection in Covariance Matrix Estimation},
  author={Newey, W.K. and West, K.D.},
  journal={Review of Economic Studies},
  volume={61},
  number={4},
  pages={631--653},
  year={1994},
  publisher={Wiley-Blackwell}
}

@article{andrewsMonahan1992,
  title={An Improved Heteroskedasticity and Autocorrelation Consistent Covariance Matrix Estimator},
  author={Andrews, D.W.K. and Monahan, J.C.},
  journal={Econometrica},
  volume={60},
  number={4},
  pages={953--966},
  year={1992},
  publisher={JSTOR}
}

@article{durbin1970testing,
  title={Testing for Serial Correlation in Least-Squares Regression when some of the Regressors are Lagged Dependent Variables},
  author={Durbin, J.},
  journal={Econometrica},
  pages={410--421},
  year={1970},
  publisher={JSTOR}
}

@article{lewis1985prediction,
  title={Prediction of Multivariate Time Series by Autoregressive Model Fitting},
  author={Lewis, R. and Reinsel, G.C.},
  journal={Journal of Multivariate Analysis},
  volume={16},
  number={3},
  pages={393--411},
  year={1985},
  publisher={Elsevier}
}

@article{baillie2023new,
  title={A New Test for Market Efficiency and Uncovered Interest Parity},
  author={Baillie, R.T. and Diebold, F.X. and Kapetanios, G. and Kim, K.H.},
  journal={Journal of International Money and Finance},
  volume={130},
  pages={102765},
  year={2023},
  publisher={Elsevier}
}

@article{anna,
  title={Linear Regression with Weak Exogeneity},
  author={Mikusheva, A. and S{\o}lvsten, M.},
  note={Working Paper at arXiv:2308.08958},
  year={2023}
}

\end{document}